\documentclass[bibyear]{aa}
\usepackage{psfig}
\usepackage{graphicx,lscape,longtable}
\usepackage{rotating}
\usepackage{natbib}
\begin{document}

\title{Accurate classification of 75 counterparts of objects detected in the 54 month Palermo
Swift/BAT hard X-ray catalogue\thanks{
Based on observations obtained from the following observatories:  
Astronomical Observatory of Bologna in Loiano (Italy);
Observatorio Astron\'omico Nacional (San Pedro
M\'artir, Mexico), Astronomical Observatory of Asiago (Italy), Cerro Tololo Interamerican Observatory (Chile).}}

%\subtitle{}

\author{P. Parisi\inst{1}, N. Masetti\inst{2}, A.F. Rojas\inst{3}, E. Jim\'enez-Bail\'on\inst{4}, V. Chavushyan\inst{5}, E. Palazzi\inst{2},
L. Bassani\inst{2}, A. Bazzano\inst{1}, A.J. Bird\inst{6}, G. Galaz\inst{3}, D. Minniti\inst{3,}\inst{7,}\inst{8}, L. Morelli\inst{9,}\inst{10} and
P. Ubertini\inst{1}
}
\institute{
Istituto di Astrofisica e Planetologia Spaziali (INAF), Via Fosso del Cavaliere 100, Roma I-00133, Italy
\and
INAF -- Istituto di Astrofisica Spaziale e Fisica Cosmica di 
Bologna, Via Gobetti 101, I-40129 Bologna, Italy
\and
Instituto de Astrofisica y Astrof\'{i}sica, Facultad de Fisica, Pontificia Universidad 
Cat\'olica de Chile, Casilla 306, Santiago 22, Chile   
\and 
Universidad Nacional Aut\'onoma de M\'exico,
Apartado Postal 70-264, 04510 M\'exico D.F., M\'exico
\and
Instituto Nacional de Astrof\'{i}sica, \'Optica y Electr\'onica,
Apartado Postal 51-216, 72000 Puebla, M\'exico
\and
School of Physics \& Astronomy, University of Southampton, Southampton, Hampshire, SO17 1BJ, UK    
\and
Specola Vaticana, V-00120 Citt\`a del Vaticano
\and
Departamento de Ciencias F\'isicas, Universidad Adnr\'es Bello, Av. Rep\'ublica 220, Santiago, Chile
\and
Dipartimento di Fisica ed Astronomia ``G.Galilei'', Univerista\`a di Padova, vicolo dell'Osservatorio 3, I-35122 Padova, Italy
\and
INAF -- Osservatorio Astronomico di Padova, Vicolo dell'Osservatorio 5, I-35122 Padova, Italy
%Physics \& Astronomy, University of Southampton, Southampton, Hampshire, SO171BJ, United Kingdom             
%\and
%European Southern Observatory, Alonso de Cordova 3107, Vitacura,
%Santiago, Chile
%Vicolo dell'Osservatorio 3, I-35122 Padua, Italy
%\and
%South African Astronomical Observatory, P.O. Box 9, Observatory 7935,
%South Africa
%\and
%University of Cape Town, Private Bag X3, Rondebosch 7701, South Africa
}

\offprints{P. Parisi (\texttt{pietro.parisi@iaps.inaf.it)}}

\abstract{Through an optical campaign performed at 4 telescopes 
located in the northern and the southern hemispheres, we have obtained 
optical spectroscopy for 75 counterparts of unclassified or 
poorly studied hard X-ray emitting objects detected with 
{\it Swift}/BAT and listed in the 54 month Palermo BAT catalogue.
All these objects have also observations taken with {\it Swift}/XRT, {\it ROSAT} or {\it Chandra} satellites which allowed us to reduce the high energy error box and pinpoint the
most likely  optical  counterpart/s.
We find that 69 sources in our sample are Active Galactic Nuclei (AGNs); of them, 35 are classified as type 1 (with broad and narrow emission lines), 33 are classified as
type 2 (with only narrow emission lines) and one is an high redshift QSO; the remaining 6 objects are galactic cataclysmic variables (CVs). Among  type 1 AGNs, 
32 are objects of intermediate Seyfert type (1.2-1.9) and 
one is Narrow Line Seyfert 1 galaxy; for 29 out of 35 type 1 AGNs, we have been able to estimate the central black hole mass and the Eddington ratio. Among type 2 AGNs, 
two display optical features typical  of the LINER class, 3 are classified as transition objects, 1 is a starburst galaxy and 2 are instead X-ray bright, optically normal galaxies.
All galaxies classified in this work are relatively nearby objects (0.006 - 0.213) except for one at redshift 1.137.
}%{}{}{}{}


\titlerunning{Classification of 75 objects in the Palermo
Swift/BAT catalogue}
\authorrunning{P. Parisi et al.}

\maketitle
\section{Introduction}
A critically important region of the astrophysical spectrum is the hard X-ray band, from 15 keV to 200 keV; this band
is being explored in detail by two satellites,  {\it INTEGRAL} (Winkler et al. 2003) and {\it Swift} (Gehrels et al. 2004), which carry on board the high energy instruments 
IBIS (Ubertini et al. 2003) and BAT (Barthelmy 2004), respectively. 
These spacecrafts permit the study of a variety of processes that take place in this observational window thus providing a deeper look into the physics of hard X-ray sources.

Both instruments have detected a 
large number of known and new objects, discovered new classes of sources, and allowed us find and study highly absorbed objects.
In particular, the nature of many sources detected above 20 keV by both satellites is often unknown, the sources being optically unclassified 
and their types only inferred basing on few available X-ray or radio observations. 

Optical follow-up observations of these objects is therefore mandatory.
In particular, the optical spectra can provide not only an accurate source classification, but also fundamental parameters that together with multiwaveband studies,
for example in the soft X-ray band, can determine the stellar population properties (Morelli et al. 2013) giving important information on these newly detected objects.

In this paper we continue the identification work on Swift/BAT sources started seven years ago and which has allowed the identification up to now of about 60 objects through optical spectroscopy (Landi et al. 2007, Parisi et al. 2009, 2012). 
In this work we focus on the optical follow-up of a number of objects having unknown classifications and/or redshifts and reported in the 54 month
Swift/BAT survey catalogue (Cusumano et al. 2010).

This survey covers 90\% of the sky down to a flux limit of 1.1 $\times$ 10$^{-11}$ erg cm$^{-2}$ s$^{-1}$ and 
50\% of the sky down to a flux limit of 0.9 $\times$ 10$^{-11}$ erg cm$^{-2}$ s$^{-1}$ in the 15-150 keV band. It lists 1256 sources, 
of which 57\% are extragalactic, 19\% are galactic, 
and 24\% are of unknown type.

From this BAT survey, we selected a sample of 73 objects (one BAT source has 3 possible optical counterparts) either without optical identification,
or that had not been deeply studied before, or without published optical spectra.
For all  these sources, we first performed the analysis of the available X-ray data to reduce the source positional uncertainty from arcmin- to arcsec-sized radii. Within the reduced X-ray error boxes, we then identified the putative optical counterpart/s to the BAT object and then performed 
optical spectroscopic follow-up work. Following the method applied by Masetti et al. (2004, 2006a,b,c,d, 2008, 2009, 2010, 2012, 2013) and Parisi et al. (2009, 2012), 
we determine the nature of all selected objects, estimating also the central black hole mass for type 1 AGNs.

The paper is structured as follows:  
in Sect. 2, we provide information on the X-ray data reduction in order to obtain the X-ray coordinates of the likely counterparts;
in Sect. 3, we describe the optical observations, the telescope employed, 
and provide information on the optical data reduction method used. 
Sect. 4 reports and discusses the main optical results  (line fluxes, distances, Galactic and local extinction, 
central black hole masses etc.). In Sect. 5, we discuss some peculiar sources and finally in Sect. 6 we summarize the main conclusions of our work. 

\section{X-ray astrometry}

In this section, we provide information on the search for the soft X-ray counterparts of the BAT objects. 
To obtain the soft X-ray coordinates, we cross-correlated the BAT positions with those in the catalogues of soft (< 10 keV) X-ray sources and/or analysed archival observations. 
More specifically, for the present sample, we selected BAT objects which have, within their BAT error box, source detections by either ROSAT (Voges et al. 1999), Swift/XRT, or Chandra (http://cxc.harvard.edu). This approach was proven by Stephen et al. (2006) to be very effective in associating, with a high degree of probability, 
hard X-ray sources with a softer X-ray counterpart and in turn drastically reducing their positional error circles to a few arcsec in radius, thus shrinking the 
search area by a factor of $\sim 10^4$.

For 73 of the 75 objects studied in this work, we used X-ray data acquired with the X-ray Telescope (XRT, 
0.3--10 keV, Burrows et al. 2004) onboard the {\it Swift} satellite.
The XRT data reduction was performed using the XRTDAS standard data pipeline package ({\sc xrtpipeline} 
v. 0.12.6), to produce screened event files. All data were extracted in photon counting 
(PC) mode (Hill et al. 2004), only adopting the standard grade filtering (0--12 for PC) according to the XRT 
nomenclature.  Depending on the source nature (bright or dim), we either used the longest exposure or coadded multiple observations 
to enhance the signal-to-noise ratio (S/N).
For each BAT detection, we then analysed, with {\sc XIMAGE} v. 4.5.1, the 3-10 keV  image of interest (single or added over more XRT pointings) to search for 
sources detected (at a confidence level $>$ 3$\sigma$) within the 90$\%$ {\it Swift}/BAT error circles; this 3-10 keV image choice ensured 
that we selected the hardest sources, hence the most likely counterparts to the BAT objects.     
We estimated the X-ray positions and relative uncertainties using the task {\sc xrtcentroid v.0.2.9} (coordinates of the likely counterparts are reported in Table \ref{log}).
For the remaining 2 sources we used the positional information from {\it ROSAT} Bright all-sky survey (Voges et al. 1999) and {\it Chandra} archival data, respectively (see Table \ref{log}).
The \textit{Chandra} data reduction was performed using the CIAO v4.5 software with the
calibration database CALDB v4.5.6, provided by the \textit{Chandra}
X-ray Center and following the science threads listed on the CIAO
website\footnote{Available at http://cxc.harvard.edu/ciao/}.
The resulting X-ray coordinates are all reported in Table \ref{log} (second and third columns) together with their relative uncertainties (fourth column).

For 6 sources the association process has been more complicated than in the other cases: indeed the putative counterparts of
PBC J0030.5$-$5902, PBC J0243.9+5323, PBC J0818.5$-$1420 and PBC J0855.8$-$2855 are all outside the 90$\%$, but inside the 99$\%$ BAT error circle; 
since no other object is detected inside the 90\% positional uncertainty we assume these associations to be correct.
For PBC J1020.5$-$0235 no soft X-ray source is detected in the 90$\%$ and 99$\%$ BAT error circles. We checked possible associations in the recent 70 month BAT Catalogue (Baumgartner et al. 2013), 
finding two possible sources (SWIFT J1020.5$-$0237A and SWIFT J1020.5$-$0237B) with comparable 14-195 keV X-ray fluxes. Taking into account their confidence levels in the energy range 3-10 keV and choosing the hardest one, it is more reasonable to associate the 54 month BAT source PBC J1020.5$-$0235 with SWIFT J1020.5$-$0237B, even though we cannot exclude a possible contribution with SWIFT J1020.5$-$0237A; we have therefore performed optical spectroscopy of the suggested counterpart, but have highlighted this source in Table \ref{log} to remind the reader that this is the only case where no straightforward X-ray association has been found.
Concerning PBC J1540.3+1415 two soft X-ray sources lie within the BAT error circle, one of the two has two optical counterparts (see Table \ref{log} where the 3 associations are reported
as PBC J1540.3+1415-1, PBC J1540.3+1415-2 and PBC J1540.3+1415-3). In this case, the association of the real optical counterpart
is not trivial (see Sect. 4.2.2).  

Since we use different satellites to pinpoint the soft X-ray objects associated with the chosen BAT sources, we have estimated
 the probability that a soft X-ray source detected by XRT, {\it Chandra} or {\it ROSAT} is associated by chance with a BAT hard X-ray object.
Concerning associations of BAT sources with XRT and {\it Chandra} objects, we used the method of Tomsick et al. (2012); assuming an average 2-10 keV flux of 10$^{-12}$ erg s$^{-1}$ cm$^{-2}$ and a 
mean BAT error radius of 4.5 arcmin, we estimated a probability of 2\% of spurious association.
For the single BAT source association with a {\it ROSAT} object we followed the method of Stephen et al. (2006) finding a probability of 1\%, comparable with the value found for XRT and {\it Chandra}.
This value is also comparable with that found by Stephen et al. (2006) for the associations of {\it INTEGRAL} sources and {\it ROSAT} bright objects.
This means that our associations can be considered reliable, independently from which of the three soft X-ray satellites we used.

Although the error box of the {\it ROSAT} object associated with PBC J0325.6$-$0820 is substantially greater than that of XRT or {\it Chandra}, the fact that this source is outside the galactic plane and it is associated with only one optical object in its error box, this does not affect our source association.

\section{Optical spectroscopy}
In this section we describe the optical follow-up studies that we performed for 
all 75 objects. In Table \ref{log}  we list the optical coordinates of all likely counterparts as obtained from the 2MASS catalogue\footnote{available 
at {\tt http://www.ipac.caltech.edu/2mass/}} (Skrutskie et al. 2006), except for one object whose position was extracted from the USNO-A2.0 catalogue (Monet et al. 2003).

The detailed log of all optical  measurements is also reported in Table~\ref{log}: we list in column 7  the telescope and  instrument 
used for the observation, while the characteristics of each spectrograph are given in columns 8 and 9. Column 10 provides the 
observation date and the UT time at mid-exposure, while column 11 reports 
the exposure times and the number of spectral pointings.

The following telescopes were used for the optical spectroscopic study presented here:

\begin{itemize}
\item the 1.5 m at the Cerro Tololo Interamerican Observatory (CTIO), Chile;
\item the 1.52m ``Cassini'' telescope of the Astronomical Observatory of 
Bologna, in Loiano, Italy;
\item the 1.82 m ``Copernicus'' telescope of the Astronomical Observatory of Asiago, Italy;
\item the 2.1m telescope of the Observatorio Astr\'onomico Nacional in San Pedro Martir, Mexico;
\end{itemize}
The data reduction was performed with the standard procedure (optimal extraction, Horne 1986) using IRAF\footnote{
IRAF is the Image Reduction and Analysis 
Facility made available to the astronomical community by the National 
Optical Astronomy Observatories, which are operated by AURA, Inc., under 
contract with the U.S. National Science Foundation. It is available at 
{\tt http://iraf.noao.edu/}}.
Calibration frames (flat fields and bias) were taken on the day preceding or following 
the observing night. The wavelength calibration was obtained using lamp spectra 
acquired soon after each on-target spectroscopic acquisition. The uncertainty in the 
calibration was $\sim$0.5~\AA~in all cases; this was checked using the positions of 
background night-sky lines. Flux calibration was performed using 
catalogued spectrophotometric standards.
Objects with more than one observation had their spectra stacked 
together to increase the S/N ratio.

The identification and classification approach we adopt in the analysis of the optical spectra is the following:
for the emission-line AGN classification, we used the criteria 
of Veilleux \& Osterbrock (1987) and the line-ratio diagnostics 
of Ho et al. (1993, 1997) and Kauffmann et al. (2003) to distinguish among the Seyfert 2, low-ionization nuclear emission-line regions (LINERs; Heckman 1980), H{\sc ii} regions and 
transition objects (LINERs whose integrated spectra are diluted or contaminated by neighboring H II regions, Ho et al. 1997).
In the LINER class, some lines ([O{\sc ii}]$_{\lambda 3723}$, [O{\sc i}]$_{\lambda 6300}$,
and [N{\sc ii}]$_{\lambda 6584}$) are stronger than in typical Seyfert 2 galaxies; the permitted emission-line luminosities are weak; and
the emission-line widths are comparable with those of type 2 AGNs.
In particular, as mentioned in Ho et al. (1993), all  sources
with [O{\sc ii}] $>$ [O{\sc iii}], [N{\sc ii}]/H$_{\alpha}$ $>$ 0.6 and [O{\sc i}] $>$1/3 [O{\sc iii}] can be
considered as LINERs. 
For the subclass assignation to Seyfert 1 galaxies, we used the 
\mbox{H$_\beta$/[O {\sc iii}]$\lambda$5007} line flux ratio criterion presented in
Winkler et al. (1992). Moreover, the criteria of Osterbrock \& Pogge (1985) allowed us to discriminate between `normal' Seyfert 1 and narrow-line Seyfert 1 (NLS1):
the latter are galaxies with a full width at half-maximum (FWHM) of the H$_{\beta}$ line lower than 2000 km s$^{-1}$, with 
permitted lines which are only slightly broader than their forbidden lines, with a [O{\sc iii}]$_{\lambda 5007}$/H$_{\beta}$ ratio $<$ 3, and finally with evident
Fe{\sc ii} and other high-ionization emission-line complexes.

We note that the spectra of all extragalactic objects are not corrected for starlight contamination 
(see, e.g., Ho et al. 1993, 1997), because of their limited S/N ratio and  
spectral resolution. However, this does not affect our results and conclusions.

To estimate the E(B-V) local optical absorption in our AGNs sample, when possible,
we first dereddened the  H$_\alpha$ and H$_\beta$ line 
fluxes by applying a correction for the Galactic absorption along the line of sight to the source. This was done 
using the galactic colour excess $E(B-V)_{\rm Gal}$ given by Schlegel et al. (1998) and
the Galactic extinction law obtained by Cardelli et al. (1989).
We then estimated the colour excess  \mbox{$E(B-V)_{\rm AGN}$} local to the AGN host galaxy by comparing the intrinsic 
line ratio and corrected that for Galactic reddening using the relation for type 2 AGNs derived from Osterbrock (1989) 
$$E(B-V)= a\; \mbox{Log} \left(\frac{H_{\alpha}/H_{\beta}}{(H_{\alpha}/H_{\beta})_0}\right).$$ 

In the above relation, H$_{\alpha}/H_{\beta}$ is the observed Balmer decrement, $(H_{\alpha}/H_{\beta})_0$ is the intrinsic one (2.86), and
{\it a} is a constant with a value of 2.21.
For type 1 objects, where the H$_{\alpha}$ is strongly blended with the forbidden narrow [N{\sc ii}] lines, it is not easy to obtain 
a reliable H$_{\alpha}$/H$_{\beta}$
estimate. In these cases, we used the H$_{\gamma}$/H$_{\beta}$ ratio, albeit H$_{\gamma}$ may also be blended with 
the [O {\sc iii}]$_{\lambda 4363}$ line; the AGN reddening was evaluated using the same relation described above but with the intrinsic $(H_{\gamma}$/H$_{\beta})_0$ ratio of 0.474 
and an {\it a} value of -5.17.

To provide extra information, we also estimated the mass of the central black hole for 29 type 1 AGNs found in the sample\footnote{We could not estimate the mass of the central black hole of PBC J0116.3+3102, PBC J0602.5+6522  and  PBC J1926.6+4131 because they all lack the H$_{\beta}$ emission line, and
for PBC J0000.9$-$0708, PBC J0917.2$-$6454 and PBC J1824.2+1846 because only the narrow component of the H$_{\beta}$ line was observed in their spectra.}.
The method used here follows the prescription of Wu et al. (2004) and Kaspi et al. (2000), 
where we used the H$_{\beta}$ emission line flux, corrected for the Galactic colour excess (Schlegel et al. 1998), and the broad-line region
(BLR) gas velocity ($v_{FWHM}$). 
Using Eq. (2) of Wu et al. (2004), we estimated the BLR size, which is used with  $v_{FWHM}$ in Eq. (5) of Kaspi et al. (2000)
to calculate the AGNs black hole mass. The results are reported in Table \ref{blr} where we also list the observed BAT X-ray luminosities  in the 15-150 keV band and 
the Eddington ratios for each AGN considered.  To calculate  the luminosity distances, we considered a \mbox{cosmology} with $H_{\rm 0}$ = 70
km s$^{-1}$ Mpc$^{-1}$, $\Omega_{\Lambda}$ = 0.7, and $\Omega_{\rm m}$ = 0.3 and used the Cosmology Calculator of Wright (2006).

The errors on black hole masses reported in Table \ref{blr}, generally come from the emission lines flux estimate that correspond to about 15\%, and also from the scatter
in the R$_{BLR} -$ L$_{H_{\beta}}$ scaling relation (Vestergaard 2004). This implies a typical error of about 50\% of the value of the black hole masses.

To derive the distance of the 6 CVs in our sample, we used the distance modulus assuming
an absolute magnitude M$_{V} \sim$ +9 and an intrinsic colour index (V-R)$_{0} \sim$ 0 mag (Warner 1995).
Although this method basically provides an order-of-magnitude value for the distance of these Galactic sources, 
our past experience (Masetti et al. 2004, 2006a,b,c,d, 2008, 2009, 2010, 2012, 2013) tells us that these estimates are in general correct to within 50\% 
of the refined value subsequently determined with more precise approaches.
\vspace{-0.43cm}
\section{Optical classification}
In this section we discuss the optical classifications found and highlight
the most interesting or peculiar objects discovered. The {\it R} magnitudes if not otherwise stated, are all extracted
from the USNO-A2.0 catalogue. Of the 75 objects studied, the majority are of extragalactic nature (69 AGNs) and only a few are of galactic origin (6 Cataclysmic variables).

Concerning the optical class, 6 objects in the sample had the optical type
already reported in the Veron-Cetty \& Veron 
13th catalogue edition (V\&V13, Veron-Cetty \& Veron 2010 and references therein), in the SIMBAD Astronomical Database and in Halpern (2013).  
Despite this, we choose to report our own data of these 6 sources in order to confirm/disclaim their classification and also to provide line flux information.

\subsection {CVs}
Six sources in our BAT sample display 
emission lines of the Balmer complex (up to at least H$_{\epsilon}$), as well as He {\sc i}  and He {\sc ii}, 
consistent with z = 0, indicating that these objects lie within our Galaxy (see Fig. \ref{cv}). 
The analysis of their optical features indicates that all are CVs (see Tab. \ref{cvo}).
Through the equivalent widths (EW) of the H$_{\beta}$ and He {\sc ii}$_{\lambda4686}$ lines we investigated their magnetic or non magnetic nature.
In the case of PBC J0746.2$-$1610 and PBC J0927.8$-$6945 the He {\sc ii}$_{\lambda4686}$/H$_{\beta}$ EW ratio is larger than 0.5, and the EW of both emission lines 
is larger than 10 \AA$\,$  implying that both objects are magnetic CVs belonging to the intermediate polar (IP) subclass (see Warner 1995, and references therein).
A tentative IP classification can also be made for PBC J0325.6$-$0820, PBC J0706.7+0327 and PBC J0820.4$-$2801 since also in these cases the He {\sc ii}$_{\lambda4686}$ 
and the H$_{\beta}$ EW are larger than 10 \AA, although their ratio is smaller than 0.5.
For PBC J2124.5+0503 (classified as LMXB in Halpern 2013) the weakness of its emission lines (< 10 \AA) and the He {\sc ii}$_{\lambda4686}$/H$_{\beta}$ ratio smaller than 0.5 imply a non magnetic nature.
The H$_{\alpha}$ to H$_{\beta}$ flux ratio is smaller than 2 for all the CVs found in this work, which allow to assume that the absorption
along the line of sight is negligible in all cases.
We also estimated their distances {\bf (see sect. 3)}, assuming no Galactic extinction along the line of sight (see Tab. \ref{cvo}). 
All CVs of likely magnetic nature are at relatively high distance ($>$ 150 pc) while the only non magnetic CV is located quite close to earth.

\subsection {Extragalactic objects}
\begin{figure*}
%\begin{center}
\hspace{-.1cm}
\centering{\mbox{\psfig{file=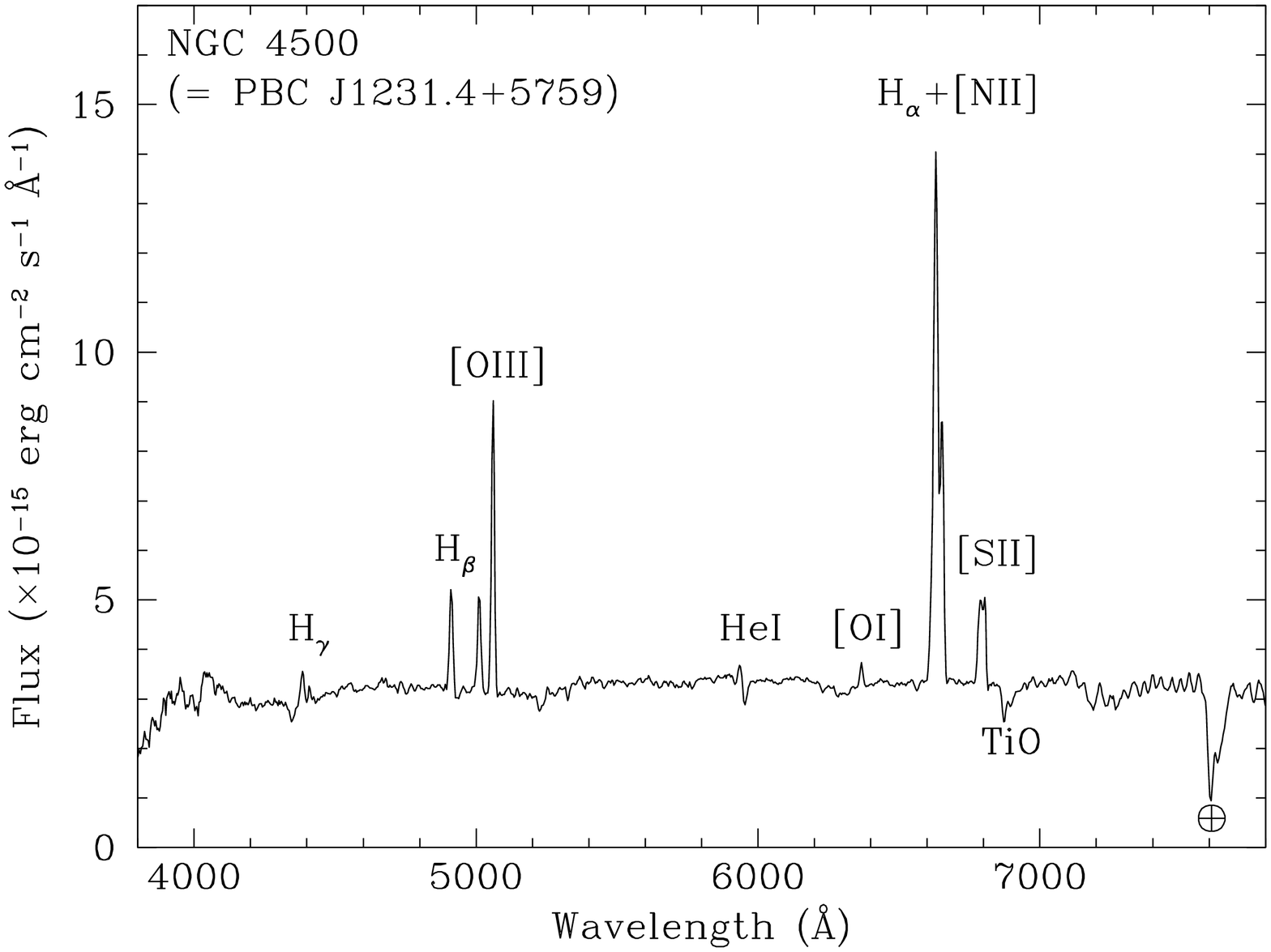,width=6.0cm,angle=270}}}
\caption{Spectrum (not corrected for the intervening Galactic absorption) of the optical counterpart of the starburst galaxy belonging to the sample of BAT sources presented in this paper. For this spectrum, the main spectral features are labeled.
}\label{region}
%\end{center}
\end{figure*}

\begin{figure*}
%\begin{center}
\hspace{-.1cm}
\centering{\mbox{\psfig{file=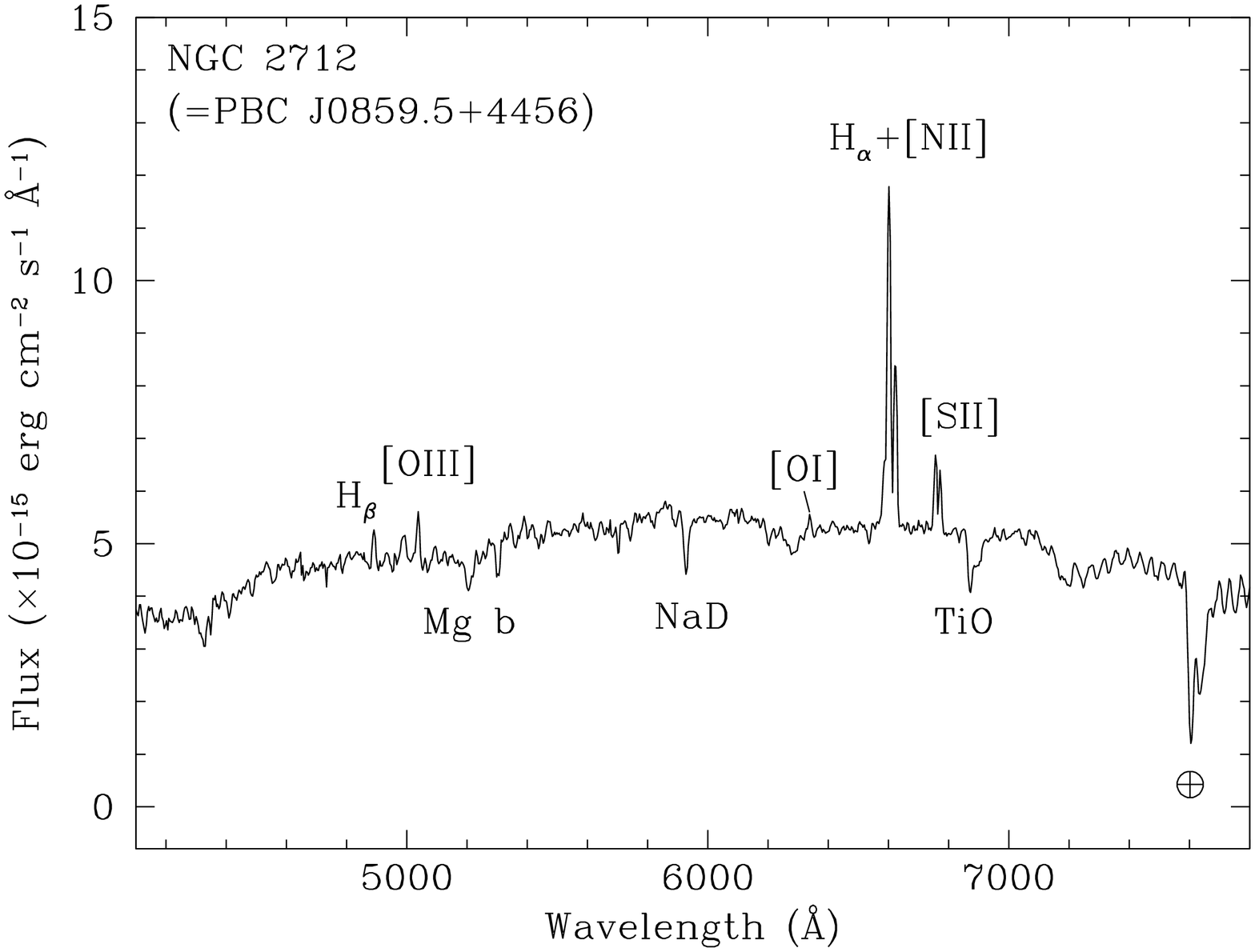,width=6.0cm,angle=270}}}
\centering{\mbox{\psfig{file=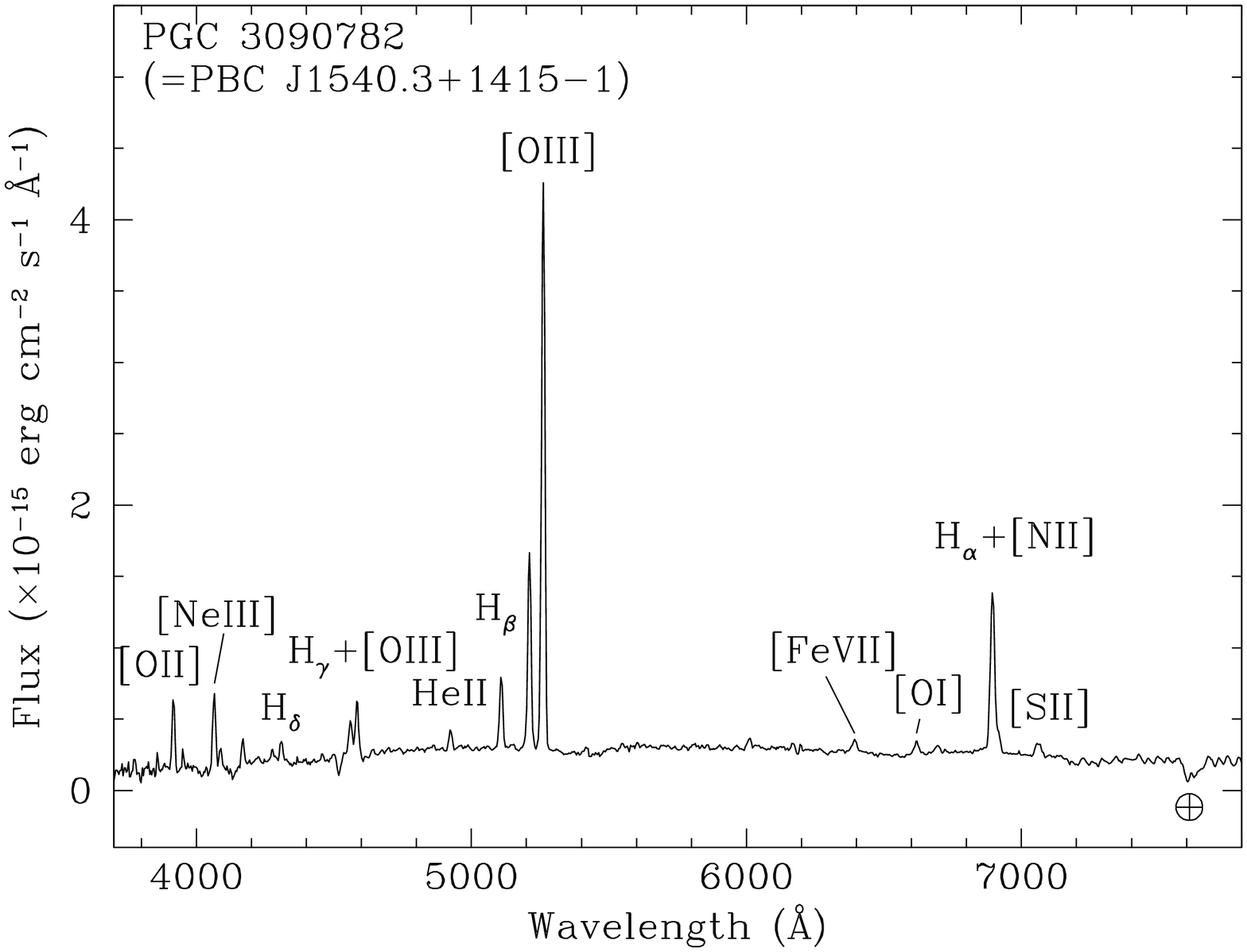,width=6.0cm,angle=270}}}
\centering{\mbox{\psfig{file=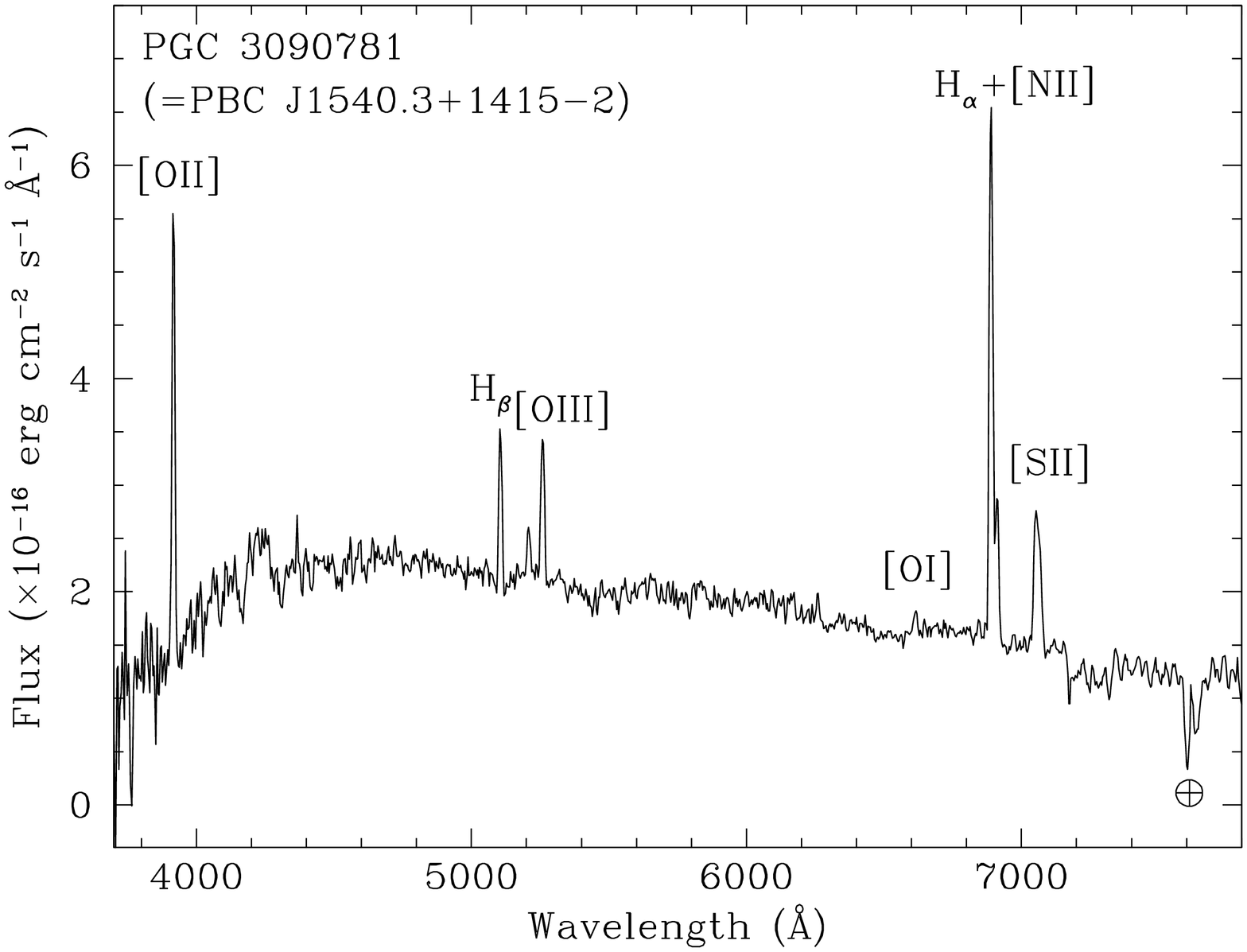,width=6.0cm,angle=270}}}
\caption{Spectra (not corrected for the intervening Galactic absorption) of the optical counterparts of the 3 transition objects belonging to the sample of BAT sources presented in this paper. For each spectrum, the main spectral features are labeled.
}\label{trans}
%\end{center}
\end{figure*}

\begin{figure*}
%\begin{center}
\hspace{-.1cm}
\centering{\mbox{\psfig{file=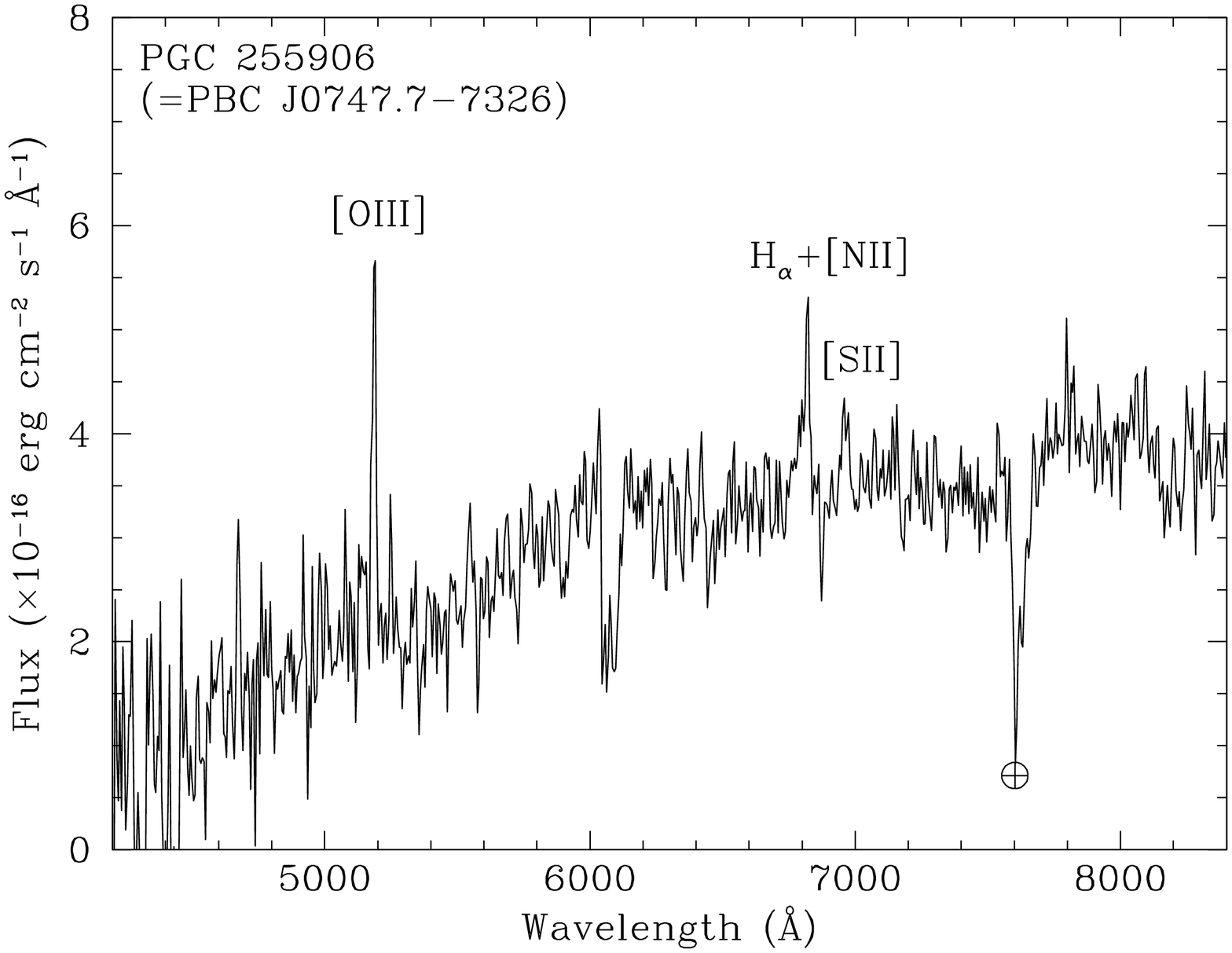,width=6.0cm,angle=270}}}
\centering{\mbox{\psfig{file=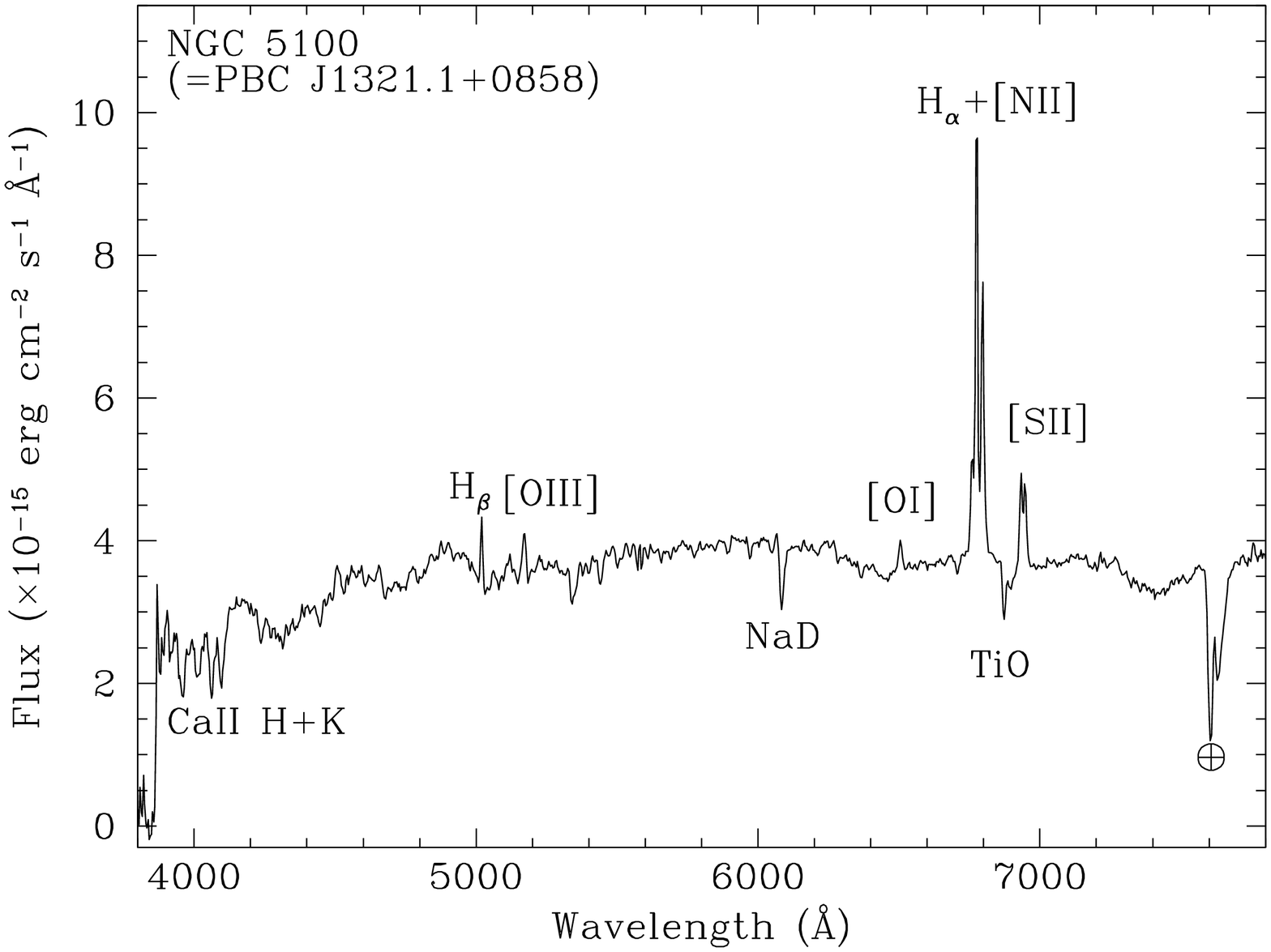,width=6.0cm,angle=270}}}
\caption{Spectra (not corrected for the intervening Galactic absorption) of the optical counterparts of the 2 LINERs belonging to the sample of BAT sources presented in this paper. For each spectrum, the main spectral features are labeled.
}\label{liner}
%\end{center}
\end{figure*}

The results of extragalactic sources are reported in Tables \ref{agn1} and \ref{agn2} where for each source we list 
the H$_{\alpha}$, H$_{\beta}$, and [O{\sc iii}] fluxes, the classification, the estimated redshift, 
the luminosity distance given in Mpc, the Galactic colour excess and the colour excess local to the AGN host.
In Table \ref{qso} instead, we report information of the only QSO found in this sample. 
All the extragalactic optical spectra are displayed in Figures \ref{spectra1}, \ref{region}, \ref{trans} and \ref{liner}. 

\subsubsection {Redshifts}
We confirm the redshift estimates reported in SIMBAD and V\&V13 for 29 AGNs. 
For the remaining 40 sources, the redshifts derived from our low-resolution optical spectra are published here for the first time (see Figs. \ref{spectra1}, \ref{region}, \ref{trans} and \ref{liner}).
Redshifts values are in the range 0.006-1.137, i.e. all AGNs are located in the local Universe (z < 0.3) except for one source which is at redshift 1.137. 

We note that all redshifts were estimated using the [O{\sc iii}] narrow emission line and when this line was unavailable, from other forbidden narrow emission lines or absorption features.

\subsubsection {Optical class}

For the first time, we provide the classification of 64 sources in the sample.
For the remaining 5 AGNs, our results agree with the classifications listed in the literature except for one object:
PBC J1002.3+0304 also known as IC 0588, is classified as a Seyfert 1 in the V\&V13 catalogue, but according to our analysis it is a Seyfert 1.8/1.9.
Concerning PBC J2157.4$-$0611, Halpern (2013) in his preliminary work classifies this source as a QSO, we refine this classification stating that it is a Seyfert 1.5.

Summarizing our extragalactic results, we found that out of 69 AGNs, 36 have strong redshifted broad and narrow emission-lines that are typical of Seyfert 1 galaxies,
while the remaining 33 display only the strong and redshifted narrow emission-lines that are indicative of a type 2 AGN nature (for the subclass assignment see Tables \ref{agn1} and \ref{agn2}).
 
Note that among broad line AGNs, only 3 are pure type 1 objects including the QSO at z $> 1$. One is a NLS1 while the remaining objects are all of intermediate type with  9  
belonging to to Seyfert 1.8-1.9, i.e. they are more  similar to type 2 AGNs due to the progressive disappearance of broad line regions.
PBC J2035.2+2604 is the only NLS1 found; it has an optical counterpart in the USNO-A2.0 catalogue (USNO-A2.0 1125$-$16409384) with magnitude R = 13.8, a redshift of 0.05 and 
significant optical extinction (E(B-V) = 0.834 mag). NLS1 are rare among galaxies  detected above 20 keV since their fraction is typically $5$\% of  type 1  (Panessa et al. 2011) 
and  $2$\% of all AGNs; this match perfectly with our findings.

Among the 33 type 2 AGNs, 22 are Seyfert 2 galaxies, 2 are LINERs, one is a starburst galaxy, 3 are transition objects and 2 are XBONG (X-ray bright, optically normal galaxies, Comastri et al. 2002).
PBC J0747.7$-$7326 and PBC J1321.1+0858 are classified as LINERs. The first object is a pure LINER, according to Ho et al. (1997);
the second source is a less clear case since in the Kauffmann et al. (2003) diagram it is placed in the LINERs region, but its nature is ambiguous according to the Ho et al. (1997) diagrams. 
We decided to classify this source as a LINER because it has some features typical of this class.  
Clear detection of both these objects at high energies points to an AGN (and not a burst of star formation), as the source that excites the ionized gas in these galaxies.

PBC J1231.4+5759 is instead classified as a starburst galaxy. 
In the diagrams of Ho et al. (1997) it is placed in the left part with a low [O {\sc iii}]/H$_{\beta}$ ratio and a very low value [N {\sc ii}]/H$_{\alpha}$ ratio.
It is a blue compact galaxy also known as NGC 4500. The source is characterized by intense far Infrared and UV emission typical of a recent burst of star formation. 
The soft X-ray counterpart of this BAT source is quite absorbed, which suggests that a low luminosity AGN maybe hidden behind the strong signature of the starburst emission.

PBC J0859.5+4457, PBC J1540.3+1415-1 and PBC J1540.3+1415-2 are instead transition objects. These sources are likely LINERs whose integrated spectra are diluted or contaminated 
by neighboring H II regions. Recent radio and mostly X-ray observations have suggested that these objects are likely to harbour low luminosity AGNs (Ho 2008); 
they could represent a late stage of the AGNs activity when the accretion rate is low
compared to the Seyfert stage. The  detection of two such objects among BAT sources, further support the likely presence of a low luminosity AGN in transition galaxies.
PBC J1540.3+1415-1 and PBC J1540.3+1415-2  have about the same redshift ($\sim 0.05$) and are likely in interaction; only PBC J1540.3+1415-1 has some associated radio emission
which makes it a more likely counterpart for the BAT source. We remind the reader that the other likely counterpart for the same BAT source is PBC J1540.3+1415-3, 
which is a Seyfert 1.2 at a redshift of 0.12 displaying strong radio emission (see following section). Using the statistical method of Tomsick et al. (2012), we estimate the probability for each of these three sources 
to be contained by chance within  the BAT hard X-ray error box and found that
PBC J1540.3+1415-1 and PBC J1540.3+1415-2 have a probability of $\sim$0.14, with a 2-10 keV flux of 0.1 $\times$10$^{-12}$ erg s$^{-1}$ cm$^{-2}$ (power law with $\Gamma$ = 1.8), while PBC J1540.3+1415-3 has a probability below 0.02, with a 2-10 keV flux of 1.4$\times$10$^{-12}$ erg s$^{-1}$ cm$^{-2}$ (power law with $\Gamma$ = 1.8), which makes it the most likely soft X-ray 
counterpart. On the other hand analysis of the XRT image at energies above 3 keV suggests that the X-ray source associated with optical objects PBC J1540.3+1415-1 and PBC J1540.3+1415-2 
is the harder of the two objects detected inside the BAT positional uncertainty and also the only one still visible above 6 keV.
PBC J1540.3+1415 is clearly a complicated object which deserve further studies possibly multiwavebands if one wants to understand which source is responsible for the hard X-ray emission 
and ultimately the real nature of the BAT object. 
 
Finally, PBC J1034.2+7301 and PBC J1355.5+3523 are both classified as XBONGs, which are X-ray bright galactic nuclei with no emission lines in their optical spectra. 
While the presence of weak H$_{\alpha}$ and [O {\sc iii}] emission lines in the first source makes it more similar to a Seyfert 2, the second shows 
no evidence of emission lines in its optical spectrum.
From a quick analysis of their X-ray spectra, both objects seem to be  absorbed in X-rays with column densities above 10$^{22}$ cm$^{-2}$, which suggest that they may be
AGNs where the obscuration is covering almost 4$\pi$ of the nuclear source (see Malizia et al. 2012 for details).
Further investigation in X-rays, but also in other wavebands can help in clarifying the nature of both sources. 

\section{Further comments}
We note that a number of AGNs in our sample display evidence of interaction/clustering: 
they either belong to a small group of objects (such as galaxy pair/triple or a compact group)
or are being disturbed by nearby galaxies. Beside PBC J1540.3+1415-1 and PBC J1540.3+1415-2 already discussed in the previous section,  also the counterpart of 
PBC J0116.3+3102 (also known as NGC 452) makes a pair with
NGC 444 around 131.8 kpc away. 

Other examples of galaxy pairs discovered in the present sample are  
NGC 5100 NED 02 and MCG +09$-$19$-$015 NED 02, the optical counterparts of PBC J1321.1+0858 
and PBC J1115.3+5425; both objects are associated with a nearby galaxy 
having similar redshift and located respectively at 23.7 kpc and 113.7 kpc of distance. 
The AGNs associated with PBC J0223.4+4549 belong instead to a triple system (VZW232) according to NED 
although no redshift is available for the other two members of the group. 
Finally, the active galaxy associated with PBC J1254.8$-$2655 belongs to a group of objects and more specifically
is listed as AM 1252$-$264 in the Catalogue of southern peculiar galaxies and associations made by Arp \& Madore (1987); 
interestingly also another member of this group, 2MASX J12544294$-$2657107, is 
visible in X-rays, but is softer and less luminous than the BAT counterpart. We have also a few cases
where the galaxy associated with the high energy source seems in interaction with a nearby 
companion; one such clear case is that of the counterpart of PBC J2035.2+2604. 

Overall we estimate that the fraction of AGNs in the present sample that display  evidence of interaction/clustering
is around 20\%, i.e. a value very close to that found by Koss et al. (2010) in a more accurate analysis
of a set of known active galaxies detected by BAT. This high rate of apparent mergers  suggests that AGNs activity
and merging are critically linked for the moderate luminosity AGNs in the BAT sample.
We also note that a few  sources display the properties of radio loud AGN: at least 4 (the counterparts of PBC 
J0459.8+2705, PBC J0602.5+6522, PBC J0654.5+0703 
and PBC J1540.3+1415 (specifically object N.3)) display strong radio emission and are characterized by a flat  
spectrum at these frequencies;
their radio morphology is that of a compact source. Two remaining objects are clearly radio galaxies.
PBC J0709.2$-$3601 also known as PKS 0707$-$35 is a complex and extended source, an edge-brightened
double, with two compact outer components, the south-east one stronger;
two weaker inner components extend away from
the axis joining the outer components and give a slight twist to the
structure (Jones \& McAdam 1992). PBC J0950.0+7315 also named 4C 73.08 is also a giant double-lobed radio-galaxy, 
with 13 arcmin angular size between hotspots and a clear FRII morphology (Aretxaga et al. 2001, Hardcastle \& Worrall 1999).

We conclude that around 9\% of the AGNs in the present sample are radio loud, a percentage which is close 
to what generally observed among active galaxies. Moreover, we note that the black hole masses listed in Table \ref{blr} cover 
quite a large range from a few 10$^6$ to around 7$\times$10$^8$ M$_{\odot}$; the Eddington ratio as well
span from 0.001 to 0.2, suggesting that BAT AGNs tend to cover a broad range in the parameter space.

Finally, even if this issue is beyond the scope of this paper as it would need a more careful selection
of the sample of sources, we checked possible correlations between optical emission line fluxes (H$_{\alpha}$, H$_{\beta}$ and O{\sc iii}) and hard X-ray flux (15-150 keV).
Using the least-squares bisector method (Isobe et al. 1986) we did not find any evident correlation between the above optical quantities and the 15-150 keV hard X-ray flux, obtaining 
correlations coefficients R$^2$ $<$ 0.15, as already suggested by Winter et al. (2010).
We do not expect any correlation between line strength and source class,
as the latter is defined through the ratio of lines and through their width,
rather than through their strength. 
Moreover, to the best of our knowledge, there is no known correlation between line
strength and redshift, as AGNs are not standard candles; also, the fact that
almost all sources in our sample lie at low z (<0.2) does not grant a wide
enough baseline for the redshift range to properly explore this point.

\section{Conclusions}

We have either provided for the first time, confirmed, or corrected the optical spectroscopic identifications of 73  
sources belonging to the Palermo 54 month {\it Swift}/BAT catalogue (Cusumano et al. 2010). 
 
This has been achieved by performing a multisite observational campaign in Europe, Central and South America.  Only in one case, that of PBC J1540.3+1415, we find more than one  optical  counterpart, specifically 3 objects likely to emit at high energies.

We have found that our sample is dominated by extragalactic object with only 6 sources of galactic nature.
The extragalactic sample is composed of 69 AGNs (35 of type 1, 33 of type 2 and 1 QSO), with redshifts between 0.006 and 1.137.
Among them we highlighted some peculiar objects, such as 2 galaxies displaying LINERs features, 1 starburst galaxy, 2 XBONGs, 3 transition objects and 1 object with the properties of a NLS1. 
For 29 type 1 AGNs, we have estimated the BLR size, velocity, and the central black hole mass as well as their Eddington ratio. 
The AGNs sample presented in this work show a large fraction (around 20\%) of objects displaying evidence of interaction/clustering while only 9\% show indication of being radio loud.

Concerning the 6 galactic sources we found 2 likely magnetic CVs belonging to the IP subclass, 3 objects with tentative IP classification and only 1 source with non magnetic properties.
Finally we checked possible correlations between  optical emission line fluxes (H$_{\alpha}$, H$_{\beta}$ and O{\sc iii}) and hard X-ray flux (15-150 keV), as well as for source class or redshift, but no evident correlation was found.

These results point out the efficiency of using catalog cross-correlation and/or follow-up  X-ray observations from satellites 
such as {\it Chandra}, {\it ROSAT} or {\it Swift}) capable of providing arcsec-sized error boxes followed by optical spectroscopy to determine the actual nature of still 
unidentified BAT sources.

 \begin{acknowledgements}
We thank John Stephen for the useful comments and suggestions; Silvia Galleti for Service Mode observations at the Loiano telescope, and Roberto Gualandi for night assistance; Giorgio Martorana for Service Mode observations at the Asiago telescope and Luciano Traverso for coordinating them; Manuel Hern\'andez, Rodrigo Hern\'andez and Jos\`e Velasquez for Service Mode observations at the CTIO telescope and Fred Walter for coordinating them.
We also acknowledge the use of public data from the {\it Swift} data archive.
This research has made use of the ASI Science Data Center Multimission 
Archive, of the NASA Astrophysics Data System Abstract Service, 
the NASA/IPAC Extragalactic Database (NED), of the NASA/IPAC Infrared 
Science Archive, which are operated by the Jet Propulsion Laboratory, 
California Institute of Technology, under contract with the National 
Aeronautics and Space Administration and of data obtained from the High Energy 
Astrophysics Science Archive Research Center (HEASARC), provided by NASA's GSFC.
This publication made use of data products from the Two Micron All 
Sky Survey (2MASS), which is a joint project of the University of 
Massachusetts and the Infrared Processing and Analysis Center/California 
Institute of Technology, funded by the National Aeronautics and Space 
Administration and the National Science Foundation.
This research has also made use of data extracted from the 6dF 
Galaxy Survey and the Sloan Digitized Sky Survey archives;
the SIMBAD database operated at CDS, Strasbourg, 
France, and of the HyperLeda catalogue operated at the Observatoire de 
Lyon, France.
The authors acknowledge the ASI and INAF financial support via grants No. I/033/10/0, I/009/10/0;
P.P. is supported by the INTEGRAL ASI-INAF grant No. I/033/10/0.
L.M. is supported by the University of Padua through grant No. 
CPDR061795/06. G.G. is supported by FONDECYT 1085267.
V.C. is supported by the CONACyT research grants 54480
and 151494 (M\'exico).
D.M. is supported by the Basal CATA PFB 06/09, and FONDAP Center for 
Astrophysics grant No. 15010003.

\end{acknowledgements}

\begin{landscape}
\begin{table}[th!]
\caption[]{Log of the spectroscopic observations presented in this paper
(see text for details). Optical source coordinates
are extracted from the 2MASS catalog, if not indicated otherwise, and have an accuracy better than 0$\farcs$1. Soft X-ray coordinates are extracted from XRT, Chandra or ROSAT observations.}
\label{log}
\scriptsize
\begin{center}
\resizebox{25cm}{!}{
\begin{tabular}{lllcllccclr}
\noalign{\smallskip}
\hline
\hline
\multicolumn{1}{c}{{\it (1)}} & \multicolumn{1}{c}{{\it (2)}} & \multicolumn{1}{c}{{\it (3)}} & \multicolumn{1}{c}{{\it (4)}} &  \multicolumn{1}{c}{{\it (5)}}&\multicolumn{1}{c}{{\it (6)}}&
{\it (7)} & {\it (8)} & {\it (9)} & \multicolumn{1}{c}{{\it (10)}}& \multicolumn{1}{c}{{\it (11)}}\\
\multicolumn{1}{c}{Object} & \multicolumn{1}{c}{RA X-ray} & \multicolumn{1}{c}{Dec X-ray} &\multicolumn{1}{c}{Error radius} & \multicolumn{1}{c}{RA Opt} &\multicolumn{1}{c}{Dec Opt}&\multicolumn{1}{c}{Telescope+instrument} & $\lambda$ range & Disp. & \multicolumn{1}{c}{UT Date \& Time}  & Exposure \\
& \multicolumn{1}{c}{} & \multicolumn{1}{c}{} &(arcsec) &\multicolumn{1}{c}{{\it (J2000)}}  &\multicolumn{1}{c}{{\it (J2000) }} & & (\AA) & (\AA/pix) & 
\multicolumn{1}{c}{at mid-exposure} & time (s)  \\
\noalign{\smallskip}
\hline
\noalign{\smallskip}
PBC J0000.9$-$0708   &  00 00 48.57 & $-$07 09 13.2   & 3.6   &00 00 48.81  &$-$07 09 11.6    & SPM 2.1m+B\&C Spc. &3450-7650& 4.0 & 01 Nov 2010, 06:42 & 2$\times$1800  \\
PBC J0030.5$-$5902$^{=+}$   &00 30 01.29 &$-$59 02 45.4 & 4.9   &00 30 01.30 &$-$59 02 45.0    & CTIO 1.5m+RC Spec. & 3300-10500 & 5.7 & 24 Aug 2009, 05:01 & 2$\times$1800  \\                                                                                                                                                                                                                                                                                                                                                                                                                                                                                                                                                                                                                                                                                                                                                                                                                                                                                                                                                                                                                                                                                                                                                                                                                                                                                                                                                                                                                                                                                                                                 
PBC J0034.6$-$0424    &00 34 32.93 &$-$04 24 10.4  & 4.4   & 00 34 32.81 & $-$04 24 12.2  & SPM 2.1m+B\&C Spc. & 3450-7650  & 4.0 & 11 Dec 2009, 04:37 & 3$\times$1800    \\
PBC J0038.5+2336         &00 38 31.83  &+23 36 46.9  & 4.4   & 00 38 32.11 &+23 36 48.3 & SPM 2.1m+B\&C Spc. & 3450-7650  & 4.0 & 18 Sep 2009, 08:59 & 2$\times$1800  \\
PBC J0050.8+7648          &00 51 06.18&+76 50 33.1  & 3.9   & 00 51 06.65 &+76 50 35.9	   & Copernicus+AFOSC &4000-8000& 4.2 & 10 Feb 2011, 20:51 & 2$\times$1800   \\
PBC J0116.3+3102         &01 16 14.47 &+31 01 58.3  & 4   & 01 16 14.81 &+31 02 01.8	   &  SPM 2.1m+B\&C Spc. &3450-7650& 4.0 & 10 Oct 2010, 07:40 & 2$\times$1800\\
PBC J0128.6$-$6038      &01 29 06.70 & $-$60 38 45.0  & 5.5   & 01 29 07.62 &$-$60 38 42.4  &CTIO 1.5m+RC Spec. & 3300-10500 & 5.7 & 14 Oct 2010, 23:59 & 2$\times$1000\\
PBC J0149.3$-$5017         &01 49 22.12 & $-$50 15 06.4  & 3.6   & 01 49 22.29 &$-$50 15 07.4	   & CTIO 1.5m+RC Spec. & 3300-10500 & 5.7 & 12 Nov 2010, 00:00 & 2$\times$1200 \\
PBC J0157.3+4715      &01 57 11.05 &+47 15 59.1  & 3.5    & 01 57 10.98 &+47 15 58.9    &  SPM 2.1m+B\&C Spc. &3450-7650& 4.0  &31 Oct 2010, 07:45 & 2$\times$1800  \\
PBC J0223.4+4549          &02 23 32.81 &+45 49 12.1   &3.8    & 02 23 33.09 &+45 49 16.2	  &Cassini+BFOSC  & 3500-8700 & 4.0 & 17 Nov 2009, 18:15 & 1800  \\
PBC J0238.3$-$6116     &02 38 42.95 &$-$61 17 20.3   & 3.7   & 02 38 43.13 &$-$61 17 22.7	&   CTIO 1.5m+RC Spec. & 3300-10500 & 5.7 & 17 Nov 2009, 05:35 &2$\times$1200\\
PBC J0243.9+5323$^{+}$          &02 44 03.03 &+53 28 28.7  & 4   & 02 44 02.96 &+53 28 28.2	  & SPM 2.1m+B\&C Spc. &3450-7650& 4.0   & 03 Nov 2010, 08:11 & 2$\times$1800  \\
PBC J0311.9+5029          &03 12 03.14 &+50 29 13.0   & 3.6  & 03 12 02.91 &+50 29 14.7	  &Copernicus+AFOSC &4000-8000& 4.2 &18 Aug 2009, 02:44  & 1800  \\
PBC J0325.6$-$0820$^R$          & 03 25 40.00&$-$08 14 42.5    & 9   & 03 25 39.42 &$-$08 14 42.8	   &  SPM 2.1m+B\&C Spc. &3450-7650& 4.0 & 31 Oct 2010, 09:40 & 2$\times$1800	    \\
PBC J0359.0$-$3017     &03 59 08.85 &$-$30 18 13.2     & 4.4   & 03 59 08.85 &$-$30 18 10.2   & SPM 2.1m+B\&C Spc. &3450-7650& 4.0  & 11 Dec 2009, 06:19 &  1800 \\ 
PBC J0429.7$-$6703     &04 29 47.14 &$-$67 03 20.6   & 3.8   & 04 29 47.36 &$-$67 03 20.5 & CTIO 1.5m+RC Spec. & 3300-10500 & 5.7 & 17 Oct 2010, 05:40 & 2$\times$1500  \\
PBC J0440.8+2739       &04 40 47.58 &+27 39 47.5   & 3.7   & 04 40 47.71 & +27 39 46.7	&  SPM 2.1m+B\&C Spc. &3450-7650& 4.0  & 03 Nov 2010, 10:45  & 2$\times$1800 \\
PBC J0459.8+2705          &04 59 55.88 &+27 06 01.2    & 4.3   &04 59 56.08 &+27 06 02.4	    & SPM 2.1m+B\&C Spc. &3450-7650& 4.0 & 01 Feb 2009, 04:56 & 2$\times$1800  \\
PBC J0515.3+1856          &05 15 19.94 &+18 54 52.5  & 4.7   & 05 15 19.79 &+18 54 51.6	    & Copernicus+AFOSC &4000-8000& 4.2 & 10 Feb 2011, 22:13  & 2$\times$1500 \\
PBC J0532.7+1346          &05 32 57.61 &+13 45 08.4   & 4.2   & 05 32 57.53 &+13 45 09.2	    & SPM 2.1m+B\&C Spc. &3450-7650& 4.0 & 03 Nov 2010, 12:52  & 2$\times$1800 \\
PBC J0535.6+4011          &05 35 31.80 &+40 11 15.7   & 3.7   & 05 35 32.11 &+40 11 15.3	    & SPM 2.1m+B\&C Spc. &3450-7650& 4.0 & 01 Feb 2009, 06:04  &  2$\times$1800\\
PBC J0543.9$-$4325          &05 44 00.31 &$-$43 25 27.9   & 4   & 05 44 00.11 &$-$43 25 26.6	    & CTIO 1.5m+RC Spec. & 3300-10500 & 5.7 & 02 Oct 2009, 08:29  &2$\times$1800  \\
PBC J0602.5+6522          &06 02 37.72 &+65 22 16.0   & 4.6   & 06 02 37.94 &+65 22 16.2	    & SPM 2.1m+B\&C Spc. &3450-7650& 4.0 & 02 Nov 2010, 12:09  &   1800\\
PBC J0609.4$-$6243          &06 10 06.28 &$-$62 43 10.9  &  3.7   & 06 10 06.52 &$-$62 43 12.5  & CTIO 1.5m+RC Spec. & 3300-10500 & 5.7 & 18 Nov 2010, 05:57 &2$\times$1200 \\
PBC J0635.0$-$7441      &06 34 03.31 &$-$74 46 37.6   & 3.7   & 06 34 03.55 &$-$74 46 37.6      & CTIO 1.5m+RC Spec. & 3300-10500 & 5.7 & 11 Nov 2010, 00:00  &2$\times$1800 \\
PBC J0654.5+0703   &06 54 33.93 &+07 03 21.4     & 3.6     & 06 54 34.18 &+07 03 21.0	  & SPM 2.1m+B\&C Spc. &3450-7650& 4.0 & 14 Mar 2010, 03:10 &1800  \\
PBC J0706.7+0327   &07 06 48.89 &+03 24 45.0     & 3.6     & 07 06 48.93 &+03 24 47.3	  & Cassini+BFOSC  & 3500-8700 & 4.0 & 09 Mar 2011, 19:35 & 3$\times$1800  \\
PBC J0709.2$-$3601   &07 09 14.23 &$-$36 01 23.9     & 3.7    & 07 09 14.08 &$-$36 01 21.7	  & CTIO 1.5m+RC Spec. & 3300-10500 & 5.7 & 16 Dec 2010, 03:06 & 2$\times$1800  \\
PBC J0746.2$-$1610   &07 46 16.87 &$-$16 11 28.2     & 3.8     & 07 46 17.12&$-$16 11 27.7	  & SPM 2.1m+B\&C Spc. &3450-7650& 4.0 & 11 Mar 2010, 06:03 & 2$\times$1800  \\
PBC J0747.7$-$7326   &07 47 38.88 &$-$73 25 47.8     & 5.2     &07 47 38.39 &$-$73 25 53.3	  & CTIO 1.5m+RC Spec. & 3300-10500 & 5.7 & 31 Dec 2009, 03:05 & 2$\times$900  \\
PBC J0749.2$-$8634   &07 50 46.15 &$-$86 32 09.5     & 3.6     &07 50 47.21 &$-$86 32 11.8	  & CTIO 1.5m+RC Spec. & 3300-10500 & 5.7 & 03 Dec 2010, 06:58 & 2$\times$1500  \\
PBC J0803.4+0840   &08 03 27.08 &+08 41 51.6     & 3.6     &08 03 27.37 &+08 41 52.3	  & Cassini+BFOSC  & 3500-8700 & 4.0 & 17 Jan 2011, 22:51 & 2$\times$1800 \\
PBC J0818.5$-$1420   &08 18 20.24 &$-$14 25 50.9      & 4.5     &08 18 20.26 &$-$14 25 52.8 & Copernicus+AFOSC &4000-8000& 4.2  & 10 Feb 2011, 23:32  & 2$\times$1800 \\
PBC J0820.4$-$2801  &08 20 34.05 &$-$28 05 00.6     & 3.6     &08 20 34.11 &$-$28 04 58.8	  & SPM 2.1m+B\&C Spc. &3450-7650& 4.0 & 13 Mar 2010, 05:55  & 2$\times$1800 \\
PBC J0855.8$-$2855$^+$  &08 55 17.59 &$-$28 54 19.2     & 3.8     &08 55 17.46 &$-$28 54 21.8	  & SPM 2.1m+B\&C Spc. &3450-7650& 4.0 & 11 Mar 2010, 05:32  & 2$\times$1800 \\
PBC J0859.5+4456  &08 59 30.41 &+44 54 49.2     & 4.1     &08 59 30.46 &+44 54 50.4	  & Cassini+BFOSC  & 3500-8700 & 4.0  & 10 Mar 2011, 21:20  & 2$\times$1200 \\
PBC J0917.2$-$6454  &09 17 27.91 &$-$64 56 25.1     & 3.7     &09 17 27.16 &$-$64 56 27.1	  & CTIO 1.5m+RC Spec. & 3300-10500 & 5.7 & 28 Jan 2010, 04:09  & 2$\times$1800 \\
PBC J0927.8$-$6945  &09 27 52.61 &$-$69 44 39.1     & 3.8     &09 27 53.09 &$-$69 44 41.9	  & CTIO 1.5m+RC Spec. & 3300-10500 & 5.7 & 03 Dec 2010, 08:06  & 2$\times$1500 \\
PBC J0929.6+6231  &09 29 37.74 &+62 32 38.1     & 4.5     &09 29 37.91 &+62 32 38.3	  & Cassini+BFOSC  & 3500-8700 & 4.0  & 21 Mar 2011, 23:57  & 2$\times$1800 \\
PBC J0942.1+2342  &09 42 04.54 &+23 41 08.3     & 3.7     &09 42 04.27 &+23 41 06.6	  & SPM 2.1m+B\&C Spc. &3450-7650& 4.0 & 10 Mar 2010, 07:54  & 2$\times$1800 \\
PBC J0950.0+7315$^C$   &09 49 46.00 &+73 14 23.1     & 0.6     &09 49 45.97 &+73 14 23.3	  & Cassini+BFOSC  & 3500-8700 & 4.0 & 18 May 2009, 20:18  & 1800 \\
PBC J1002.3+0304  &10 02 07.17 &+03 03 25.1    & 5.3    &10 02 07.01&+03 03 27.7	  & Cassini+BFOSC  & 3500-8700 & 4.0 & 17 Jan 2011, 00:09  & 2$\times$1800 \\
PBC J1017.2$-$0404  &10 17 16.89 &$-$04 04 53.5     & 3.9     &10 17 16.81&$-$04 04 55.9	  & SPM 2.1m+B\&C Spc. &3450-7650& 4.0  & 30 Jan 2009, 08:27 & 2$\times$1800 \\
{\bf PBC J1020.5$-$0235} &10 19 58.57 &$-$02 34 36.7    & 3.7     &10 19 58.56&$-$02 34 36.3	  & Cassini+BFOSC  & 3500-8700 & 4.0  & 09 Mar 2011, 23:49 & 2$\times$1800 \\
PBC J1034.2+7301  &10 34 23.14 &+73 00 47.4     & 4     &10 34 23.67 &+73 00 49.9	  & Cassini+BFOSC  & 3500-8700 & 4.0  & 10 Mar 2011, 23:13 & 1800 \\
\noalign{\smallskip}
\hline
\hline
\end{tabular}}
\end{center}
\end{table}
\end{landscape}

\begin{landscape}
\begin{table}
\setcounter{table}{0}
\caption{-- \emph{continued}}
\scriptsize
\begin{center}
\resizebox{25cm}{!}{
\begin{tabular}{lcccccccccc}
\hline
\hline
\multicolumn{1}{c}{{\it (1)}} & \multicolumn{1}{c}{{\it (2)}} & \multicolumn{1}{c}{{\it (3)}} & \multicolumn{1}{c}{{\it (4)}} &  \multicolumn{1}{c}{{\it (5)}}&\multicolumn{1}{c}{{\it (6)}}&
{\it (7)} & {\it (8)} & {\it (9)} & \multicolumn{1}{c}{{\it (10)}}& \multicolumn{1}{c}{{\it (11)}}\\
\multicolumn{1}{c}{Object} & \multicolumn{1}{c}{RA X-ray} & \multicolumn{1}{c}{Dec X-ray} &\multicolumn{1}{c}{Error radius} & \multicolumn{1}{c}{RA Opt} &\multicolumn{1}{c}{Dec Opt}&\multicolumn{1}{c}{Telescope+instrument} & $\lambda$ range & Disp. & \multicolumn{1}{c}{UT Date \& Time}  & Exposure \\
& \multicolumn{1}{c}{} & \multicolumn{1}{c}{} & (arcsec) &\multicolumn{1}{c}{{\it (J2000)}}  &\multicolumn{1}{c}{{\it (J2000)}} & & (\AA) & (\AA/pix) & 
\multicolumn{1}{c}{at mid-exposure} & time (s)  \\
\noalign{\smallskip}
\hline
\noalign{\smallskip}
PBC J1113.6+7942  &11 14 43.98 &+79 43 37.2     & 4.4     &11 14 43.86 & +79 43 35.7	  & Cassini+BFOSC  & 3500-8700 & 4.0  & 20 Apr 2010, 22:22 & 1800 \\
PBC J1115.3+5425  &11 15 20.31 &+54 23 15.3     & 4.4     &11 15 19.92 & +54 23 16.8	  & SPM 2.1m+B\&C Spc. &3450-7650& 4.0  & 30 Jan 2009, 09:49 & 2$\times$1800 \\
PBC J1145.4+5858  &11 45 32.78 &+58 58 38.4     & 3.7     &11 45 33.18 & +58 58 40.9	  & SPM 2.1m+B\&C Spc. &3450-7650& 4.0  & 11 Mar 2010, 09:14 & 2$\times$1800 \\
PBC J1231.4+5759  &12 31 21.95 &+57 57 52.9     & 5.8     &12 31 22.14 & +57 57 52.9	  & Cassini+BFOSC  & 3500-8700 & 4.0   & 15 Dec 2010, 04:12 & 900+1200 \\
PBC J1240.8+2736  &12 40 46.52 &+27 33 52.6     & 4.6    &12 40 46.41 &+27 33 53.5	         & Cassini+BFOSC  & 3500-8700 & 4.0   & 22 Mar 2011, 20:53 & 3$\times$1800 \\
PBC J1254.8$-$2655  &12 54 56.30 &$-$26 57 00.5     & 3.6    &12 54 56.37  &$-$26 57 02.1	 & SPM 2.1m+B\&C Spc. &3450-7650& 4.0    & 12 Mar 2010, 08:51  & 2$\times$1800 \\
PBC J1321.1+0858  & 13 20 59.29 &+08 58 42.9     & 7.1    &13 20 59.61  &+08 58 42.2	 & SPM 2.1m+B\&C Spc. &3450-7650& 4.0    & 11 Mar 2010, 10:23  & 2$\times$1800 \\
PBC J1349.0+4443  & 13 49 08.34 &+44 41 31.2     & 3.9    &13 49 08.42  &+44 41 29.5	 & Cassini+BFOSC  & 3500-8700 & 4.0     & 22 Mar 2011, 21:09  & 1800 \\
PBC J1355.5+3523  & 13 55 34.08 &+35 20 59.8     & 4    &13 55 33.83  &+35 20 57.4	 & Cassini+BFOSC  & 3500-8700 & 4.0     & 01 Apr 2011, 01:15  & 2$\times$1800 \\
PBC J1416.8$-$1158  & 14 16 50.08 &$-$11 59 00.8     & 3.6    &14 16 50.02 &$-$11 58 57.7 & SPM 2.1m+B\&C Spc. &3450-7650& 4.0     & 12 Mar 2010, 10:08  & 2$\times$1800 \\
PBC J1540.3+1415: &  &     & & & & &&   &   &  \\
$\;$$\;$$\;$PBC J1540.3+1415--1 & 15 40 13.24 &+14 16 41.5     & 4.6    &15 40 12.96 &+14 16 43.3 & SPM 2.1m+B\&C Spc. &3450-7650& 4.0     & 15 Jun 2012, 00:04  & 2$\times$1800 \\
$\;$$\;$$\;$PBC J1540.3+1415--2 & 15 40 13.24 &+14 16 41.5     & 4.6    &15 40 12.35 &+14 16 59.3 & SPM 2.1m+B\&C Spc. &3450-7650& 4.0     & 15 Jun 2012, 23:04  & 2$\times$1800 \\
$\;$$\;$$\;$PBC J1540.3+1415--3$^-$& 15 40 07.94 &+14 11 37.9     &3.7    &15 40 07.85 &+14 11 37.1 & SPM 2.1m+B\&C Spc. &3450-7650& 4.0     & 11 Mar 2010, 11:48  & 2$\times$1800 \\
PBC J1821.2+5957 & 18 21 26.79 &+59 55 19.9     &4.1    &18 21 26.81 &+59 55 20.9 & Cassini+BFOSC  & 3500-8700 & 4.0      & 02 Jul 2010, 21:44  & 1800 \\
PBC J1824.2+1846 & 18 24 10.76 &+18 46 09.0     &4.2    &18 24 10.83 &+18 46 08.8 & Cassini+BFOSC  & 3500-8700 & 4.0      & 01 Jul 2010, 22:22  & 2$\times$1200 \\
PBC J1826.6+3251& 18 26 32.49 &+32 51 27.0     &3.8    &18 26 32.39 &+32 51 30.1 & Cassini+BFOSC  & 3500-8700 & 4.0      & 01 Jul 2010, 20:39  & 2$\times$900 \\
PBC J1846.0+5607& 18 45 56.82 &+56 10 02.5     &5.2    &18 45 56.89 &+56 10 02.3 & Cassini+BFOSC  & 3500-8700 & 4.0      & 10 Apr 2011, 02:40  & 1800 \\
PBC J1903.7+3349& 19 03 48.96 &+33 50 38.3     &3.7    &19 03 49.16 &+33 50 40.8 & Cassini+BFOSC  & 3500-8700 & 4.0      & 08 May 2011, 00:45  & 2$\times$1200 \\
PBC J1926.6+4131& 19 26 30.48 &+41 33 01.4     &3.5    &19 26 30.18&+41 33 05.3 & SPM 2.1m+B\&C Spc. &3450-7650& 4.0     & 14 Jul 2010, 09:29  & 1800 \\
PBC J2010.2+4759& 20 10 17.20 &+48 00 23.1     &4.5    &20 10 17.40 &+38 00 21.5 &Cassini+BFOSC  & 3500-8700 & 4.0         & 17 Nov 2011, 17:30  & 2$\times$1800 \\
PBC J2035.2+2604& 20 35 05.44 &+26 03 29.2    &3.7    &20 35 05.66 &+26 03 30.2 &SPM 2.1m+B\&C Spc. &3450-7650& 4.0    & 15 Jul 2010, 08:48  & 2$\times$1800 \\
PBC J2116.2+2519& 21 16 10.36 &+25 16 58.3     &3.6    &21 16 10.28 &+25 17 01.1 &SPM 2.1m+B\&C Spc. &3450-7650& 4.0    & 17 Jul 2010, 10:30  & 1800 \\
PBC J2123.9+3407& 21 24 00.12 &+34 09 12.4     &4.4    &21 24 00.28 &+34 09 11.4 &SPM 2.1m+B\&C Spc. &3450-7650& 4.0    & 03 Nov 2010, 04:34  & 2$\times$1800 \\
PBC J2124.5+0503& 21 24 12.35 &+05 02 43.7     &3.9   &21 24 12.44 &+05 02 43.6 &Cassini+BFOSC  & 3500-8000 & 4.0        & 23 Aug 2011, 00:22  & 2$\times$600 \\
PBC J2150.7+1405& 21 50 46.92&+14 06 37.0     &4.6   &21 50 46.76&+14 06 36.9 &SPM 2.1m+B\&C Spc. &3450-7650& 4.0         & 11 Dec 2009, 03:31  & 2$\times$1800 \\
PBC J2157.4$-$0611& 21 57 26.83&$-$06 10 18.4     &3.6   &21 57 26.78&$-$06 10 17.6 &SPM 2.1m+B\&C Spc. &3450-7650& 4.0         & 31 Oct 2010, 02:58  &1800 \\
PBC J2248.8+1725& 22 48 44.27&+17 27 01.4    &3.6   &22 48 44.31&+17 27 03.5  &Cassini+BFOSC  & 3500-8000 & 4.0         & 05 Dec 2011, 19:41  &2$\times$1800 \\
PBC J2307.8+2244& 23 07 49.31&+22 42 17.4    &5.6   &23 07 48.88&+22 42 36.8   &Cassini+BFOSC  & 3500-8000 & 4.0         & 23 Aug 2011, 02:14  &2$\times$1800 \\
PBC J2307.9+4015& 23 07 57.32&+40 16 37.7    &3.8   &23 07 57.25&+40 16 39.4   &SPM 2.1m+B\&C Spc. &3450-7650& 4.0    & 31 Oct 2010, 05:24  &2$\times$1800 \\
PBC J2343.8+0539& 23 43 59.37&+05 38 22.5    &4.2   &23 43 59.56&+05 38 23.4   &Cassini+BFOSC  & 3500-8000 & 4.0      & 27 Jul 2010, 01:51  &2$\times$1800 \\
\noalign{\smallskip}
\hline
\hline
\noalign{\smallskip} 
\multicolumn{11}{l}{Note: if not indicated otherwise, source X-ray coordinates were obtained from XRT data.}\\
\multicolumn{11}{l}{$^=$ The reported optical coordinates are obtained from USN-A2.0 catalog.} \\
\multicolumn{11}{l}{$^C$ The reported X-ray coordinates are obtained from Chandra data.} \\ 
\multicolumn{11}{l}{$^R$ The reported X-ray coordinates are obtained from ROSAT bright catalogue.} \\
\multicolumn{11}{l}{$^+$ This source is outside  BAT 90\% error box, but inside the 99\% one.} \\
\multicolumn{11}{l}{$^-$ This source is at the edge of  BAT 90\% error box, inside the 99\% one.}\\
\multicolumn{11}{l}{\bf The source is outside BAT 90\% and 99\% error box, but it is associated with a 2MASS source in the 70 month BAT catalog.}\\
\noalign{\smallskip}
\noalign{\smallskip}
\end{tabular}}
\end{center}
\end{table}
\end{landscape}

\begin{table*}
\begin{center}
\caption[]{Main results obtained from the analysis of the optical spectra of the 35 type 1 AGNs.}\label{agn1}
\scriptsize
\begin{tabular}{lccccccccc}
\noalign{\smallskip}
\hline
\hline
\noalign{\smallskip}
\multicolumn{1}{c}{Object} & $F_{\rm H_\alpha}$$^*$ & $F_{\rm H_\beta}$ &
$F_{\rm [OIII]}$ & Class & $z$ & \multicolumn{1}{c}{$D_L$} & \multicolumn{2}{c}{$E(B-V)$} & \multicolumn{1}{c}{2MASS Name}\\
\cline{8-9}
\noalign{\smallskip}
& & & & & & (Mpc) & Gal. & AGN& \\
\noalign{\smallskip}
\hline
\noalign{\smallskip}

PBC J0000.9$-$0708 & 16.2$\pm$2.8 &1.67$\pm$0.53 & 26.3$\pm$2.1& Sy1.9 & 0.038 & 167.5 & 0.034  &0.869 & 2MASX J00004876$-$0709117       \\
& [25.4$\pm$5.7] &[3.59$\pm$0.45]  & [31.5$\pm$2.3] & & & & & &\\

& & & & & & & & & \\

PBC J0050.8+7648 & 129$\pm$8.3 & 28.8$\pm$5.2 & 12.8$\pm$1.5 & Sy1.2 & 0.128 & 600.1 & 0.435 & 0.074 &  2MASS J00510664+7650359     \\
& [297$\pm$31.7]& [96.1$\pm$16.9] & [41.1$\pm$4.1] & & & & & &\\

& & & & & & & & & \\

PBC J0116.3+3102 & 84.1$\pm$51.8 & abs. & 18.6$\pm$4.4 & Sy1.9 & 0.018 & 78.2 & 0.070 & -- &  2MASX J01161480+3102017   \\
& [98.2$\pm$59.7]&  [abs.] & [23.1$\pm$4.8] & & & & & &\\

& & & & & & & & & \\

PBC J0149.3$-$5017 & 125$\pm$17 & 34.1$\pm$6.3 & 16.0$\pm$1.3 & Sy1.2 & 0.0299 & 131.0 & 0.022 & 0.224 &   2MASX J01492228-5015073    \\
& [134.0$\pm$16.5]&[37.1$\pm$8.2]& [17.4$\pm$2.1] & & & & & &\\

& & & & & & & & & \\

PBC J0157.3+4715 & 378$\pm$49 & 36.9$\pm$10.1 & 11.3$\pm$2.3 & Sy1.2 & 0.049 & 217.7 & 0.165 & 0.859 &  2MASX J01571097+4715588     \\
& [563$\pm$68]&[80.4$\pm$13.9]& [18.5$\pm$3.2] & & & & & &\\

& & & & & & & & & \\

PBC J0311.9+5029 & 85.9$\pm$10.1 & 11.2$\pm$3.5 & 19.3$\pm$1.7 & Sy1.5 & 0.062 & 278.0 & 0.759 & 0 &  2MASX J03120291+5029147     \\
& [442$\pm$44]&[158$\pm$36]& [183$\pm$16] & & & & & &\\

& & & & & & & & & \\

PBC J0429.7$-$6703 & 212$\pm$18 & 53.1$\pm$7.6 & 30.9$\pm$3.2 & Sy1.5 & 0.065 & 292.1 & 0.044& 0.312 &  2MASX J04294735$-$6703205     \\
& [234$\pm$19]&[59.1$\pm$10.2]& [34.5$\pm$3.8] & & & & & &\\

& & & & & & & & & \\

PBC J0440.8+2739 & 89.2$\pm$10.8 & 5.6$\pm$1.7 & 9.3$\pm$0.9 & Sy1.5 & 0.038 & 167.5 & 0.805& 0.931 &  2MASX J04404770+2739466   \\
& [497$\pm$65]&[65.9$\pm$19.5]& [102$\pm$12] & & & & & &\\

& & & & & & & & & \\

PBC J0532.7+1346 & 33.5$\pm$7.8 & < 0.4& 1.3$\pm$0.5 & Sy1.5 & 0.024 & 104.7 & 0.843& >2.831 &  2MASX J05325752+1345092  \\
& [224$\pm$40]& [<4.1] & [17.5$\pm$5.7] & & & & & &\\

& & & & & & & & & \\

PBC J0535.6+4011& 133$\pm$13 & 20.1$\pm$5.3& 31.4$\pm$2.4  & Sy1.5 & 0.021 & 91.4 & 0.635& 0.166&  2MASX J05353211+4011152  \\
& [551$\pm$47]& [162$\pm$30] & [198$\pm$13] & & & & & &\\

& & & & & & & & & \\

PBC J0543.9$-$4325& 5.4$\pm$0.8 & 1.4$\pm$0.4& 4.8$\pm$0.5  & Sy1.8 & 0.045 & 199.3 & 0.051& 0.408&  2MASX J05440009$-$4325265  \\
& [10.5$\pm$1.7]& [2.4$\pm$0.8] & [10.8$\pm$1.4] & & & & & &\\

& & & & & & & & & \\

PBC J0602.5+6522& 460$\pm$251& <0.3& 44.5$\pm$5.8  & Sy1.9 & 0.017& 73.8 & 0.108& >5.035&  2MASX J06023793+6522161  \\
& [597$\pm$324]&[<1.1] & [62.5$\pm$8.1] & & & & & &\\

& & & & & & & & & \\

PBC J0609.4$-$6243& 87.6$\pm$14.1& 8.5$\pm$1.7& 4.9$\pm$1.2  & Sy1.5 & 0.157& 749.7 & 0.053& 1.286&  2MASX J06100652$-$6243125  \\
& [99.4$\pm$15.1]& [9.1$\pm$1.8] & [6.8$\pm$1.4] & & & & & &\\

& & & & & & & & & \\

PBC J0635.0$-$7441  & 297$\pm$28& 70.3$\pm$14.2& 12.7$\pm$4.5  & Sy1 & 0.112& 519.7 & 0.106& 0.144&  2MASX J06340354$-$7446376  \\
& [369$\pm$41]& [111$\pm$20] & [17.9$\pm$3.6] & & & & & &\\

& & & & & & & & & \\

PBC J0654.5+0703  & 317$\pm$34& 22.8$\pm$5.5& 12.6$\pm$1.9  & Sy1.5 & 0.024& 104.7 & 0.295& 1.152&  2MASX J06543417+0703210  \\
& [590$\pm$49]& [62.1$\pm$13.1] & [137$\pm$6.8] & & & & & &\\

& & & & & & & & & \\

PBC J0749.2$-$8634  & 153$\pm$14 & 25.3$\pm$4.6& 10.1$\pm$1.9  & Sy1.2 & 0.109& 504.7 & 0.180& 0.647&  2MASX J07504720$-$8632118 \\
& [220$\pm$21]& [39.2$\pm$6.7] & [16.2$\pm$3.2] & & & & & &\\

& & & & & & & & & \\

PBC J0803.4+0840  & 215$\pm$34 & 12.1$\pm$3.9 & 93.7$\pm$3.9  & Sy1.5 & 0.047& 208.5 & 0.022& 1.582&  2MASX J08032736+0841523 \\
& [217$\pm$40]& [14.6$\pm$5.1] & [98.2$\pm$5.2] & & & & & &\\

& & & & & & & & & \\

PBC J0818.5$-$1420  & 44.1$\pm$7.3 & 16.1$\pm$2.7 & 10.1$\pm$1.1  & Sy1.5 & 0.107& 494.8 & 0.074& 0.024&  2MASS J08182026$-$1425527 \\
& [47.5$\pm$8.2]& [16.2$\pm$2.8] & [12.3$\pm$1.4] & & & & & &\\

& & & & & & & & & \\

PBC J0917.2$-$6454  & 49.8$\pm$6.9 & 8.3$\pm$2.2 & 71.4$\pm$5.1  & Sy1.9 & 0.086& 393.1 & 0.224& 0.562&  2MASX J09172721$-$6456270 \\
& [85.3$\pm$11.4]& [16.6$\pm$4.1] & [132$\pm$10.7] & & & & & &\\

& & & & & & & & & \\

PBC J0942.1+2342  & 55.1$\pm$17.3 & 48.1$\pm$0.4 & 37.1$\pm$4.5  & Sy1.5 & 0.022& 95.8 & 0.027& 0&  2MASX J09420476+2341066 \\
& [81.1$\pm$19.1]& [70.8$\pm$23.5] & [37.4$\pm$4.6] & & & & & &\\

& & & & & & & & & \\

PBC J1002.3+0304  & 62.8$\pm$9.8& 5.8$\pm$02.7 & 39.1$\pm$6.7  & Sy1.8/1.9 & 0.023& 100.2 & 0.082& 1.147&  2MASX J10020701+0303277 \\
& [76.5$\pm$10.4]& [8.1$\pm$3.1] & [51.7$\pm$7.1] & & & & & &\\

& & & & & & & & & \\

PBC J1020.5$-$0235  & 94.4$\pm$32.1& 76.3$\pm$30.2 & 6.4$\pm$2.5  & Sy1 & 0.060& 268.7 & 0.044& 0  &  2MASX J10195855$-$0234363 \\
& [94.5$\pm$32.4]& [76.4$\pm$30.9] & [7.1$\pm$2.6] & & & & & &\\

& & & & & & & & & \\

PBC J1254.8$-$2655  & 523$\pm$65& 72.2$\pm$22.6 & 72.3$\pm$4.9  & Sy1.5 & 0.060& 268.7 & 0.081& 0.813  &  2MASX J12545637$-$2657021 \\
& [660$\pm$77]& [98.9$\pm$30.7] & [94.6$\pm$6.9] & & & & & &\\

& & & & & & & & & \\

PBC J1349.0+4443  & 22.6$\pm$4.4& 17.5$\pm$6.4 & 7.3$\pm$1.1  & Sy1.2 & 0.065& 292.1 & 0.015& 0  &  2MASX J13490841+4441295 \\
& [22.6$\pm$4.4]& [23.1$\pm$7.3] & [8.1$\pm$1.4] & & & & & &\\

& & & & & & & & & \\

PBC J1416.8$-$1158  & 132$\pm$31& 55.5$\pm$11.1 & 33.7$\pm$2.4  & Sy1.5 & 0.099 & 455.4 & 0.072& 0.634  &  2MASX J14165001$-$1158577 \\
& [416$\pm$72]& [75.1$\pm$15.1] & [41.2$\pm$2.9] & & & & & &\\

& & & & & & & & & \\

\hline
\hline
\end{tabular}
\end{center}
\end{table*} 

\begin{table*}
\setcounter{table}{1}
\caption{-- \emph{continued}}
\scriptsize
\begin{center}
\begin{tabular}{lccccccccc}
\hline
\hline
\noalign{\smallskip}
\multicolumn{1}{c}{Object} & $F_{\rm H_\alpha}$$^*$ & $F_{\rm H_\beta}$ &
$F_{\rm [OIII]}$ & Class & $z$ & \multicolumn{1}{c}{$D_L$} & \multicolumn{2}{c}{$E(B-V)$} & \multicolumn{1}{c}{2MASS Name}\\
\cline{8-9}
\noalign{\smallskip}
& & & & & & (Mpc) & Gal. & AGN& \\
\noalign{\smallskip}
\hline
\hline
\noalign{\smallskip}
& & & & & & & & & \\

PBC J1540.3+1415--3  & 146$\pm$15& 34.1$\pm$7.5 & 12.2$\pm$1.1  & Sy1.2 & 0.12 & 559.7  & 0.052& 0.522  &  2MASS J15400784+1411371 \\
& [168$\pm$18]& [36.3$\pm$7.9] & [14.1$\pm$1.6] & & & & & &\\

& & & & & & & & & \\

PBC J1824.2+1846  & 80.1$\pm$11.7& 9.6$\pm$2.6 & 90.9$\pm$5.1  & Sy1.9 & 0.067 & 301.5  & 0.223& 0.774  &  2MASX J18241083+1846088 \\
& [116$\pm$17]& [18.1$\pm$4.9] & [174$\pm$9.8] & & & & & &\\

& & & & & & & & & \\

PBC J1846.0+5607  & 20.9$\pm$2.7& 2.1$\pm$0.6 & 7.9$\pm$0.5  & Sy1.5 & 0.07 & 315.7  & 0.058& 0.931  &  2MASX J18455688+5610022\\
& [23.4$\pm$3.1]& [3.1$\pm$0.9] & [9.4$\pm$0.6] & & & & & &\\

& & & & & & & & & \\

PBC J1926.6+4131  & 31.1$\pm$24.2& <0.8 & <0.8  & Sy1.9 & 0.072& 325.1  & 0.112& >2.46  &  2MASX J19263018+4133053\\
& [33.4$\pm$25.9]&[<0.9] &[<1.3] & & & & & &\\

& & & & & & & & & \\

PBC J2035.2+2604  & 258$\pm$18& 28.2$\pm$7.9 & 76.5$\pm$4.5 & NLS1 & 0.05 & 222.3  & 0.283& 0.834  &  2MASX J20350566+2603301\\
& [476$\pm$33]& [69.8$\pm$19.6] & [179$\pm$10.5] & & & & & &\\

& & & & & & & & & \\

PBC J2116.2+2519  & 83.4$\pm$12.8& 23.7$\pm$2.6 & 5.4$\pm$0.6 & Sy1.2 & 0.153 & 728.8  & 0.130& 0.023  &  2MASX J21161028+2517010\\
& [102$\pm$16]& [34.8$\pm$3.8] & [7.6$\pm$0.8] & & & & & &\\

& & & & & & & & & \\

PBC J2123.9+3407  & 149$\pm$20& 18.3$\pm$5.1 & 18.8$\pm$1.1 & Sy1.5 & 0.083 & 377.7  & 0.144& 0.916  &  2MASX J21240027+3409114\\
& [202$\pm$27]& [27.2$\pm$7.6] & [28.7$\pm$1.7] & & & & & &\\

& & & & & & & & & \\

PBC J2157.4$-$0611  & 3570$\pm$772& 394$\pm$94.9 & 305$\pm$34.2 & Sy1.5 & 0.176 & 850.2  & 0.03& 1.064  &  2MASS J21572678$-$0610176\\
& [3690$\pm$798]& [426$\pm$103] & [329$\pm$36.9] & & & & & &\\

& & & & & & & & & \\

PBC J2248.8+1725  & 15.2$\pm$3.9& 1.3$\pm$0.5 & 8.9$\pm$0.8 & Sy1.8 & 0.086 & 392.1  & 0.083&   1.24&  2MASS J21572678$-$0610176\\
& [17.7$\pm$4.5]& [1.7$\pm$0.7] & [10.8$\pm$0.9] & & & & & &\\

& & & & & & & & & \\

PBC J2307.9+4015  & 441$\pm$93.8& 27.1$\pm$9.7 & 23.7$\pm$2.1 & Sy1.5 & 0.073 & 329.9  & 0.122& 1.619  &  2MASX J23075724+4016393\\
& [558$\pm$118]& [36.1$\pm$12.9] & [34.1$\pm$3.1] & & & & & &\\

& & & & & & & & & \\

\noalign{\smallskip} 
\noalign{\smallskip} 
\hline
\noalign{\smallskip} 
\multicolumn{10}{l}{Note: emission line fluxes are reported both as
observed and (between square brackets) corrected for the intervening Galactic} \\ 
\multicolumn{10}{l}{absorption $E(B-V)_{\rm Gal}$ along the object line of sight 
(from Schlegel et al. 1998). Line fluxes are in units of 10$^{-15}$ erg cm$^{-2}$ s$^{-1}$,} \\
\multicolumn{10}{l}{The typical error on the redshift measurement is $\pm$0.001}\\
\multicolumn{10}{l}{$^*$: blended with [N {\sc ii}] lines} \\
\noalign{\smallskip} 
\hline
\hline
\end{tabular} 
\end{center}
\end{table*}

\begin{table*}[th!]
\begin{center}
\caption[]{Main results obtained from the analysis of the optical spectra of the 33 type 2 AGNs.}\label{agn2}
\scriptsize
\begin{tabular}{lccccccccc}
\noalign{\smallskip}
\hline
\hline
\noalign{\smallskip}
\multicolumn{1}{c}{Object} & $F_{\rm H_\alpha}$$^*$ & $F_{\rm H_\beta}$ &
$F_{\rm [OIII]}$ & Class & $z$ & \multicolumn{1}{c}{$D_L$} & \multicolumn{2}{c}{$E(B-V)$} & \multicolumn{1}{c}{2MASS Name}\\
\cline{8-9}
\noalign{\smallskip}
& & & & & & (Mpc) & Gal. & AGN& \\
\noalign{\smallskip}
\hline
\hline
\noalign{\smallskip}
PBC J0034.6$-$0424& -- & 1.44$\pm$0.31 &17.0$\pm$0.4 & Sy 2 & 0.213 & 1051.6 & 0.039 & -- & 2MASX J00343284$-$0424117   \\
& & [1.61$\pm$0.37] & [18.7$\pm$1.1] & & & & & & \\

& & & & & & & & & \\

PBC J0038.5+2336& 3.65$\pm$1.18 & abs. &9.31$\pm$1.09 & Sy 2 & 0.025 & 109.1& 0.028 & -- & 2MASX J00383214+2336475   \\
& [3.84$\pm$2.41]& [abs.] & [9.23$\pm$2.37] & & & & & & \\

& & & & & & & & & \\

PBC J0128.6$-$6038& 98.2$\pm$45.4 & 14.2$\pm$3.7  &83.5$\pm$25.2 & Sy 2 & 0.203 & 996.5& 0.024 & 1.056 & 2MASX J01290761$-$6038423   \\
& [128.0$\pm$46.5]& [21.4$\pm$6.9] & [92.9$\pm$21.1] & & & & & & \\

& & & & & & & & & \\

PBC J0223.4+4549& 59.8$\pm$11.2 & 6.9$\pm$2.8  &107$\pm$7 & Sy 2 & 0.062 & 278.0 & 0.112 & 1.001 & 2MASX J0223330+4549162 \\
& [76.2$\pm$29.2]& [9.4$\pm$2.8] & [147$\pm$10] & & & & & & \\

& & & & & & & & & \\

PBC J0238.3$-$6116& 12.4$\pm$2.4 & abs.  &7.3$\pm$0.8 & Sy 2 & 0.054 & 240.8 & 0.024 & -- & 2MASX J02384313$-$6117227   \\
&[12.5$\pm$2.5] & [abs.] & [7.9$\pm$1.1] & & & & & & \\

& & & & & & & & & \\

PBC J0243.9+5323& 4.5$\pm$1.1 & abs. &1.6$\pm$0.3 & Likely Sy 2 & 0.036 & 158.4 & 0.409 & -- & 2MASX J02440296+5328281   \\
& [10.8$\pm$2.1]& [abs.]& [5.9$\pm$1.2] & & & & & & \\

& & & & & & & & & \\

PBC J0359.0$-$3017 & 16.5$\pm$4.5 & 4.2$\pm$0.7 &18.4$\pm$1.5 & Sy 2 & 0.094 & 430.9 & 0.01 & 0.93 & 2MASX J03590885$-$3018102   \\
& [32.7$\pm$4.7]& [4.3$\pm$0.6] & [19.5$\pm$1.5] & & & & & & \\

& & & & & & & & & \\

PBC J0459.8+2705 & 37.2$\pm$3.9 & 1.8$\pm$0.6 & 16.1$\pm$1.3 & Sy 2 & 0.061 & 273.3 & 1.212 & 1.013 & 2MASX J04595608+2706024  \\
& [448$\pm$64]& [54.5$\pm$13.3] & [484$\pm$34] & & & & & & \\

& & & & & & & & & \\

PBC J0515.3+1856 & 42.5$\pm$3.4 & 1.5$\pm$0.4 & 31.9$\pm$2.7 & Sy 2 & 0.023 & 100.2 & 0.606 & 1.714 & 2MASX J05151978+1854515  \\
& [168$\pm$16]& [9.9$\pm$4.5] & [217$\pm$16] & & & & & & \\

& & & & & & & & & \\

PBC J0709.2$-$3601 & 5.5$\pm$0.7 & <0.9 & 3.9$\pm$0.9 & Sy 2 & 0.110 & 509.7 & 0.745 & >0.786 & 2MASX J07091409$-$3601219  \\
& [25.3$\pm$3.8]&  [<3.9]& [31.7$\pm$3.1] & & & & & & \\

& & & & & & & & & \\

PBC J0747.7$-$7326 & 0.5$\pm$0.1 & abs. & 8.2$\pm$1.2 & LINER & 0.036 & 158.4 & 0.440 & -- & 2MASX J07473839$-$7325533 \\
& [7.4$\pm$1.4]&  [abs.]& [32.4$\pm$4.7] & & & & & & \\

& & & & & & & & & \\

PBC J0855.8$-$2855  & 3.4$\pm$0.3  & <0.9 & <0.3&Likely Sy 2 & 0.073& 329.9 &0.150 & >0.133&  2MASX J08551746$-$2854218\\
& [4.6$\pm$0.4]& [<1.4] & [<0.4]& & & & & &\\

& & & & & & & & & \\

PBC J0859.5+4457& 88.9$\pm$8.8 & 11.4$\pm$3.1 & 10.6$\pm$3.3 & Trans. obj & 0.006 & 26.2 & 0.022 & 0.927 & 2MASX J08593045+4454504\\
& [90.9$\pm$24.9]& [12.1$\pm$3.3] & [12.5$\pm$4.2] & & & & & & \\

& & & & & & & & & \\

PBC J0929.6+6231& 5.3$\pm$2.2 & -- & 8.1$\pm$1.1 & Sy 2 & 0.025 & 109.1 & 0.033 & -- & 2MASX J09293791+6232382\\
& [5.9$\pm$2.3]&  & [8.9$\pm$1.2] & & & & & & \\

& & & & & & & & & \\

PBC J0950.0+7315 & 13.4$\pm$3.6 & 1.6$\pm$0.5 & 26.9$\pm$3.7 & Sy 2 & 0.058 & 259.3 & 0.027 & 0.993 & 2MASX J09494596+7314232\\
& [14.5$\pm$3.7]&[1.8$\pm$0.6]  & [29.2$\pm$3.9] & & & & & & \\

& & & & & & & & & \\

PBC J1017.2$-$0404 & 12.4$\pm$1.6 & 1.3$\pm$1.2 & 9.8$\pm$1.3 & Sy 2 & 0.041 & 181.1 & 0.047 & 1.279& 2MASX J10171680$-$0404558\\
& [14.1$\pm$4.3]&[1.3$\pm$1.2]  & [11.4$\pm$1.3] & & & & & & \\

& & & & & & & & & \\

PBC J1034.2+7301 & 8.4$\pm$3.1 & abs. & 7.6$\pm$3.9 & XBONG/ & 0.022 & 95.8& 0.128 & -- & 2MASX J10342366+7300499\\
& [17.4$\pm$4.4]&[abs.]  & [11.2$\pm$5.6] &Likely Sy 2 & & & & & \\

& & & & & & & & & \\

PBC J1113.6+7942 & 13.6$\pm$3.1 & <1.1& 23.4$\pm$3.1 & Sy 2 & 0.037 & 162.9& 0.046 & >1.266 & 2MASX J11144385+7943357\\
& [13.9$\pm$3.6]& [<1.3] & [26.9$\pm$3.1] & & & & & & \\

& & & & & & & & & \\

PBC J1115.3+5425 & 16.8$\pm$1.2 & 1.9$\pm$0.6 & 28.1$\pm$1.4 & Sy 2 & 0.071 & 320.4& 0.014 & 1.083 & 2MASX J11151992+5423167\\
& [16.8$\pm$1.3]& [1.9$\pm$0.6] & [28.9$\pm$1.5] & & & & & & \\

& & & & & & & & & \\

PBC J1145.4+5858 & abs. & <1.8& abs.& Type 2 & 0.008 & 34.5& 0.022& -- & 2MASX J11453317+5858408\\
& [abs.] & [<1.9]& [abs.] & & & & & & \\

& & & & & & & & & \\

PBC J1231.4+5759 & 211$\pm$20 & 32.3$\pm$3.7 & 78.6$\pm$4.8 & starburst & 0.011 & 47.5 & 0.013&  0.782& 2MASX J12312214+5757528\\
& [217$\pm$21]& [33.6$\pm$3.9] & [83.1$\pm$4.9] & galaxy & & & & & \\

& & & & & & & & & \\

PBC J1240.8+2736 & 35.4$\pm$5.7 & 1.1$\pm$0.5 & 5.6$\pm$1.1 & Sy 2& 0.056 & 250 & 0.017&  2.338& 2MASX J12404640+2733535\\
& [36.6$\pm$6.1]& [1.1$\pm$0.5] & [5.9$\pm$1.2] & & & & & & \\

& & & & & & & & & \\

PBC J1321.1+0858 & 60.1$\pm$4.2 & 9.9$\pm$3.9 & 12.6$\pm$3.1 & LINER& 0.033 & 144.9 & 0.031& 0.636 & 2MASX J13205961+0858421\\
& [63.8$\pm$4.2]& [11.5$\pm$4.3] & [12.7$\pm$3.2] & & & & & & \\

& & & & & & & & & \\
\hline
\hline
\end{tabular}
\end{center}
\end{table*} 

\begin{table*}
\setcounter{table}{2}
\caption{-- \emph{continued}}
\scriptsize
\begin{center}
\begin{tabular}{lccccccccc}
\hline
\hline
\noalign{\smallskip}
\multicolumn{1}{c}{Object} & $F_{\rm H_\alpha}$$^*$ & $F_{\rm H_\beta}$ &
$F_{\rm [OIII]}$ & Class & $z$ & \multicolumn{1}{c}{$D_L$} & \multicolumn{2}{c}{$E(B-V)$} & \multicolumn{1}{c}{2MASS Name}\\
\cline{8-9}
\noalign{\smallskip}
& & & & & & (Mpc) & Gal. & AGN& \\
\noalign{\smallskip}
\hline
\hline
\noalign{\smallskip}
& & & & & & & & & \\
PBC J1355.5+3523 & --& -- & -- &XBONG & 0.100 & 460.3 & 0.016& -- & 2MASX J13553383+3520573\\
& & & & & & & & & \\

& & & & & & & & & \\

PBC J1540.3+1415--1 & 20.9$\pm$1.6& 7.2$\pm$0.9 & 58.7$\pm$1.3 & Trans. obj. & 0.0508 & 225.5 & 0.049& 0 & 2MASX J15401296+1416432\\
& [23.5$\pm$1.8]& [8.4$\pm$1.1] & [68.1$\pm$1.4] & & & & & & \\

& & & & & & & & & \\

PBC J1540.3+1415--2 & 8.1$\pm$0.5 & 2.3$\pm$0.3 & 2.1$\pm$0.3 & Trans. obj. & 0.05 & 222.3 & 0.049& 0.123 & 2MASX J15401234+1416592\\
& [9.1$\pm$0.6]& [2.8$\pm$0.3] & [2.5$\pm$0.3] & & & & & & \\

& & & & & & & & & \\

PBC J1821.2+5957 & 24.9$\pm$4.1 & 3.5$\pm$1.4 & 36.7$\pm$2.3 & Sy2 & 0.099& 455.4 & 0.045& 0.772 & 2MASX J18212680+5955209\\
& [9.1$\pm$0.6]& [2.8$\pm$0.3] & [2.5$\pm$0.3] & & & & & & \\

& & & & & & & & & \\

PBC J1826.6+3251 & 17.4$\pm$3.5 & <1.5 & 31.7$\pm$2.1 & Sy2 & 0.022& 95.8 & 0.115&>0.896& 2MASX J18263239+3251300\\
& [21.1$\pm$4.2]& [<2.9] & [45.2$\pm$2.9] & & & & & & \\

& & & & & & & & & \\

PBC J1903.7+3349 & 35.7$\pm$11.2 & 4.8$\pm$1.5 & 60.4$\pm$3.2 & Sy2 & 0.015& 65 & 0.096& 0.907 & 2MASX J19034916+3350407\\
& [44.9$\pm$14.1]& [6.1$\pm$1.9] & [82.1$\pm$4.3] & & & & & & \\

& & & & & & & & & \\

PBC J2010.2+4759 & 2.2$\pm$0.4 & abs.& 1.4$\pm$0.2 & Sy2 & 0.026 & 113.6 & 0.453& -- & 2MASX J20101740+4800214\\
& [5.9$\pm$1.1]& [abs.] & [5.6$\pm$0.8] & & & & & & \\

& & & & & & & & & \\

PBC J2150.7+1405 & 41.6$\pm$6.6 & 5.3$\pm$0.7 & 16.1$\pm$2.5 & Sy2 & 0.031& 135.9 & 0.141& 0.909 & 2MASX J21504675+1406369\\
& [59.7$\pm$9.5]& [8.1$\pm$1.1] & [26.5$\pm$4.1] & & & & & & \\

& & & & & & & & & \\

PBC J2307.8+2244 & 9.3$\pm$6.6 & 1.3$\pm$0.1 & 29.8$\pm$2.2 & Sy2 & 0.035& 153.9 & 0.269& 0.740 & 2MASX J23074887+2242367\\
& [16.7$\pm$11.8]& [2.7$\pm$0.2] & [69.1$\pm$5.1] & & & & & & \\

& & & & & & & & & \\

PBC J2343.8+0539 & 9.4$\pm$2.2 & <1.1 & 22.6$\pm$2.3 & Sy2 & 0.056& 250 & 0.137& 0.435 & 2MASX J23435956+0538233\\
& [15.3$\pm$3.6]& [<3.4]& [35.1$\pm$3.6] & & & & & & \\

& & & & & & & & & \\
\noalign{\smallskip} 
\hline
\noalign{\smallskip} 
\multicolumn{10}{l}{Note: emission line fluxes are reported both as
observed and (between square brackets) corrected for the intervening Galactic} \\ 
\multicolumn{10}{l}{absorption $E(B-V)_{\rm Gal}$ along the object line of sight 
(from Schlegel et al. 1998). Line fluxes are in units of 10$^{-15}$ erg cm$^{-2}$ s$^{-1}$,} \\
\multicolumn{10}{l}{The typical error on the redshift measurement is $\pm$0.001}\\
\multicolumn{10}{l}{$^*$: blended with [N {\sc ii}] lines} \\
\noalign{\smallskip} 
\hline
\hline
\end{tabular} 
\end{center}
\end{table*}

\begin{table*}[th!]
\begin{center}
\caption[]{Main results obtained from the analysis of the optical spectrum of the QSO PBC J0030.5$-$5902.}\label{qso}
\scriptsize
\begin{tabular}{lcccccccc}
\noalign{\smallskip}
\hline
\hline
\noalign{\smallskip}
\multicolumn{1}{c}{Object}  & $F_{\rm Mg II}$ &
$F_{\rm CIII]}$ & Class & $z$ & \multicolumn{1}{c}{$D_L$} & \multicolumn{2}{c}{$E(B-V)$} & \multicolumn{1}{c}{Optical source}\\
\cline{7-8}
\noalign{\smallskip}
& & & &  & (Mpc) & Gal. & AGN& \\
\noalign{\smallskip}
\hline
\noalign{\smallskip}
PBC J0030.5$-$5902& 4.67$\pm$0.98 & 6.82$\pm$2.28  & QSO & 1.137 & 7744.2 & 0.013 & -- & USNO-A2.0 0300$-$00143700  \\
& [4.9$\pm$1.2] & [15.6$\pm$2.4] & & & & &  \\

& & & & & & & &  \\
\noalign{\smallskip} 
\hline
\noalign{\smallskip} 
\multicolumn{9}{l}{Note: emission line fluxes are reported both as
observed and (between square brackets) corrected for the intervening Galactic} \\ 
\multicolumn{9}{l}{Line fluxes are in units of 10$^{-15}$ erg cm$^{-2}$ s$^{-1}$.} \\
\multicolumn{9}{l}{The typical error on the redshift measurement is $\pm$0.001.}\\
\noalign{\smallskip} 
\hline
\hline
\end{tabular} 
\end{center}
\end{table*}

\begin{table*}[th!]
\caption[]{Main optical results concerning sources
identified as cataclismic variables (see Fig. \ref{cv}).}
\label{cvo}
\hspace{-1.2cm}
\vspace{-.5cm}
\begin{center}
\resizebox{18cm}{!}{
\begin{tabular}{lcccccccc}
\noalign{\smallskip}
\hline
\hline
\noalign{\smallskip}
\multicolumn{1}{c}{Object} & \multicolumn{2}{c}{H$_\alpha$} & 
\multicolumn{2}{c}{H$_\beta$} &
\multicolumn{2}{c}{He {\sc ii} $\lambda$4686} &
R  & $d$  \\
\cline{2-7}
\noalign{\smallskip} 
 & EW & Flux & EW & Flux & EW & Flux & mag. & (pc)  \\
\noalign{\smallskip}
\hline
\noalign{\smallskip}
PBC J0325.6$-$0820 & 54$\pm$3 & 30.1$\pm$1.9 & 51$\pm$3 & 12.9$\pm$0.8 & 17$\pm$3 & 4.1$\pm$0.6 &
 16.6 & 330  \\
 & & & & & & & &  \\
 
 PBC J0706.7+0327 & 82$\pm$8 & 48$\pm$4 & 64$\pm$5 & 52$\pm$4 & 16$\pm$5 & 14$\pm$4 &
 15.8 & 230  \\
 & & & & & & & &  \\
 
 PBC J0746.2$-$1610 & 35.5$\pm$2.1 & 128$\pm$7 & 32.5$\pm$2.4 & 105$\pm$8 & 19.6$\pm$2.2 & 55.7$\pm$6.1 &
 15.0 & 158  \\
 & & & & &  & & &  \\
 
  PBC J0820.4$-$2801 & 52$\pm$5& 55.1$\pm$4.8 & 48$\pm$8 & 60.1$\pm$10.4 & 12.2$\pm$1.6 & 14.2$\pm$1.9 &
 15.5 & 200  \\
 & & & & & & & &  \\
 
PBC J0927.8$-$6945 & 34$\pm$3 & 83.8$\pm$6.2 & 11$\pm$2 & 38.7$\pm$5.8 & 16$\pm$3 & 60.5$\pm$11.3 &
 15.9 & 240  \\
 & & & & & & & &  \\
 
 PBC J2124.5+0503 & 20$\pm$1.9& 629$\pm$60 & 9$\pm$0.8 & 579$\pm$52 & 5$\pm$0.9 & 335$\pm$63 &
 12.2 & 44  \\
 & & & & & & & &  \\

\noalign{\smallskip}
\hline
\multicolumn{9}{l}{Note: EWs are expressed in \AA, line fluxes are
in units of 10$^{-15}$ erg cm$^{-2}$ s$^{-1}$}\\
\hline
\hline
\end{tabular}}
\end{center} 
\end{table*}

\begin{figure*}[h!]
%\begin{center}
\hspace{-.1cm}
\centering{\mbox{\psfig{file=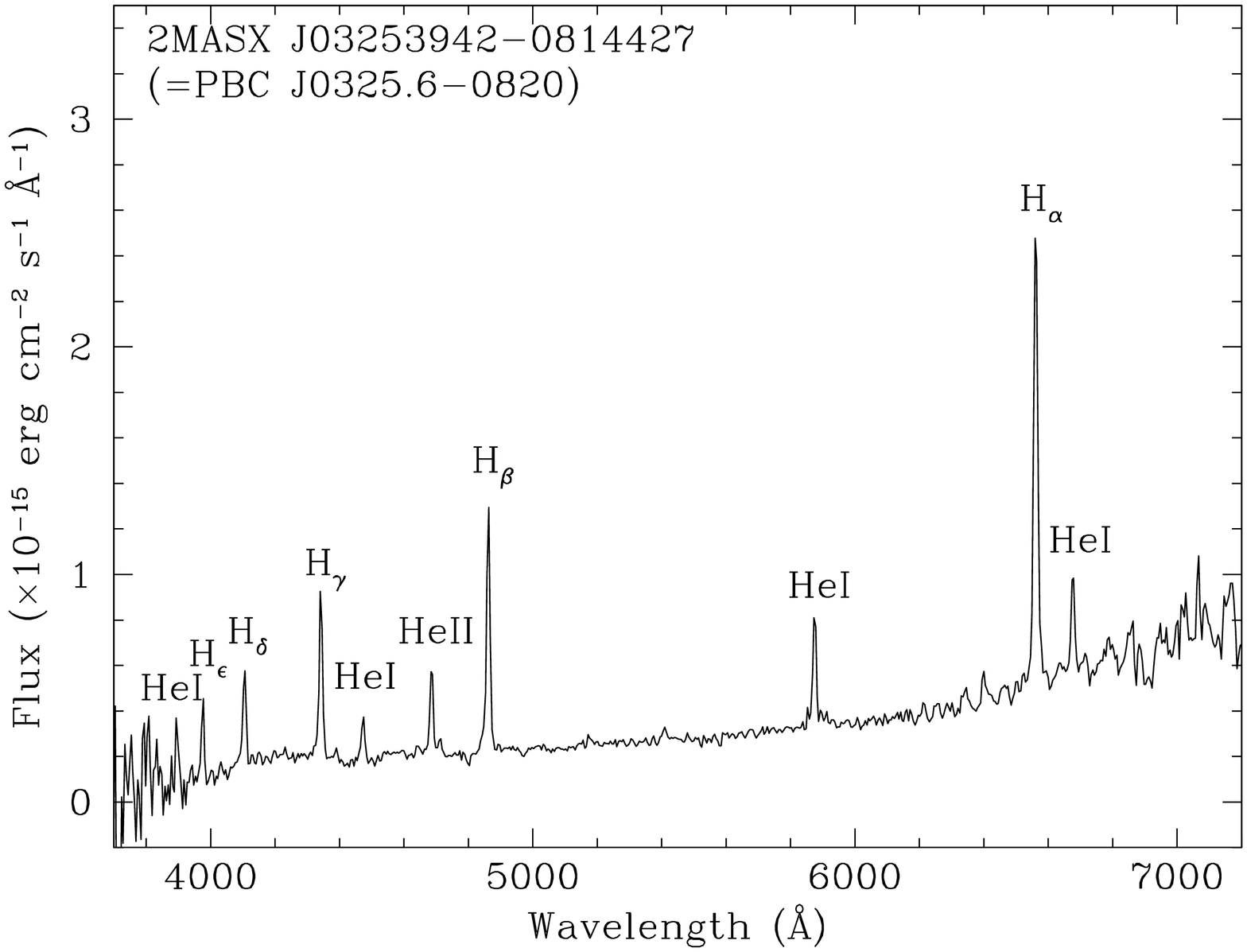,width=6.0cm,angle=270}}}
\centering{\mbox{\psfig{file=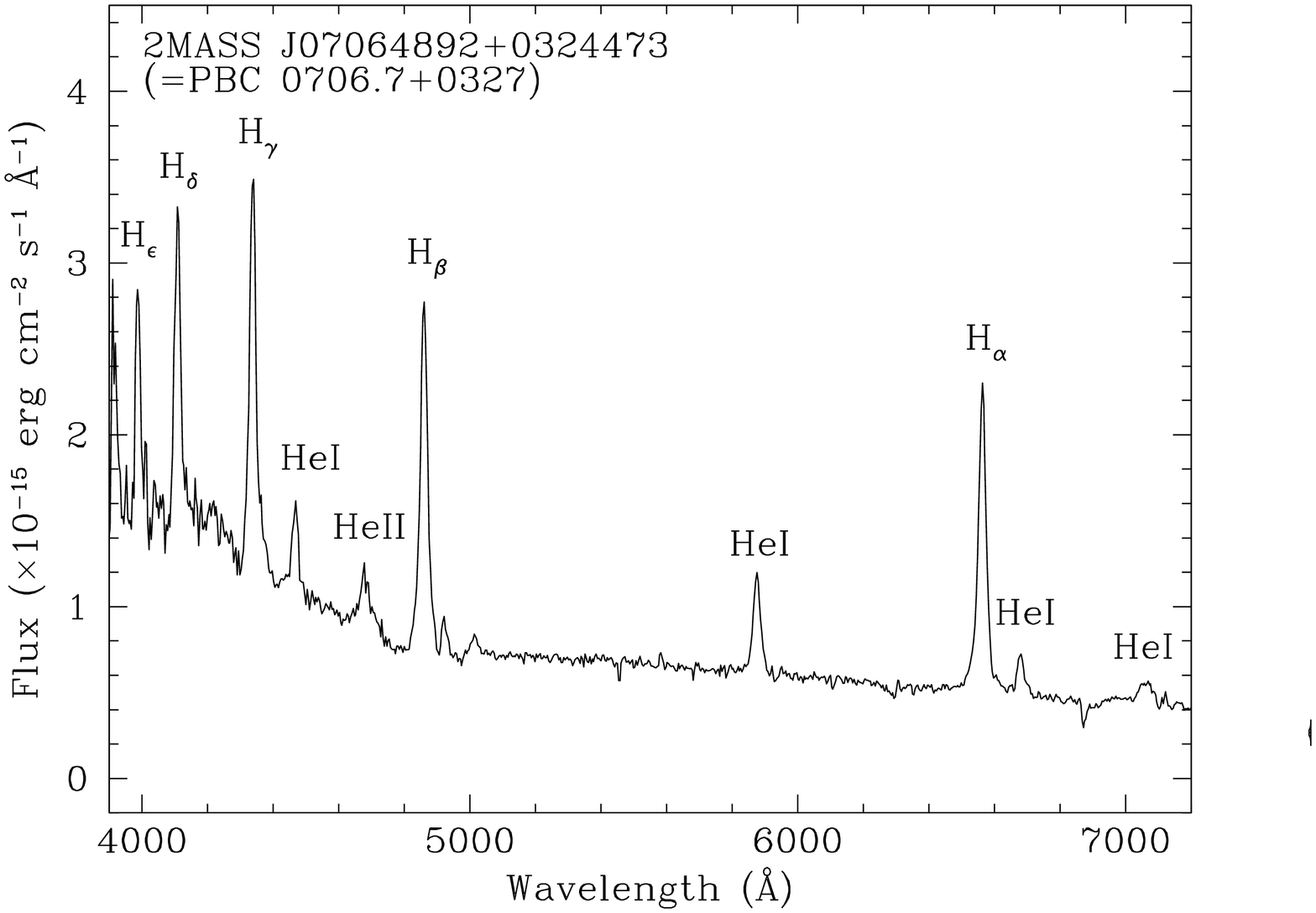,width=6.0cm,angle=270}}}
\centering{\mbox{\psfig{file=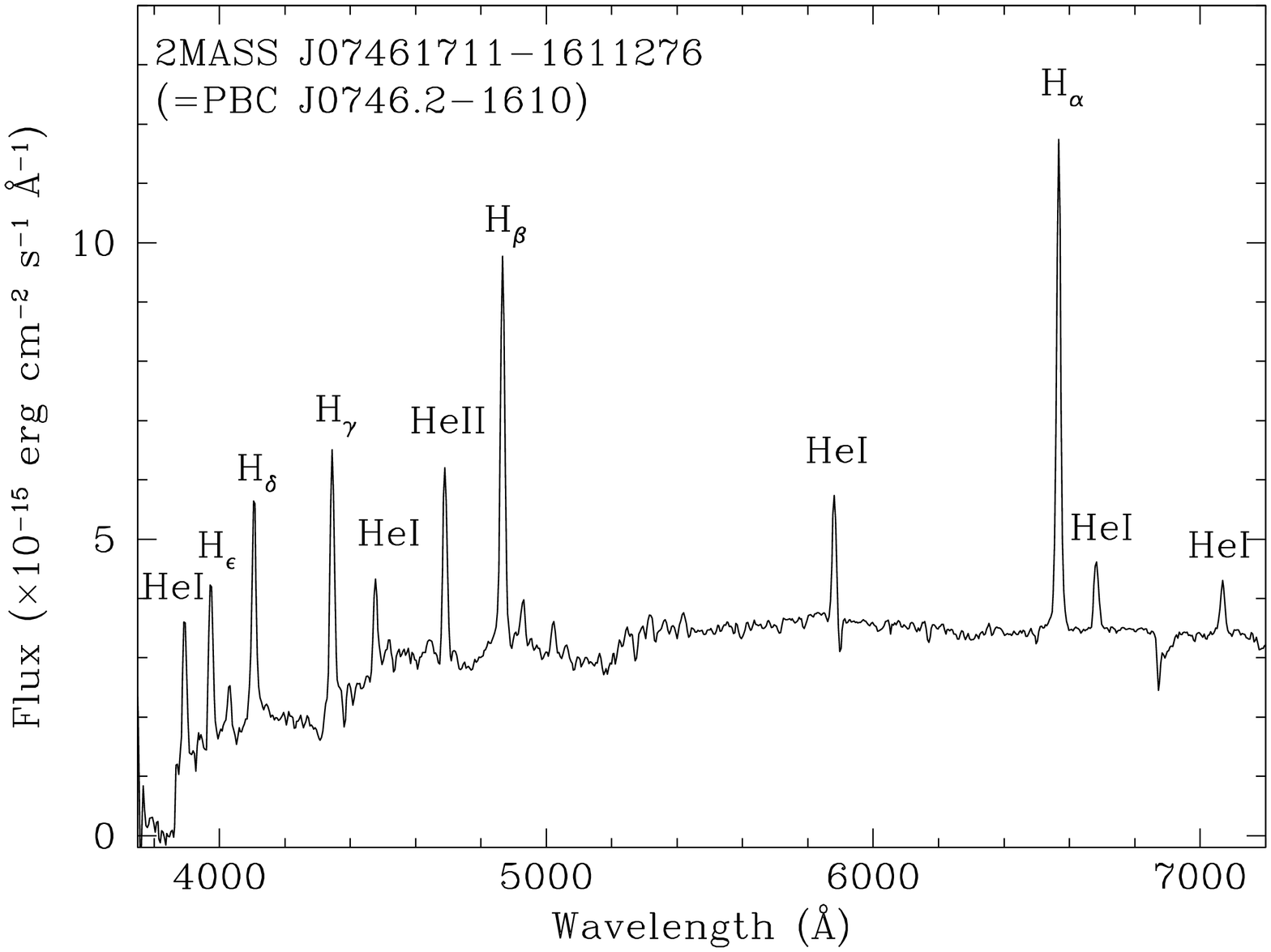,width=6.0cm,angle=270}}}
\centering{\mbox{\psfig{file=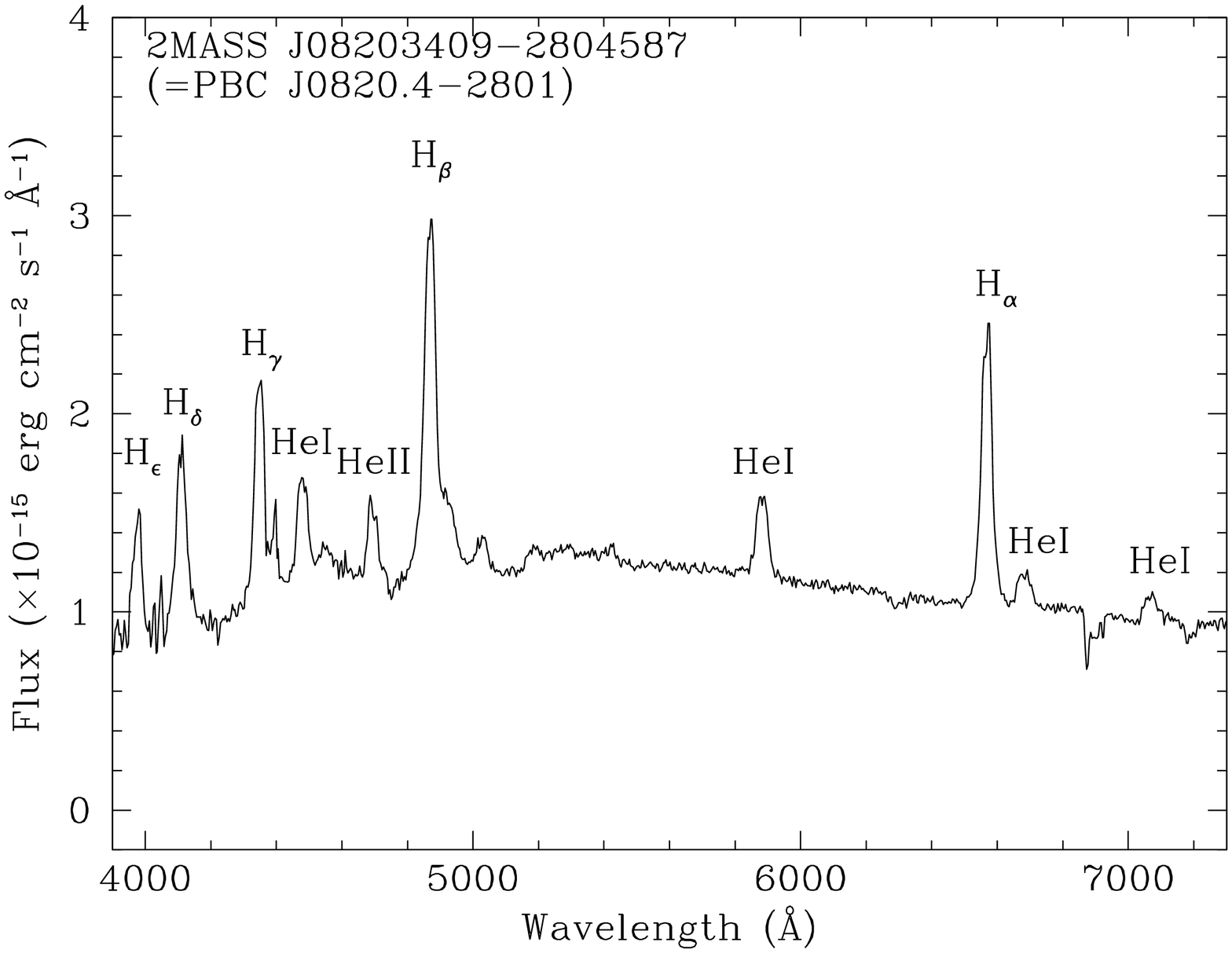,width=6.0cm,angle=270}}}
\centering{\mbox{\psfig{file=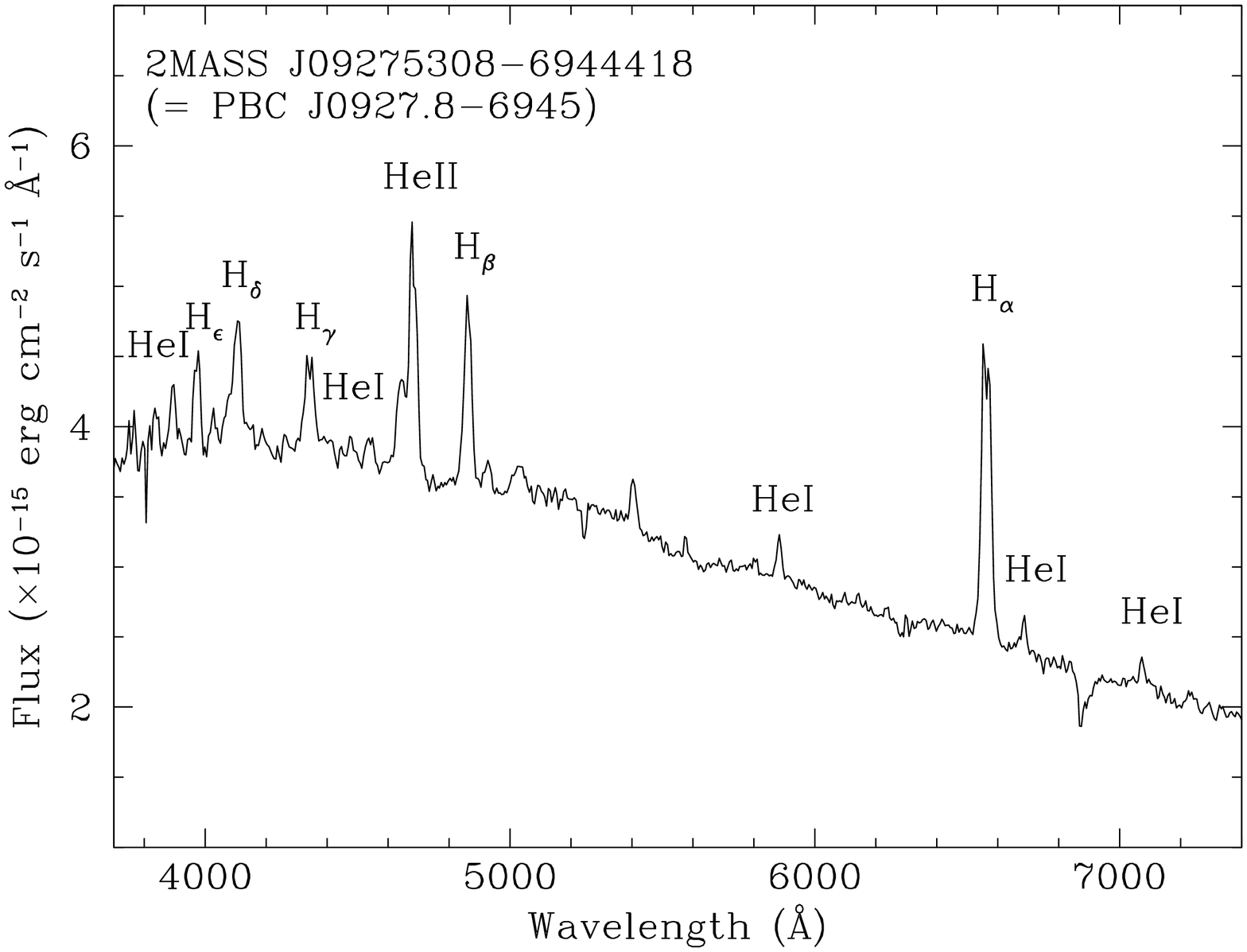,width=6.0cm,angle=270}}}
\centering{\mbox{\psfig{file=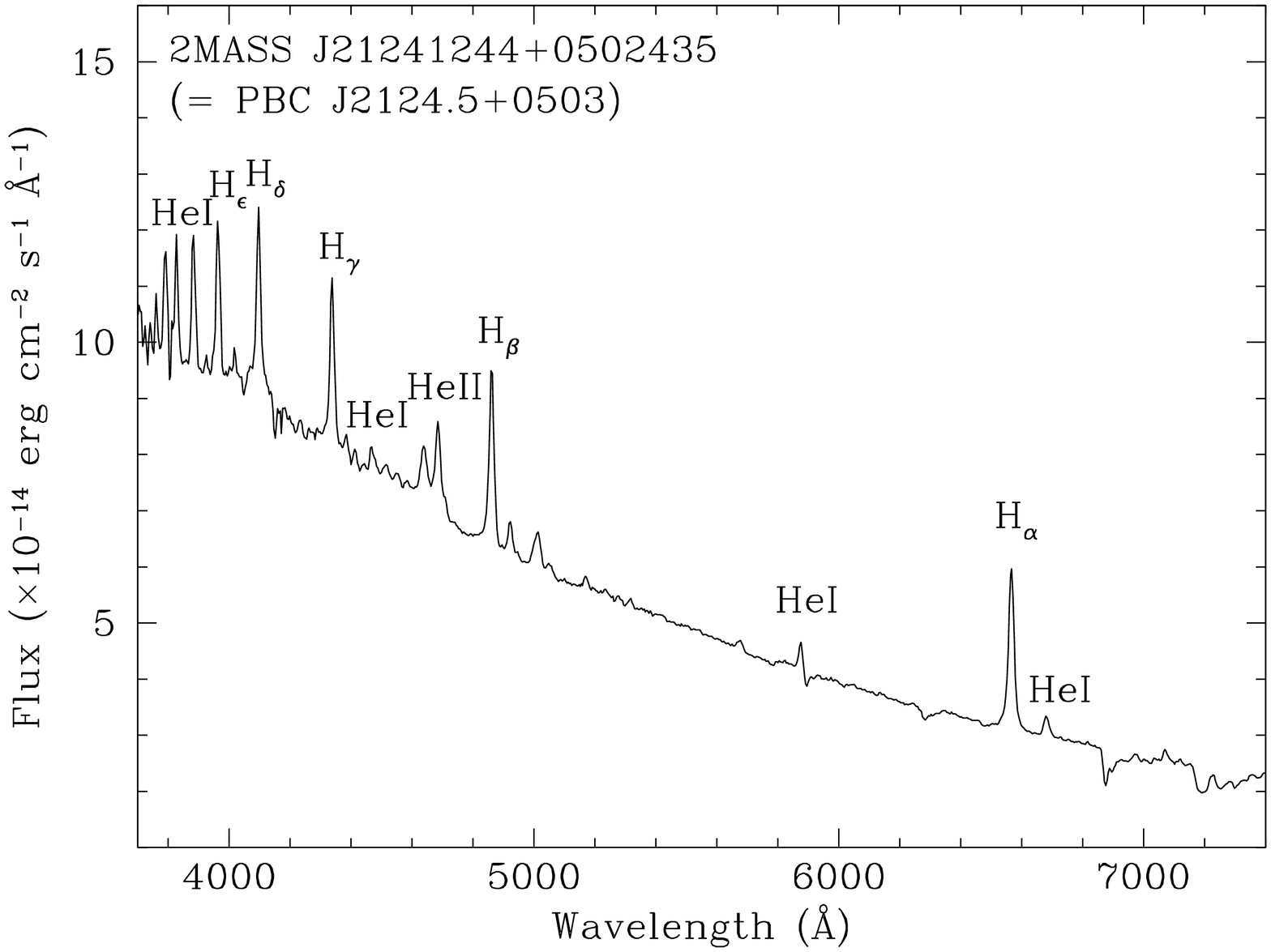,width=6.0cm,angle=270}}}
\caption{Spectra (not corrected for the intervening Galactic absorption) of the optical counterparts of the 6 CVs belonging to the sample of BAT sources presented in this paper. For each spectrum, the main spectral features are labeled.
}\label{cv}
%\end{center}
\end{figure*}

\begin{figure*}
\setcounter{figure}{1}
\caption{Spectra of the optical counterpart of AGNs presented in this work (not corrected for the intervening galactic absorption). For each spectrum, the main spectral features are labeled.
}
%\begin{center}
\hspace{-.1cm}
\centering{\mbox{\psfig{file=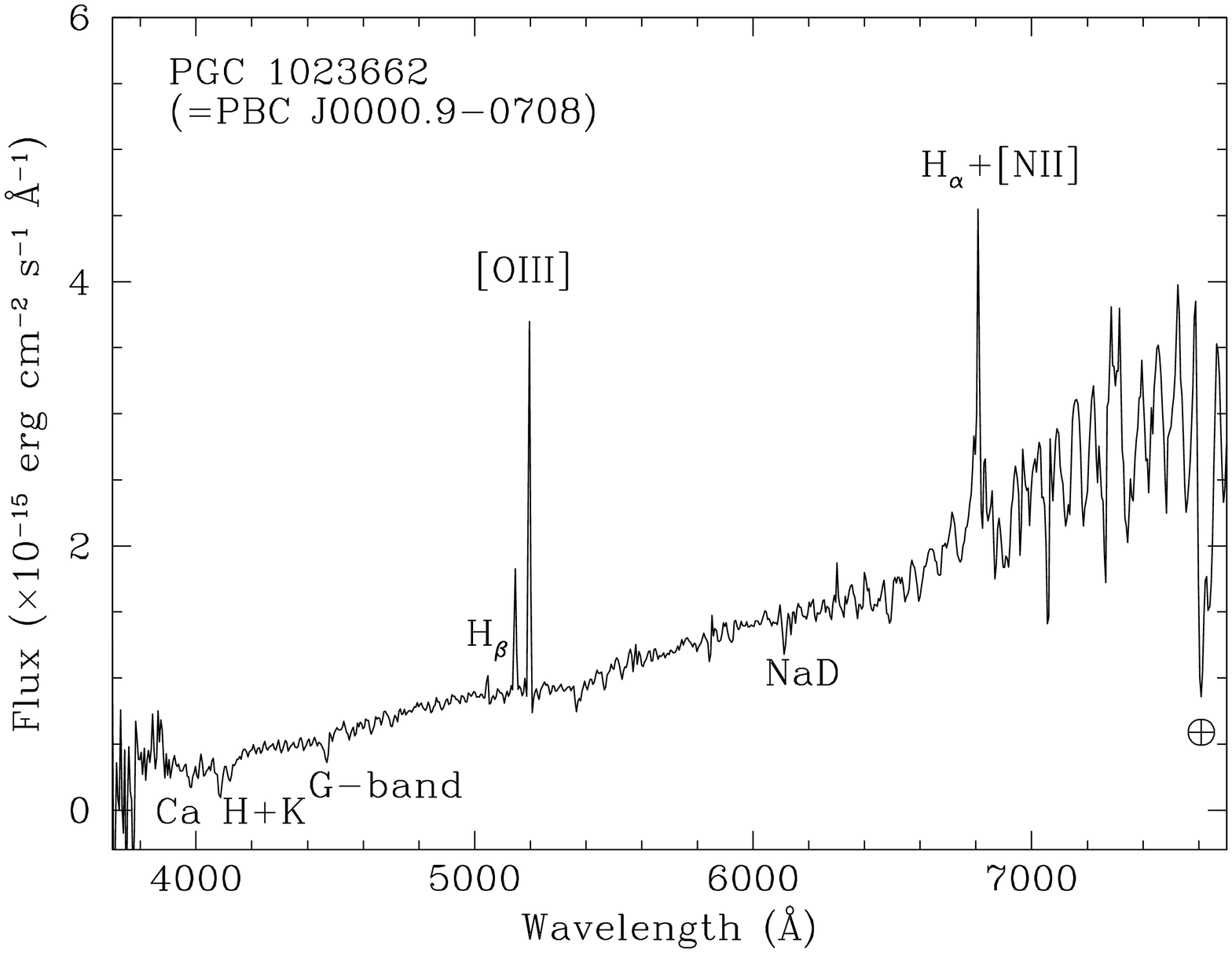,width=6.0cm,angle=270}}}
\centering{\mbox{\psfig{file=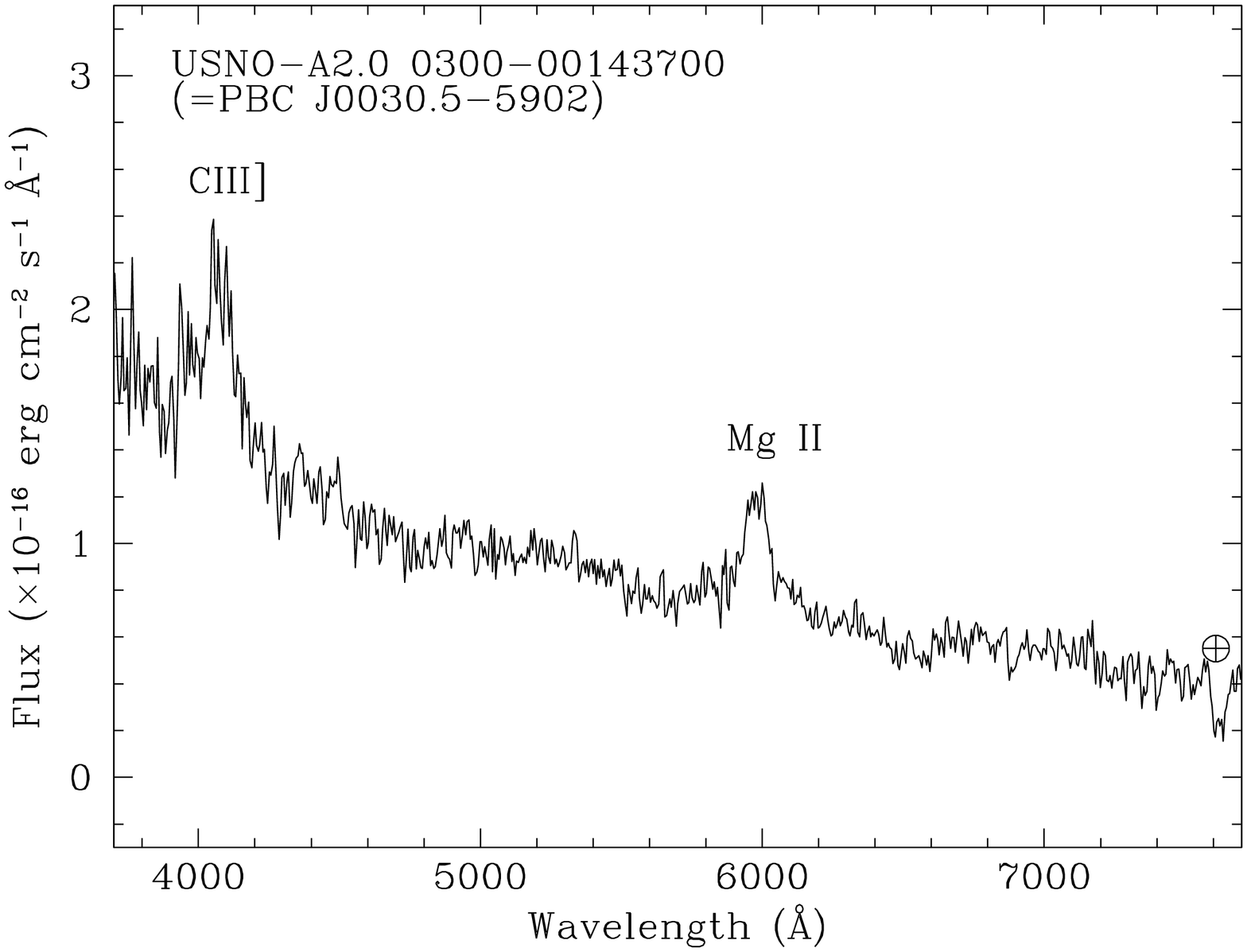,width=6.0cm,angle=270}}}
\centering{\mbox{\psfig{file=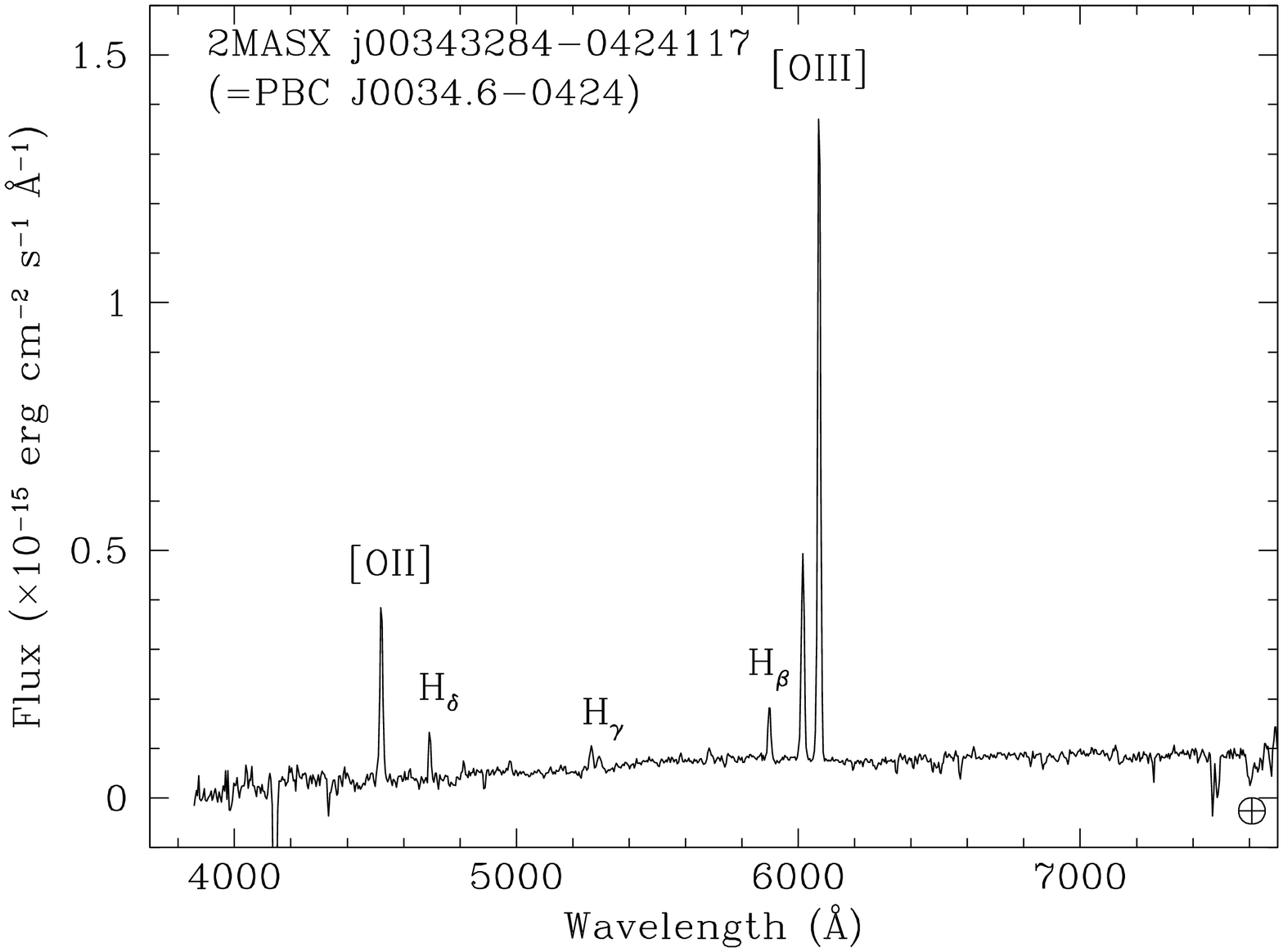,width=6.0cm,angle=270}}}
\centering{\mbox{\psfig{file=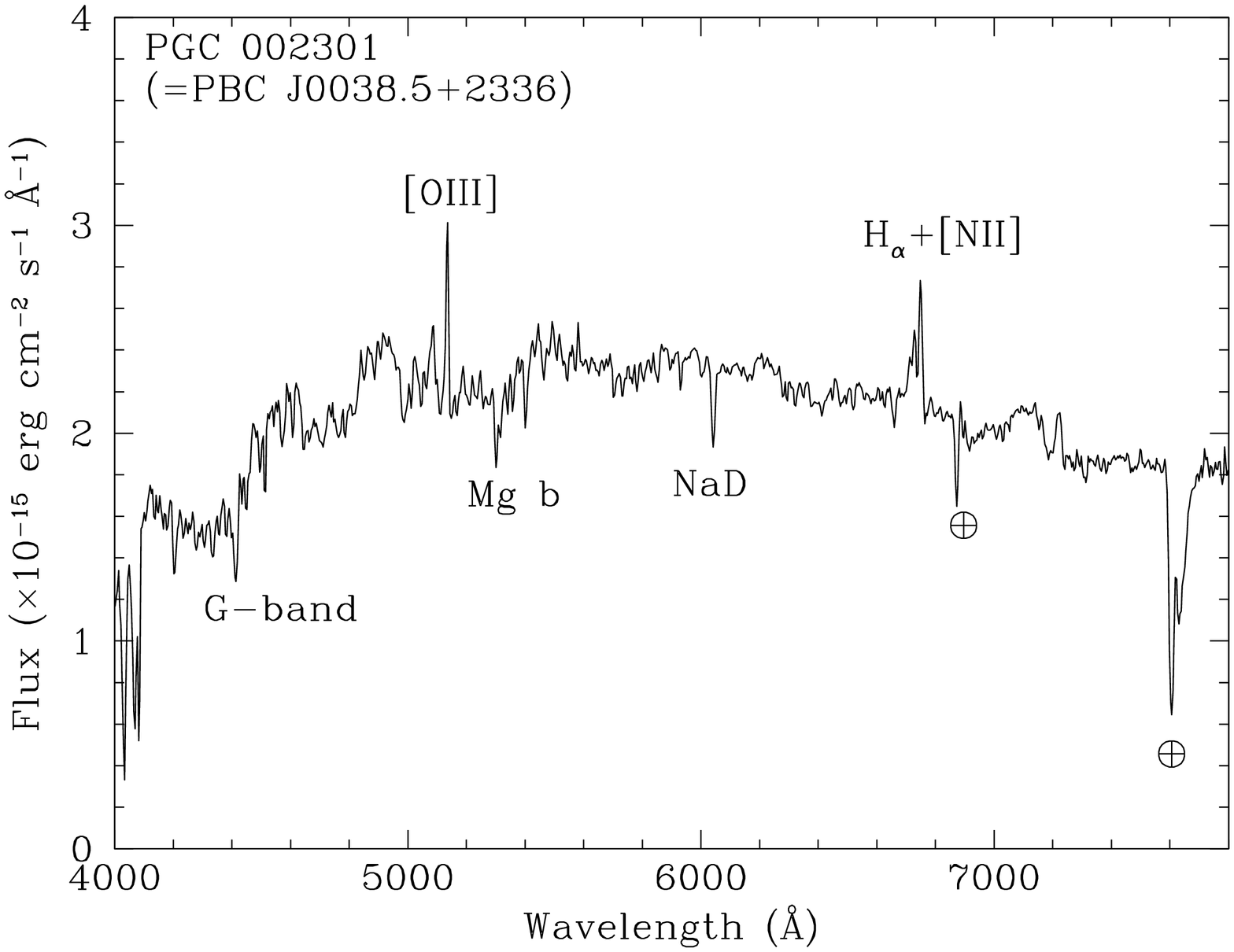,width=6.0cm,angle=270}}}
\centering{\mbox{\psfig{file=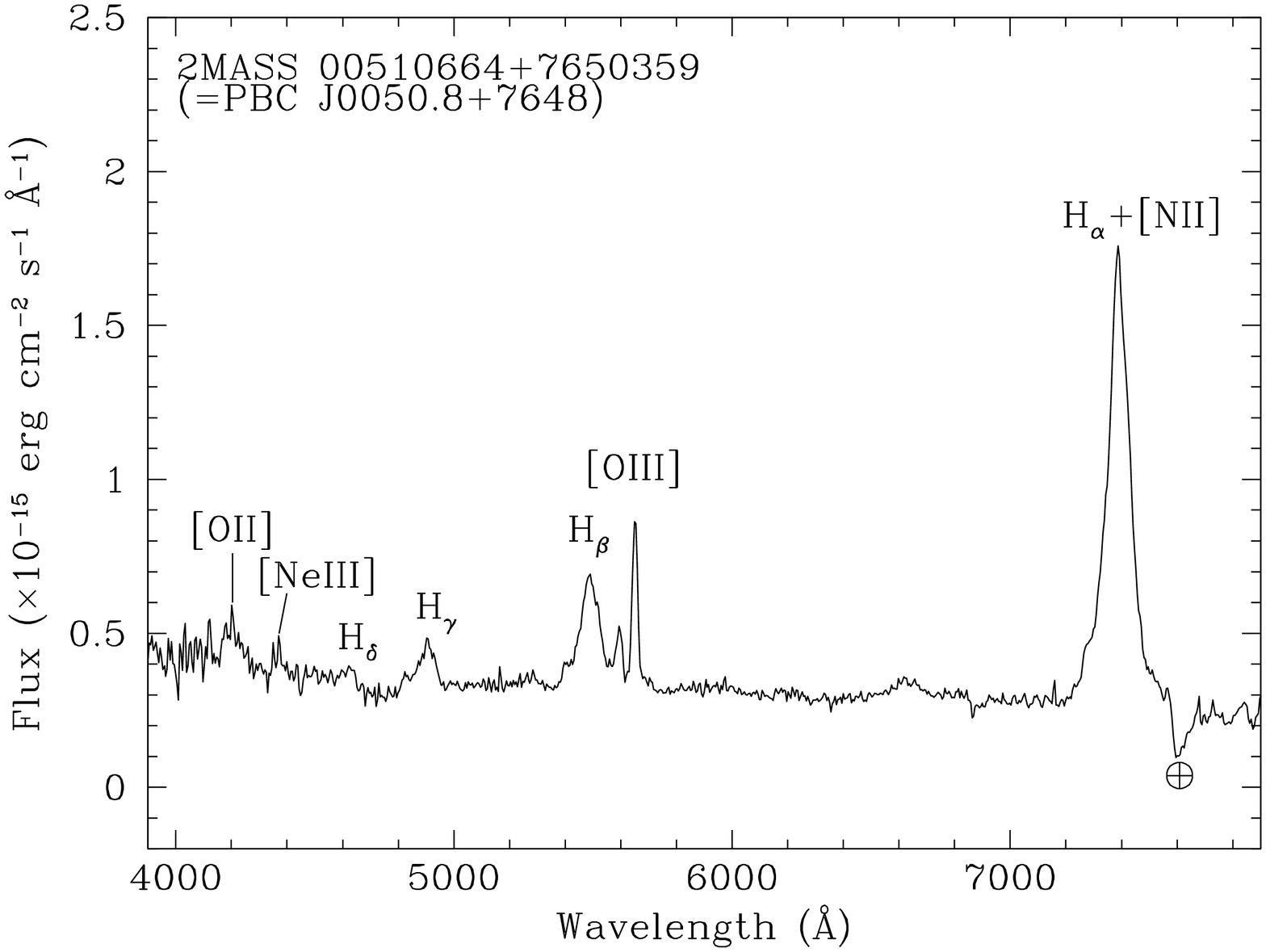,width=6.0cm,angle=270}}}
\centering{\mbox{\psfig{file=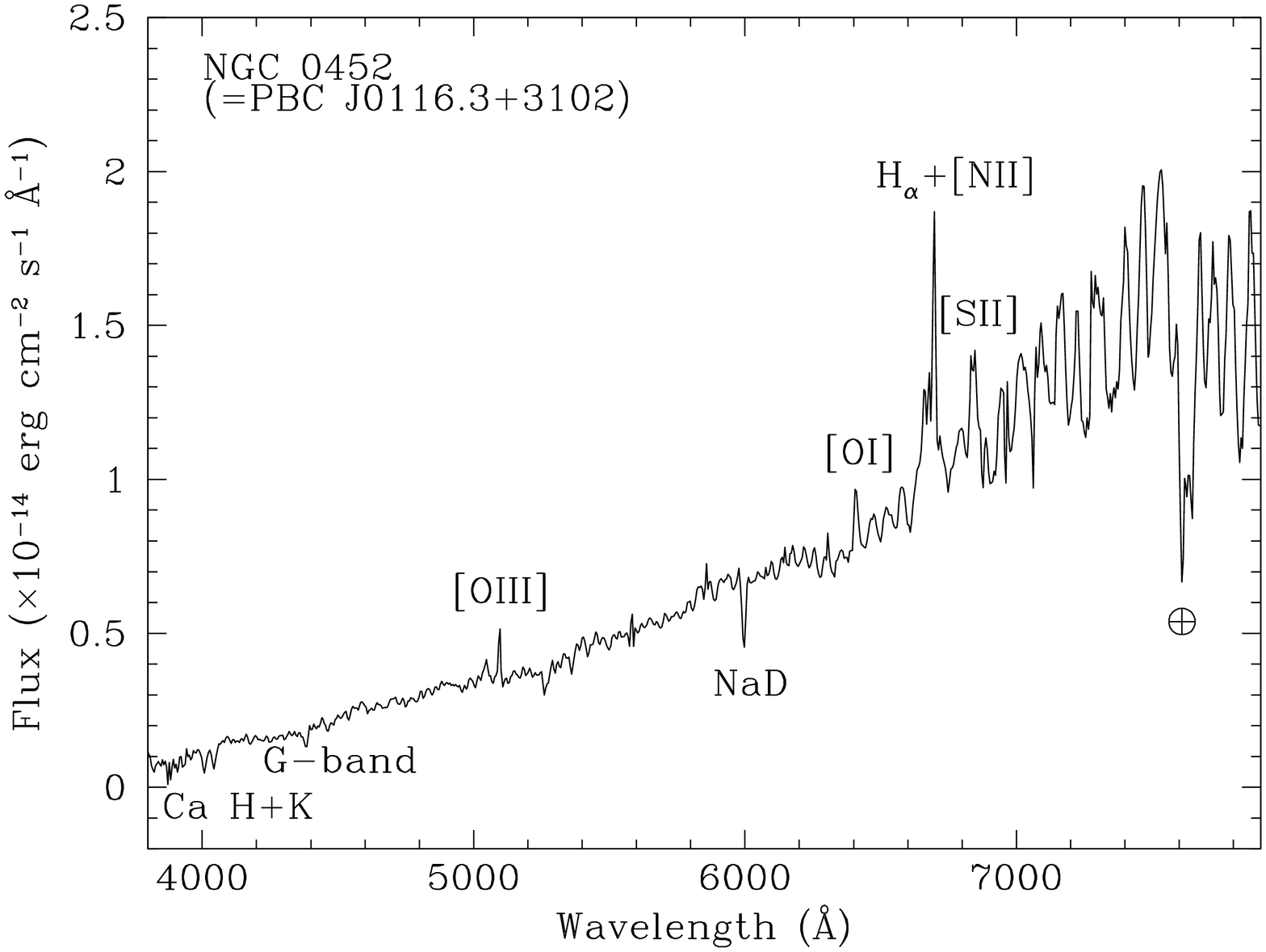,width=6.0cm,angle=270}}}
\centering{\mbox{\psfig{file=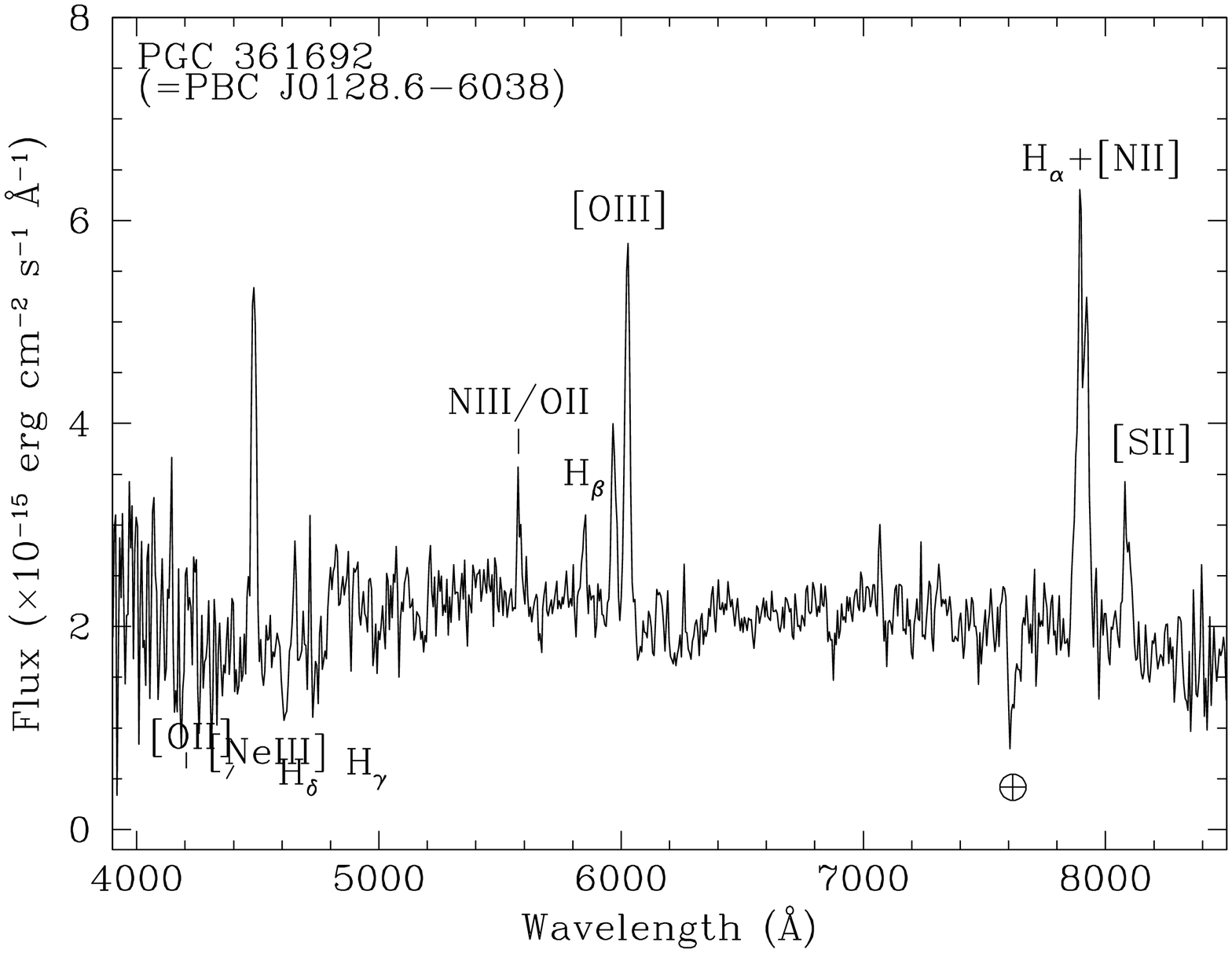,width=6.0cm,angle=270}}}
\centering{\mbox{\psfig{file=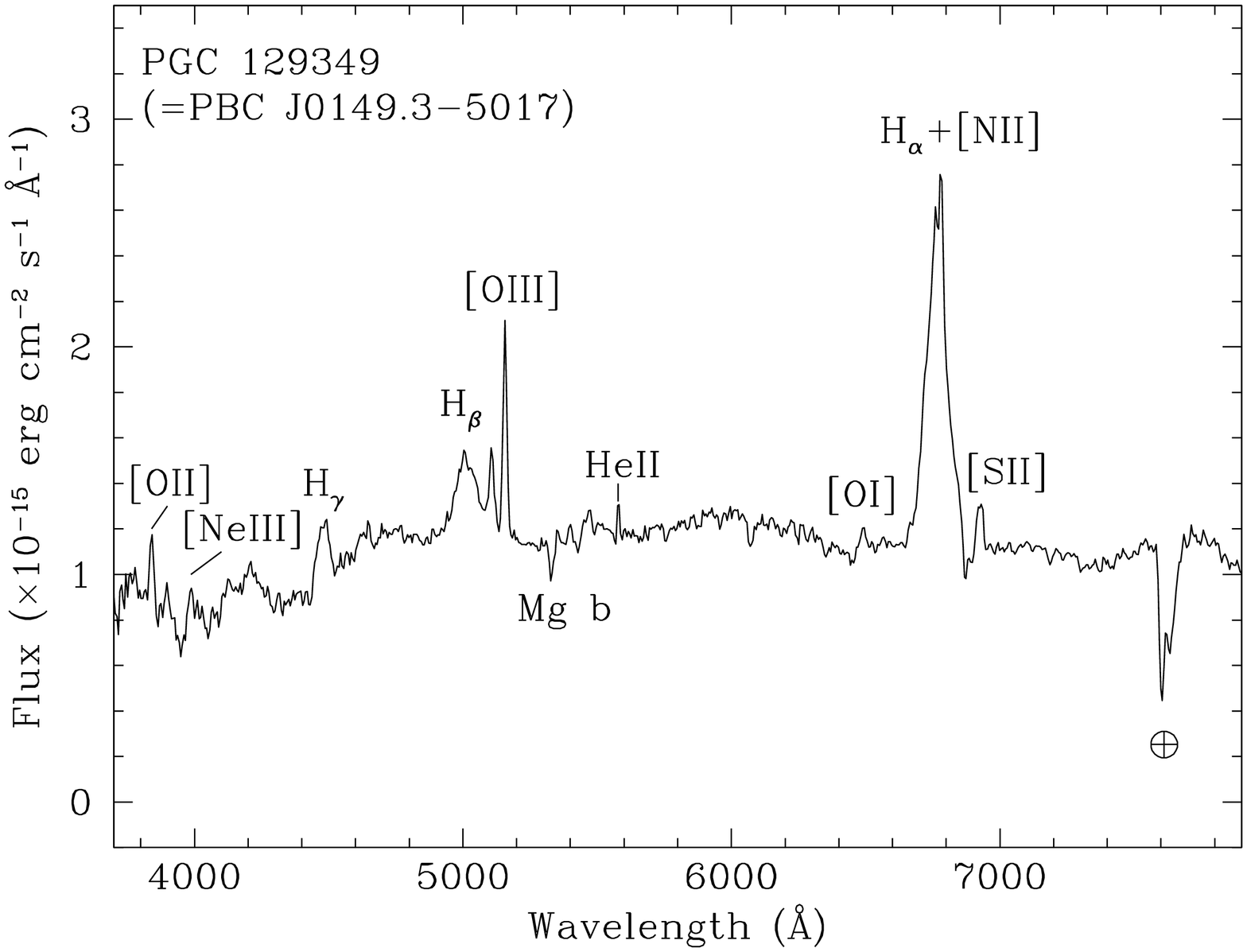,width=6.0cm,angle=270}}}
\centering{\mbox{\psfig{file=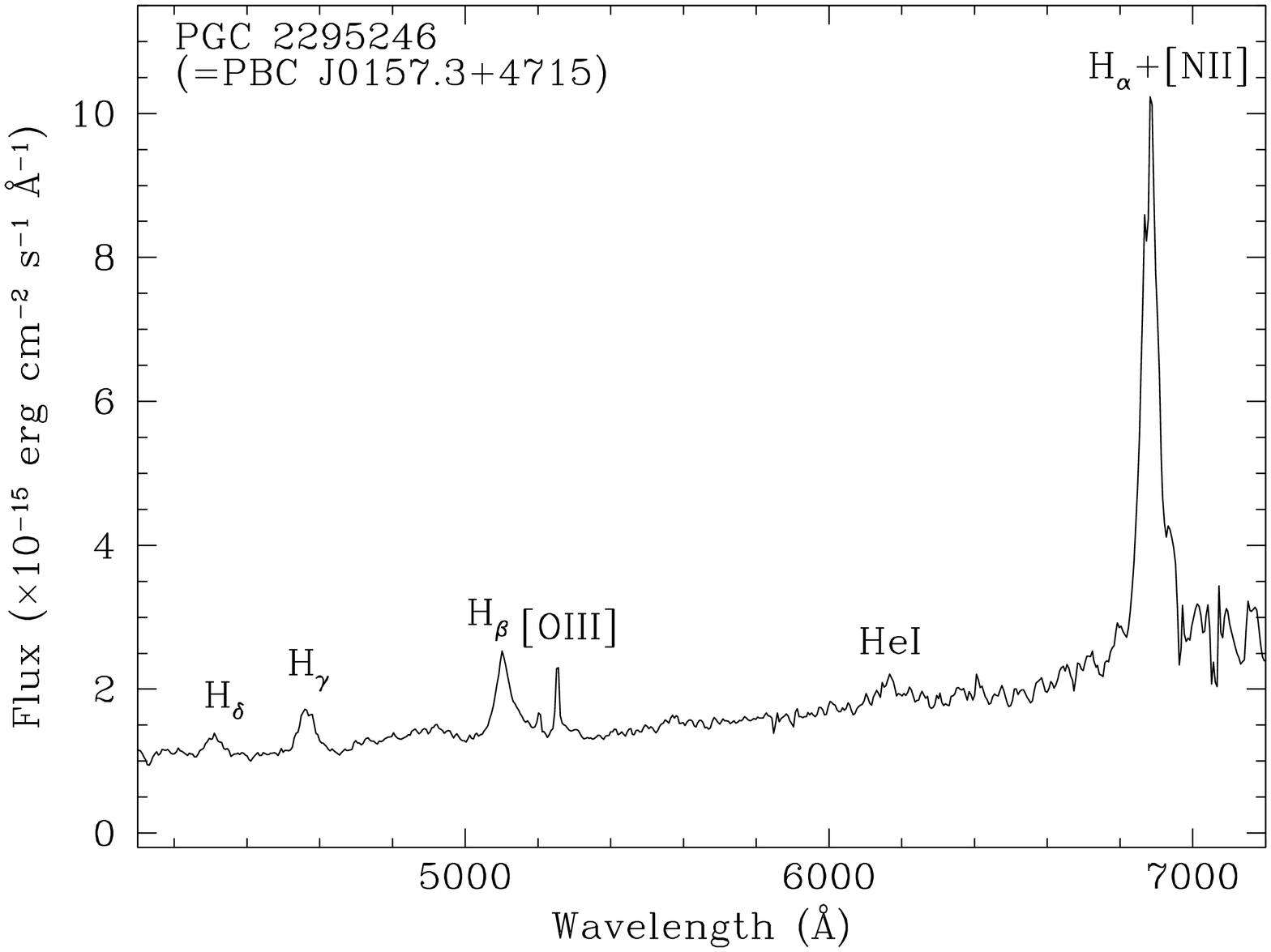,width=6.0cm,angle=270}}}
\centering{\mbox{\psfig{file=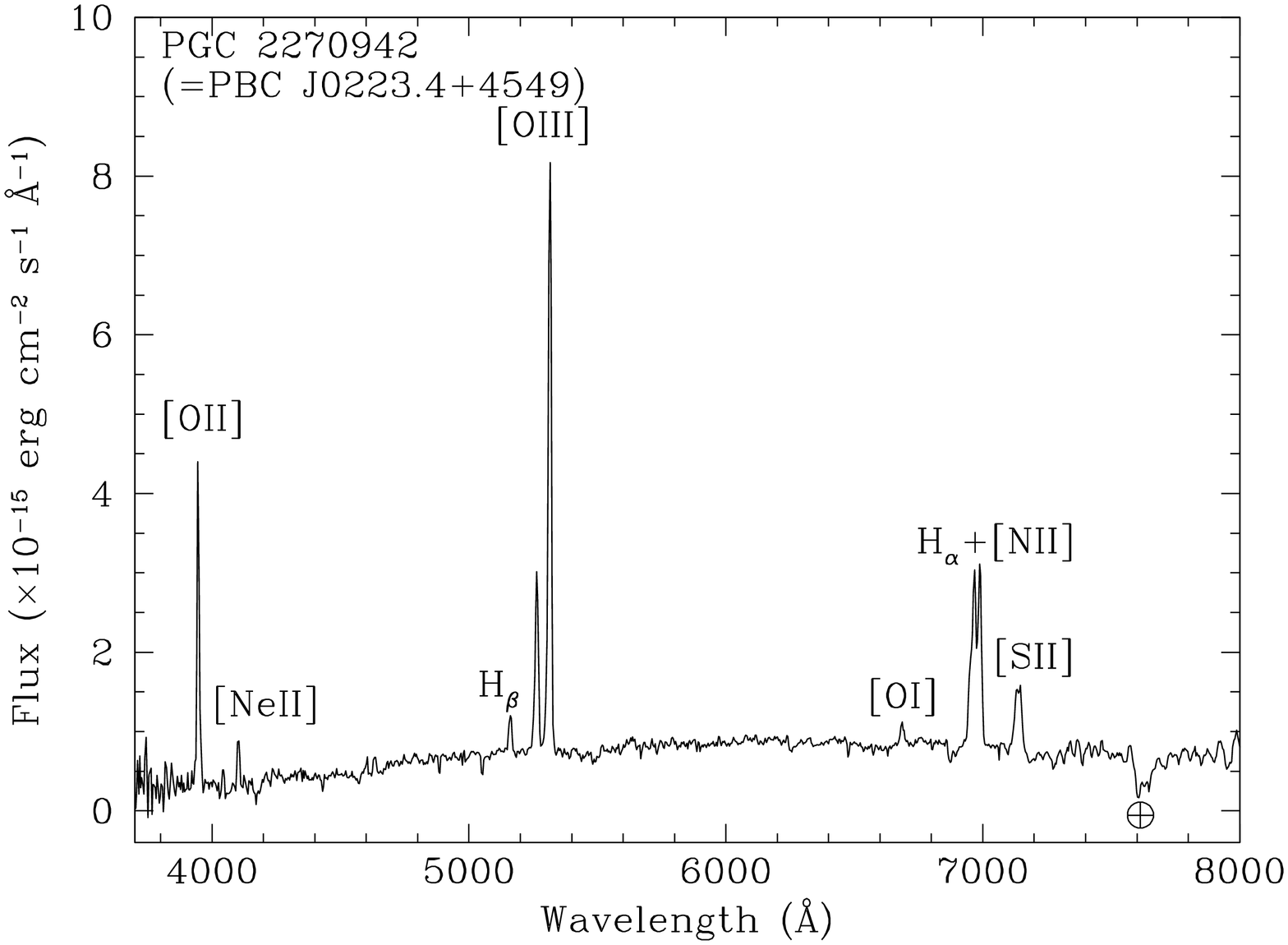,width=6.0cm,angle=270}}}
\centering{\mbox{\psfig{file=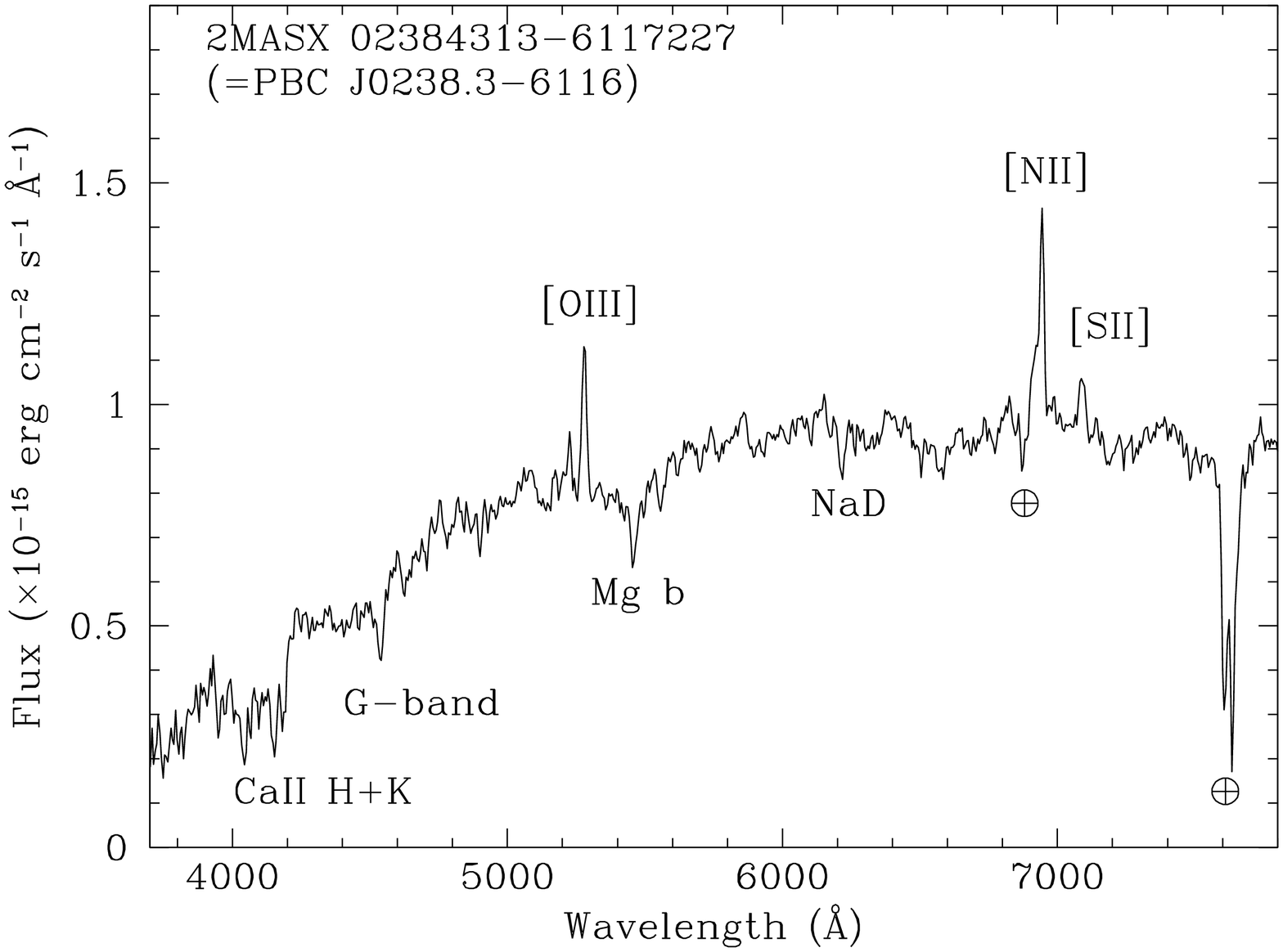,width=6.0cm,angle=270}}}
\centering{\mbox{\psfig{file=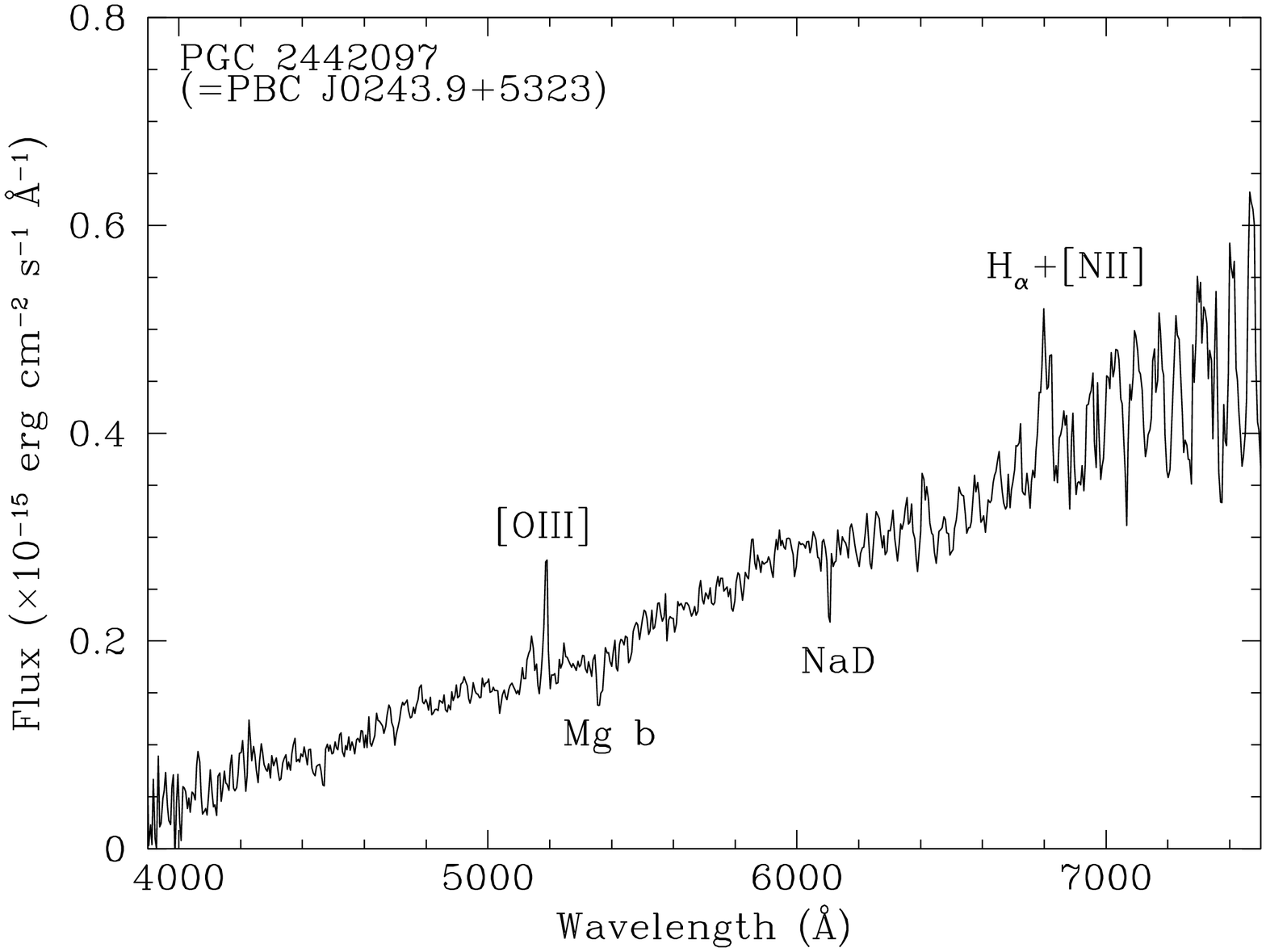,width=6.0cm,angle=270}}}
\centering{\mbox{\psfig{file=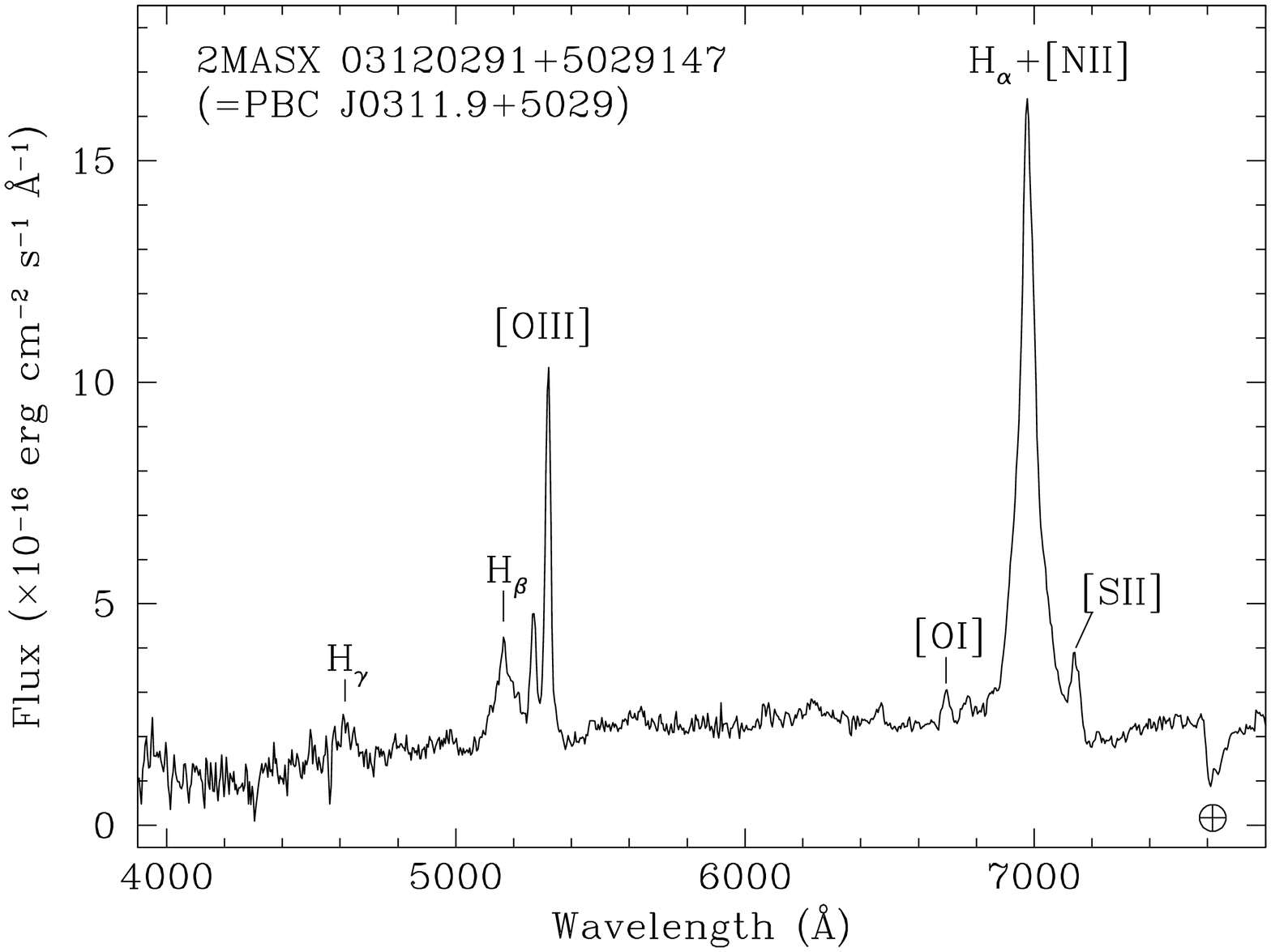,width=6.0cm,angle=270}}}
\centering{\mbox{\psfig{file=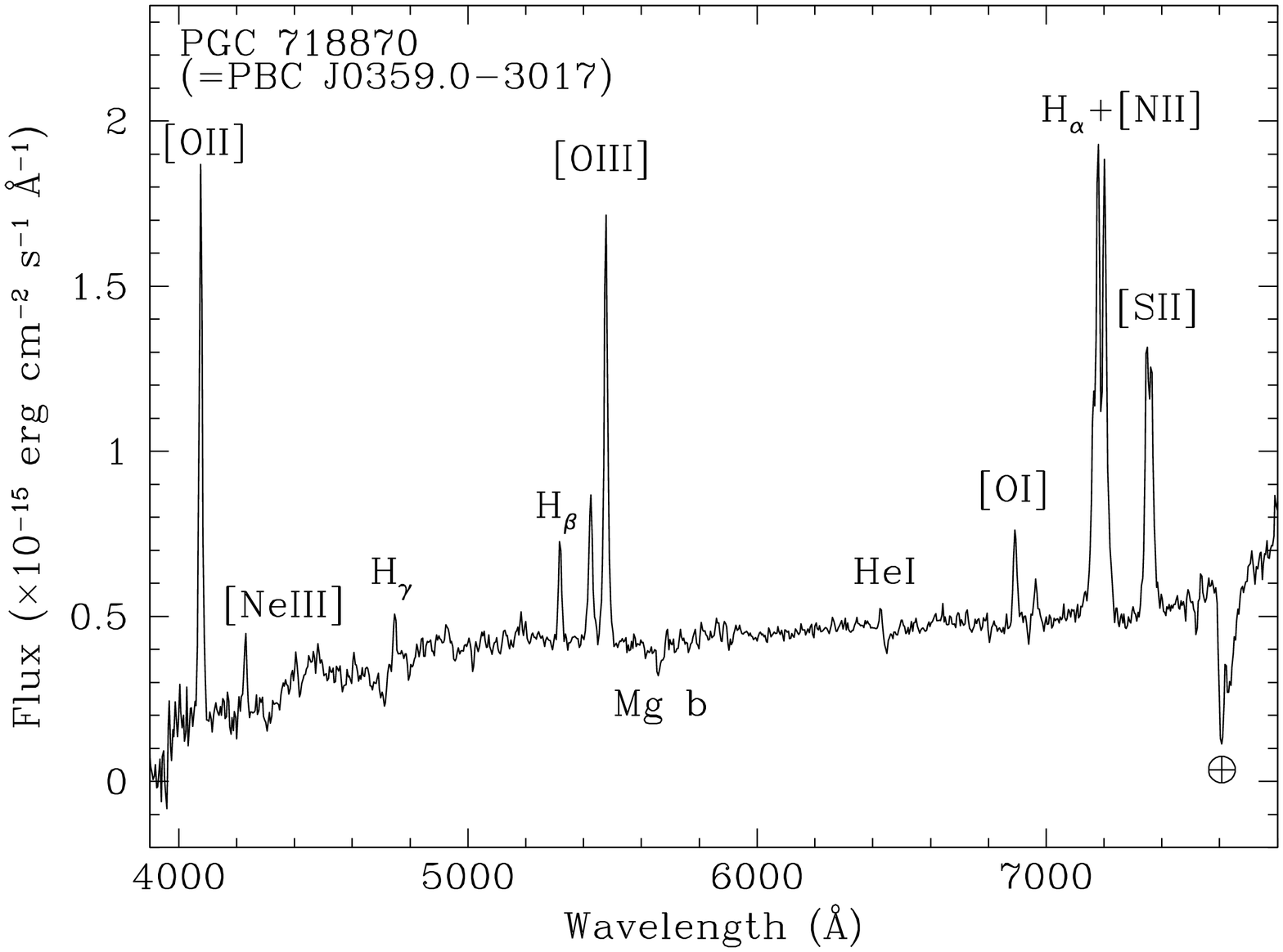,width=6.0cm,angle=270}}}
\centering{\mbox{\psfig{file=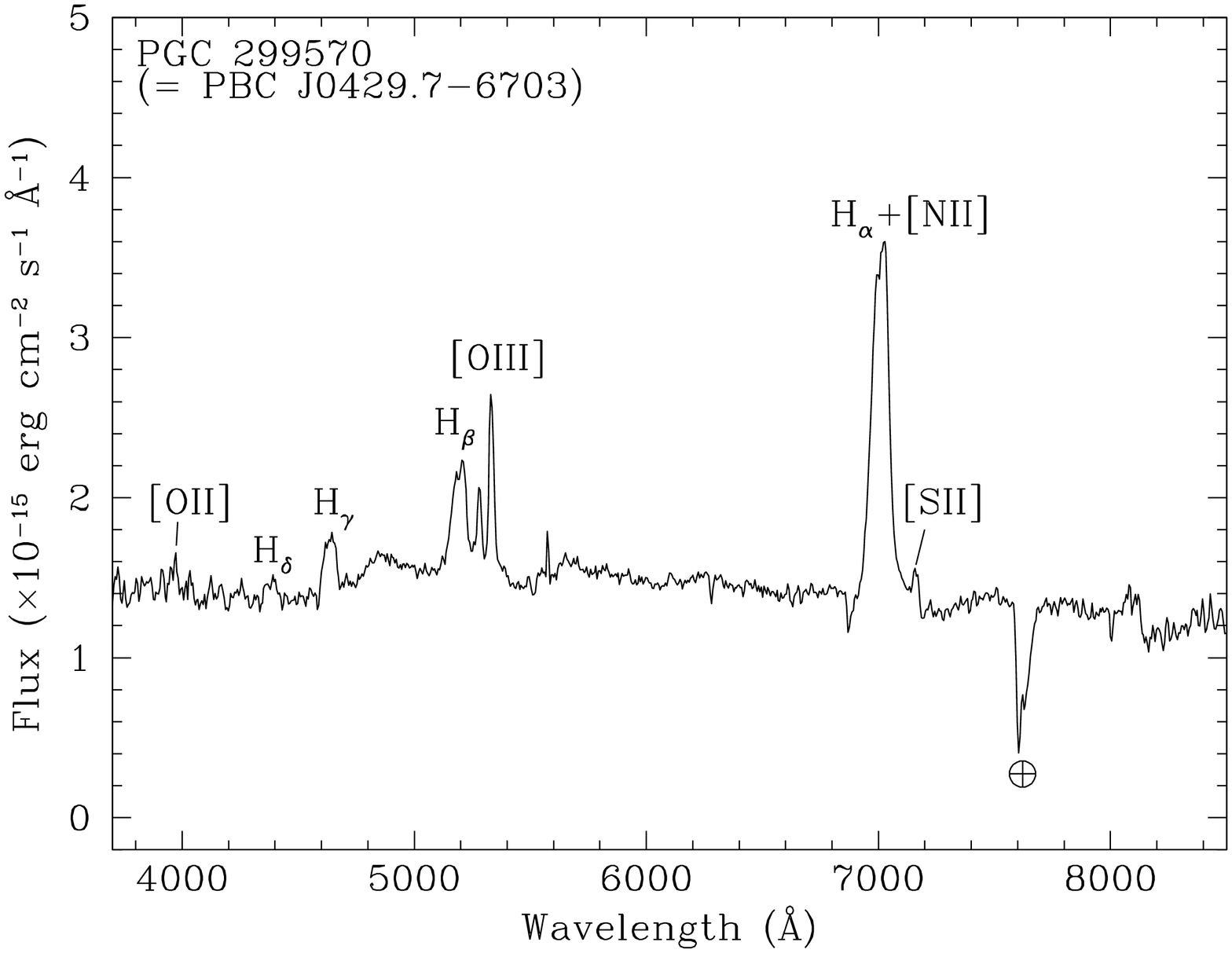,width=6.0cm,angle=270}}}
\label{spectra1}
%\end{center}
\end{figure*}

\begin{figure*}
\setcounter{figure}{1}
%\begin{center}
\hspace{-.1cm}
\centering{\mbox{\psfig{file=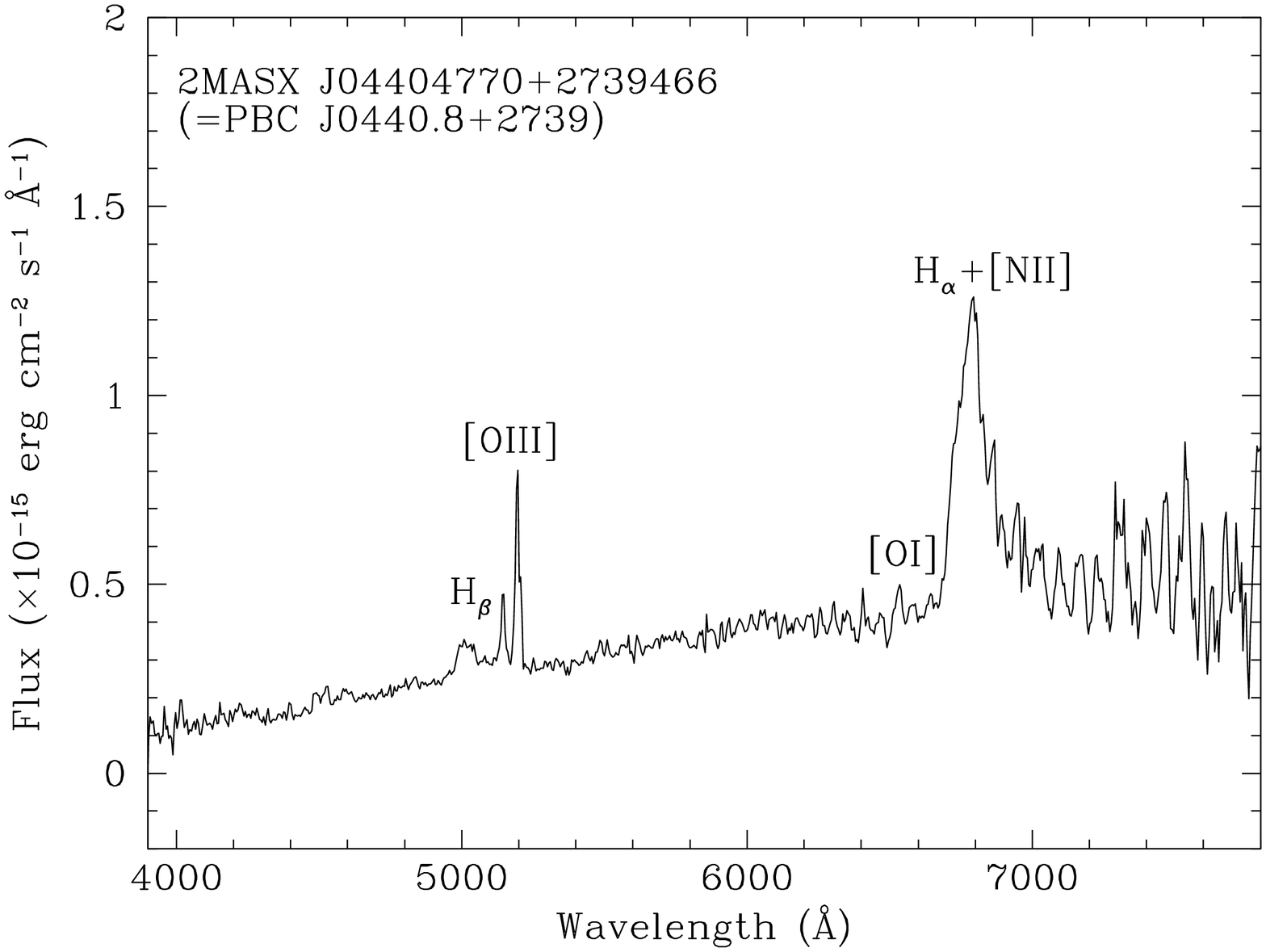,width=6.0cm,angle=270}}}
\centering{\mbox{\psfig{file=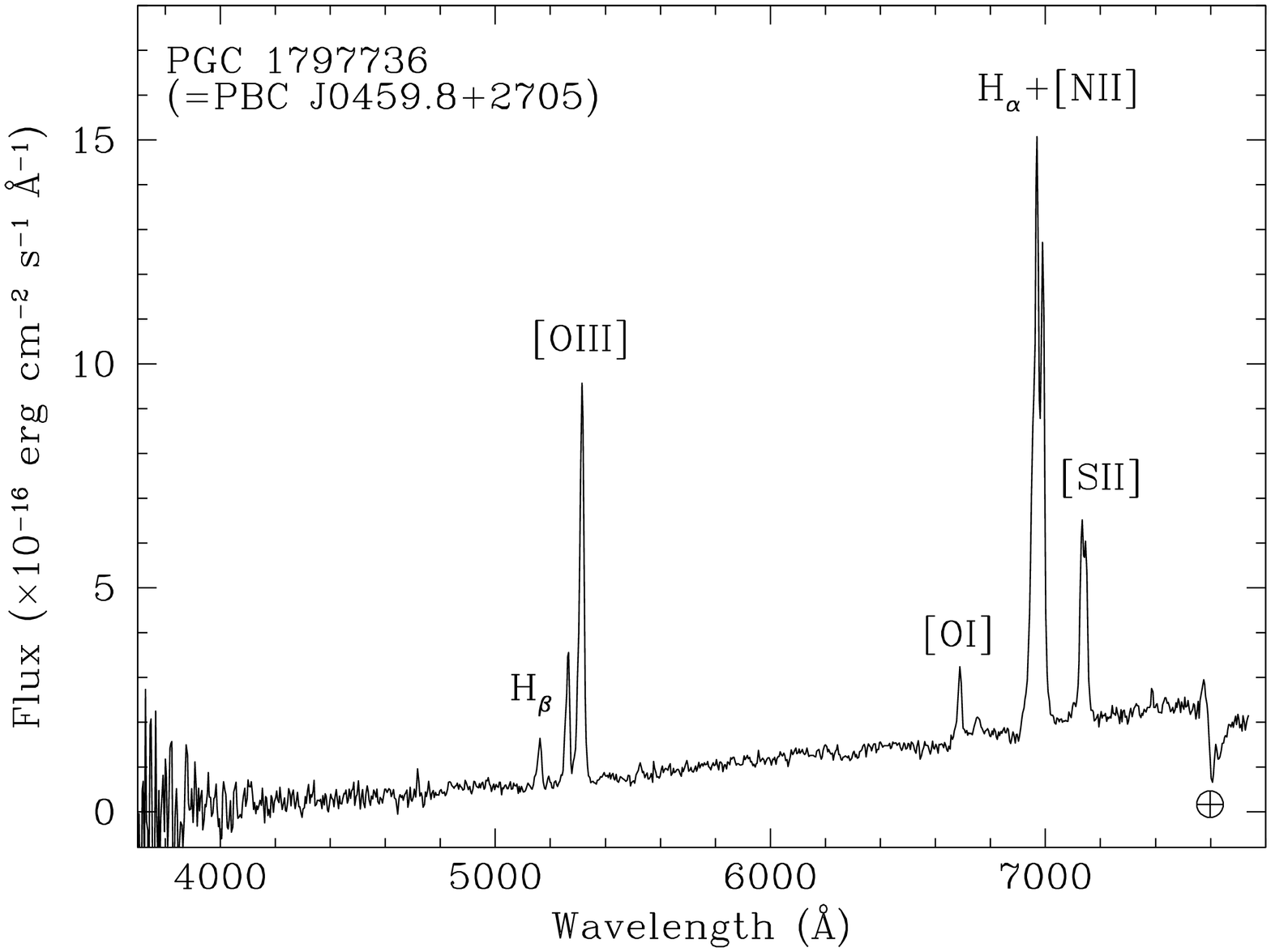,width=6.0cm,angle=270}}}
\centering{\mbox{\psfig{file=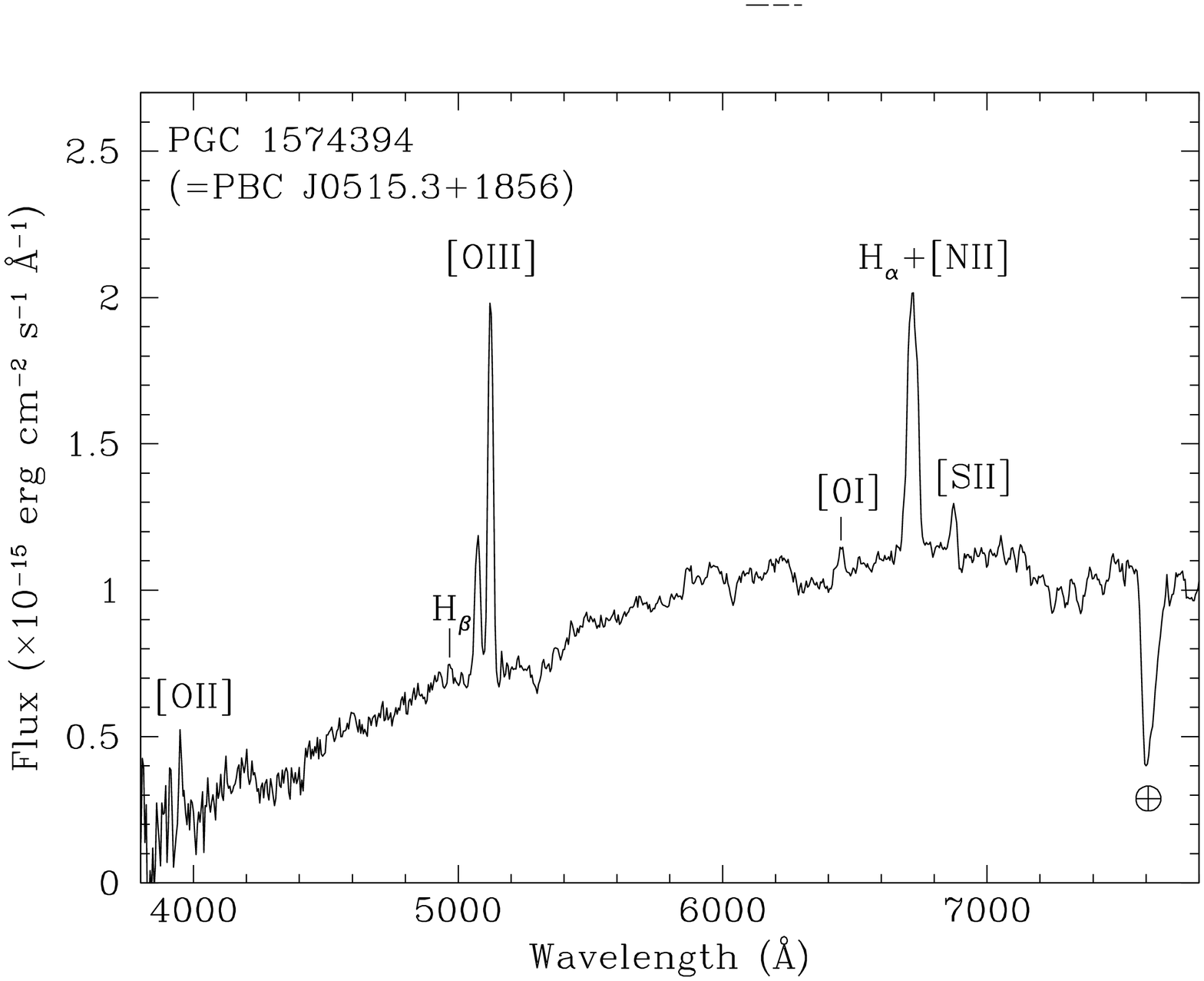,width=6.0cm,angle=270}}}
\centering{\mbox{\psfig{file=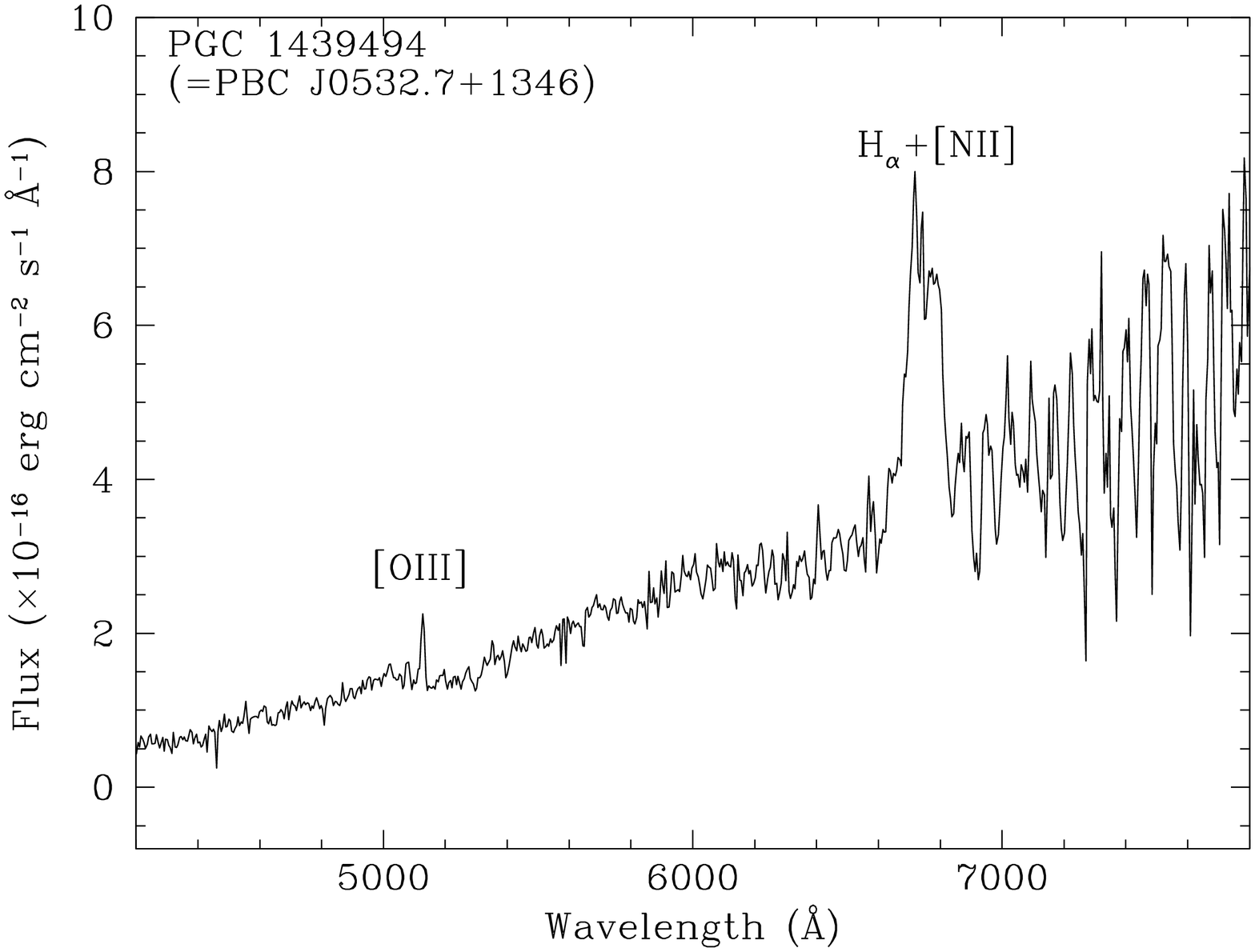,width=6.0cm,angle=270}}}
\centering{\mbox{\psfig{file=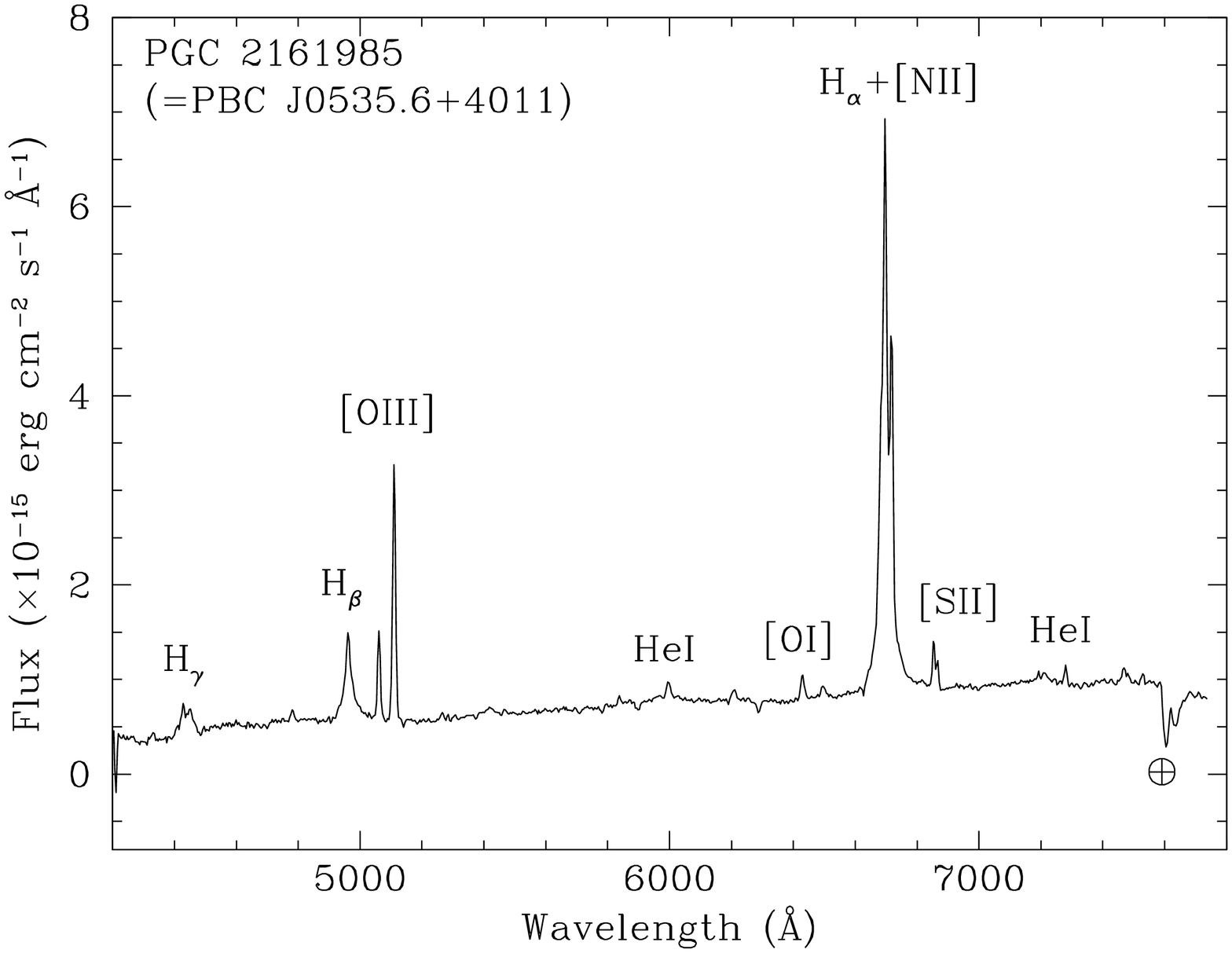,width=6.0cm,angle=270}}}
\centering{\mbox{\psfig{file=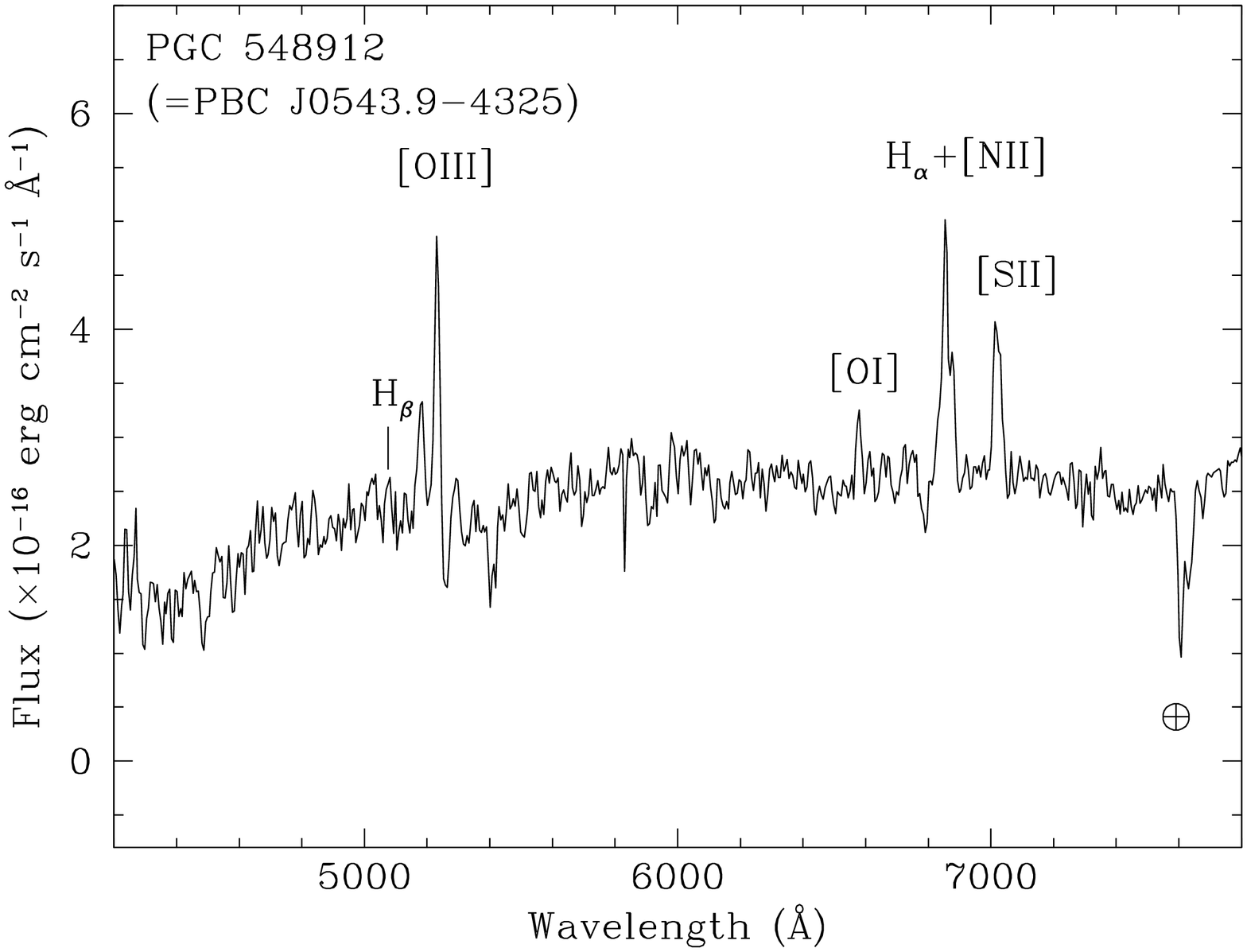,width=6.0cm,angle=270}}}
\centering{\mbox{\psfig{file=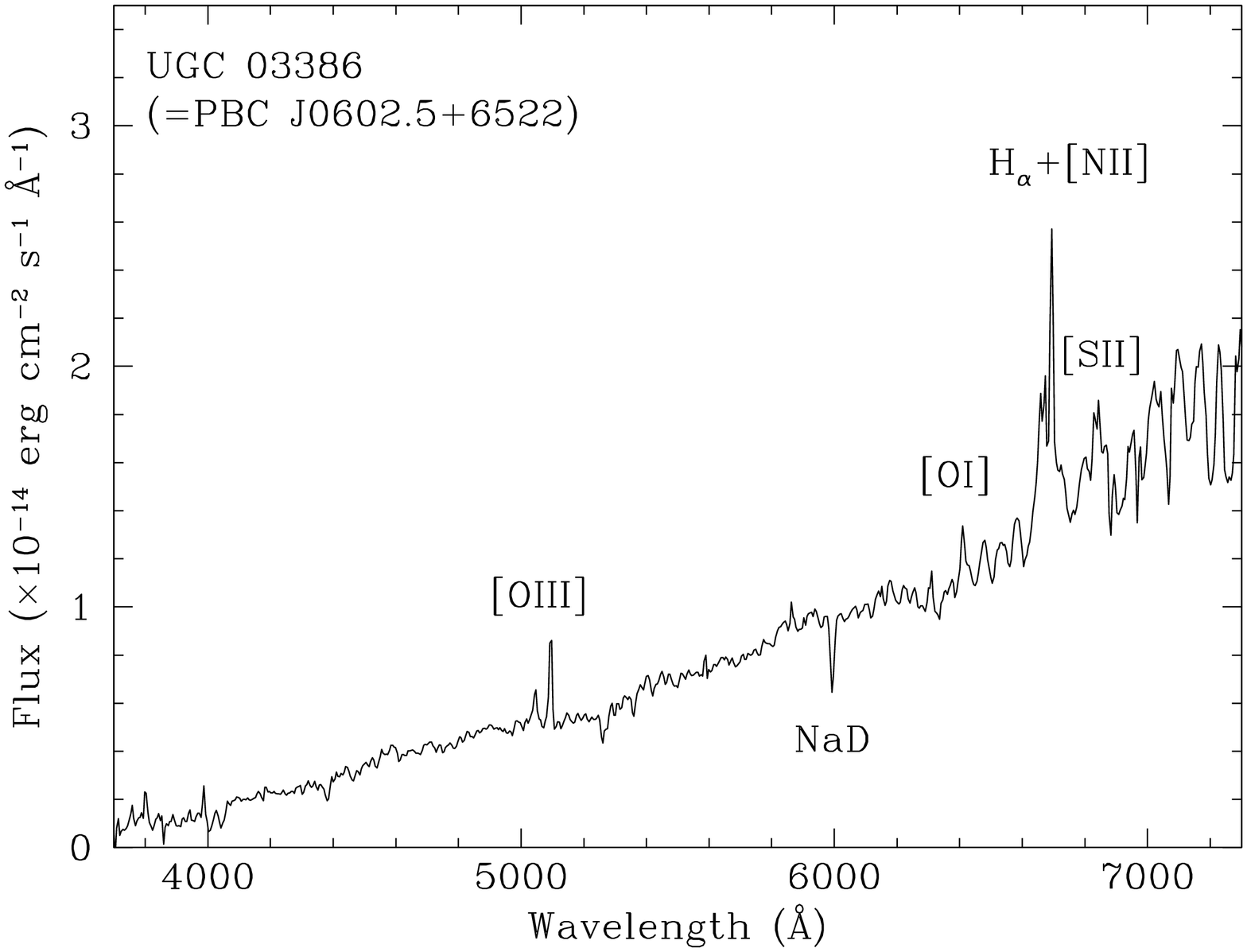,width=6.0cm,angle=270}}}
\centering{\mbox{\psfig{file=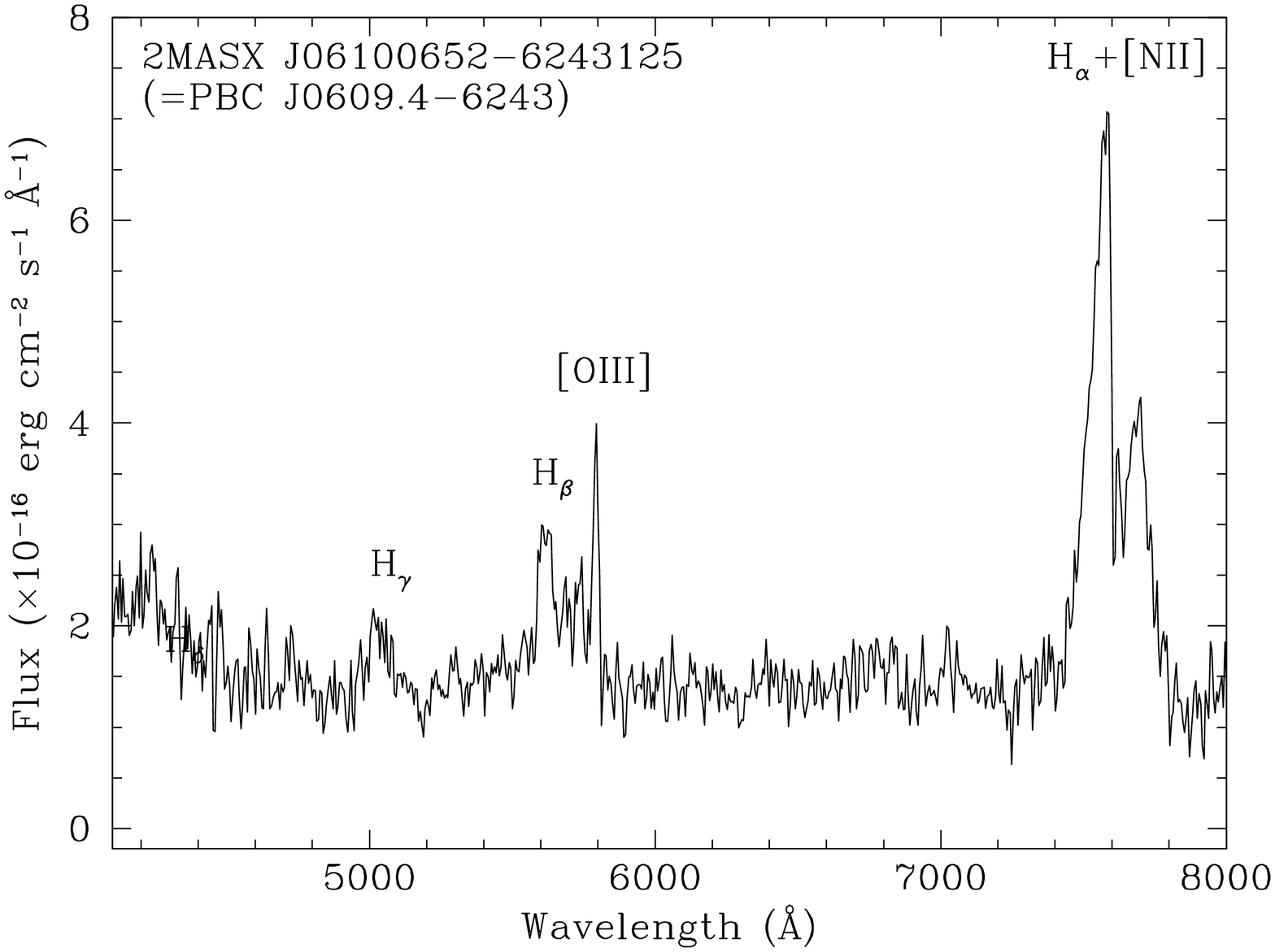,width=6.0cm,angle=270}}}
\centering{\mbox{\psfig{file=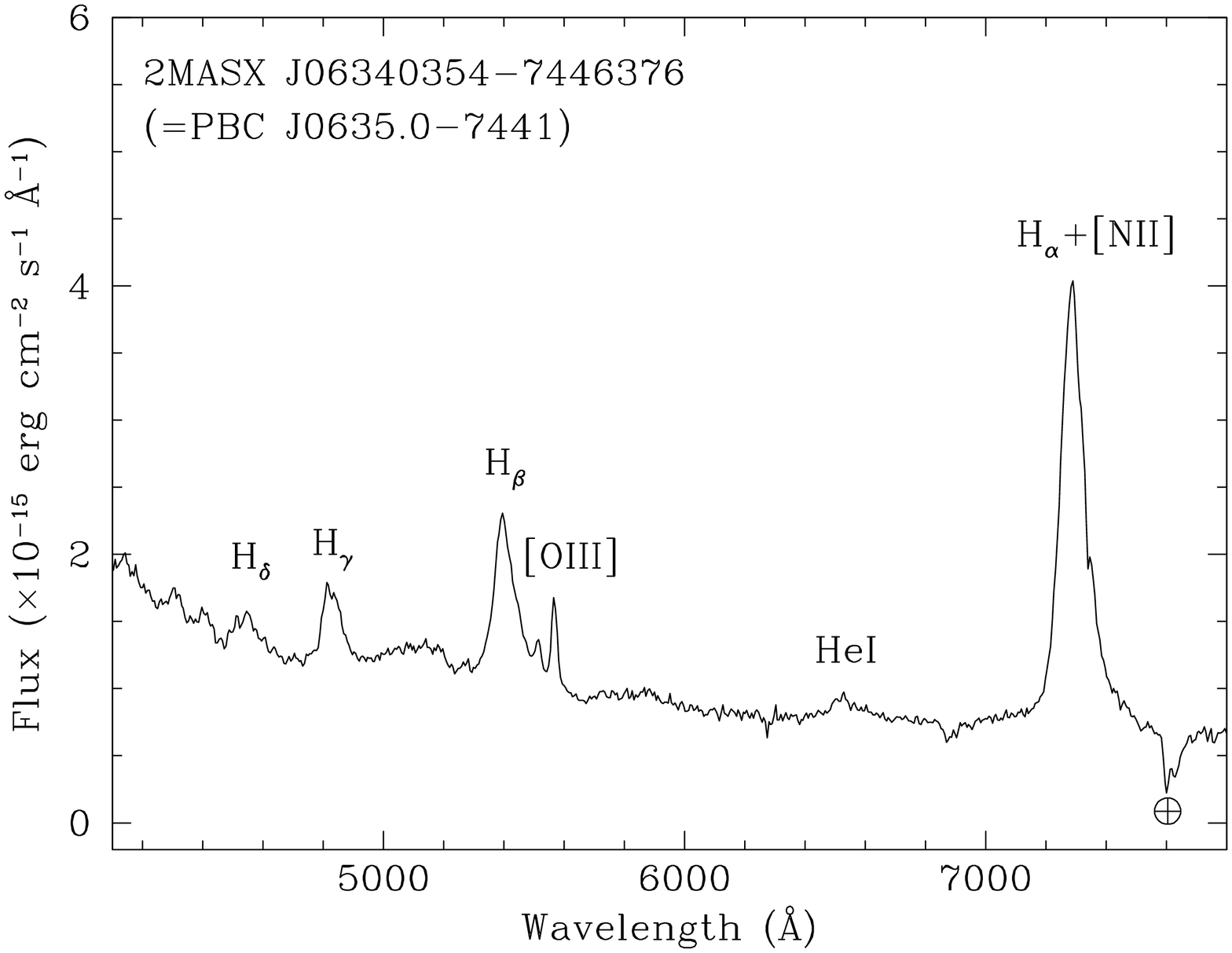,width=6.0cm,angle=270}}}
\centering{\mbox{\psfig{file=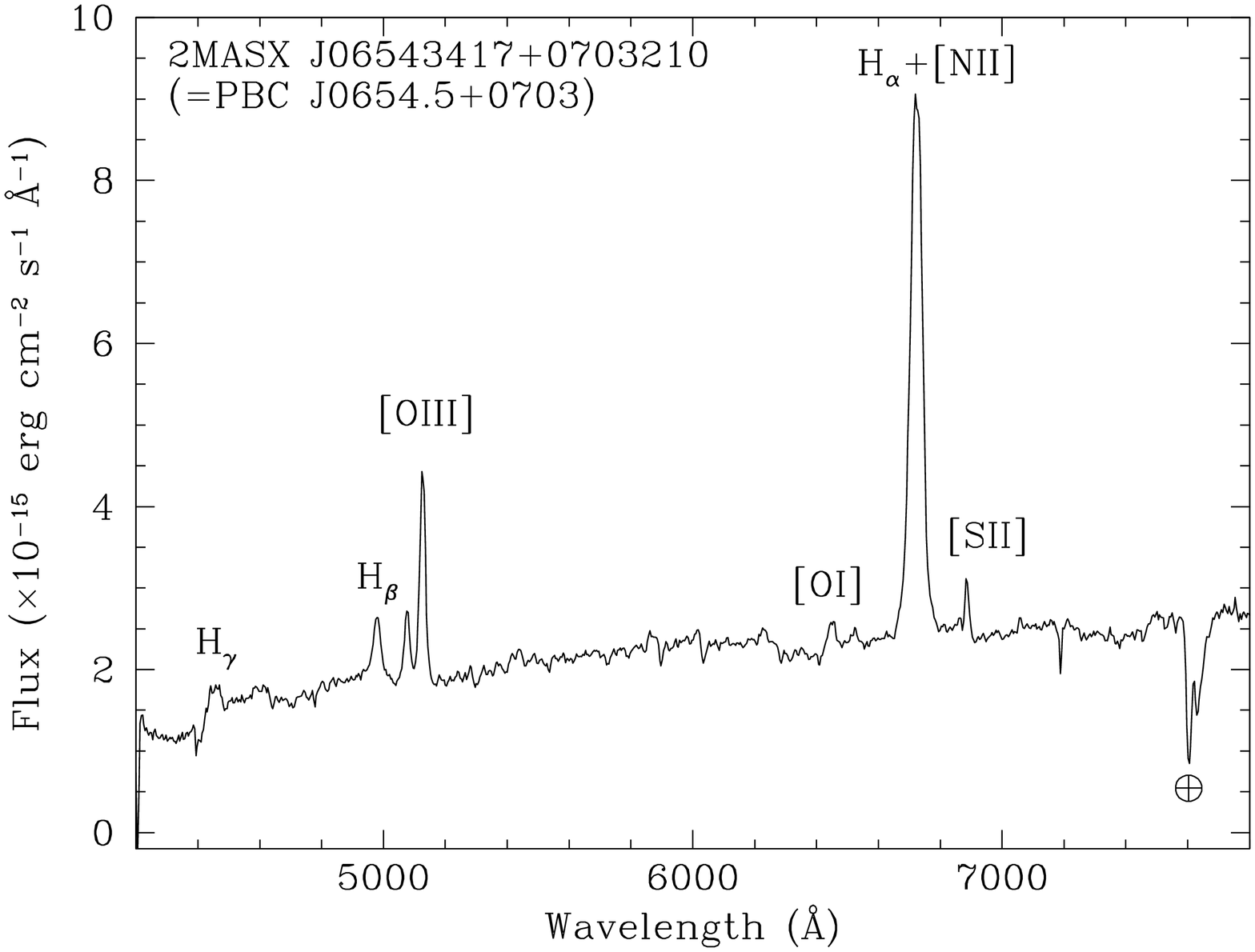,width=6.0cm,angle=270}}}
\centering{\mbox{\psfig{file=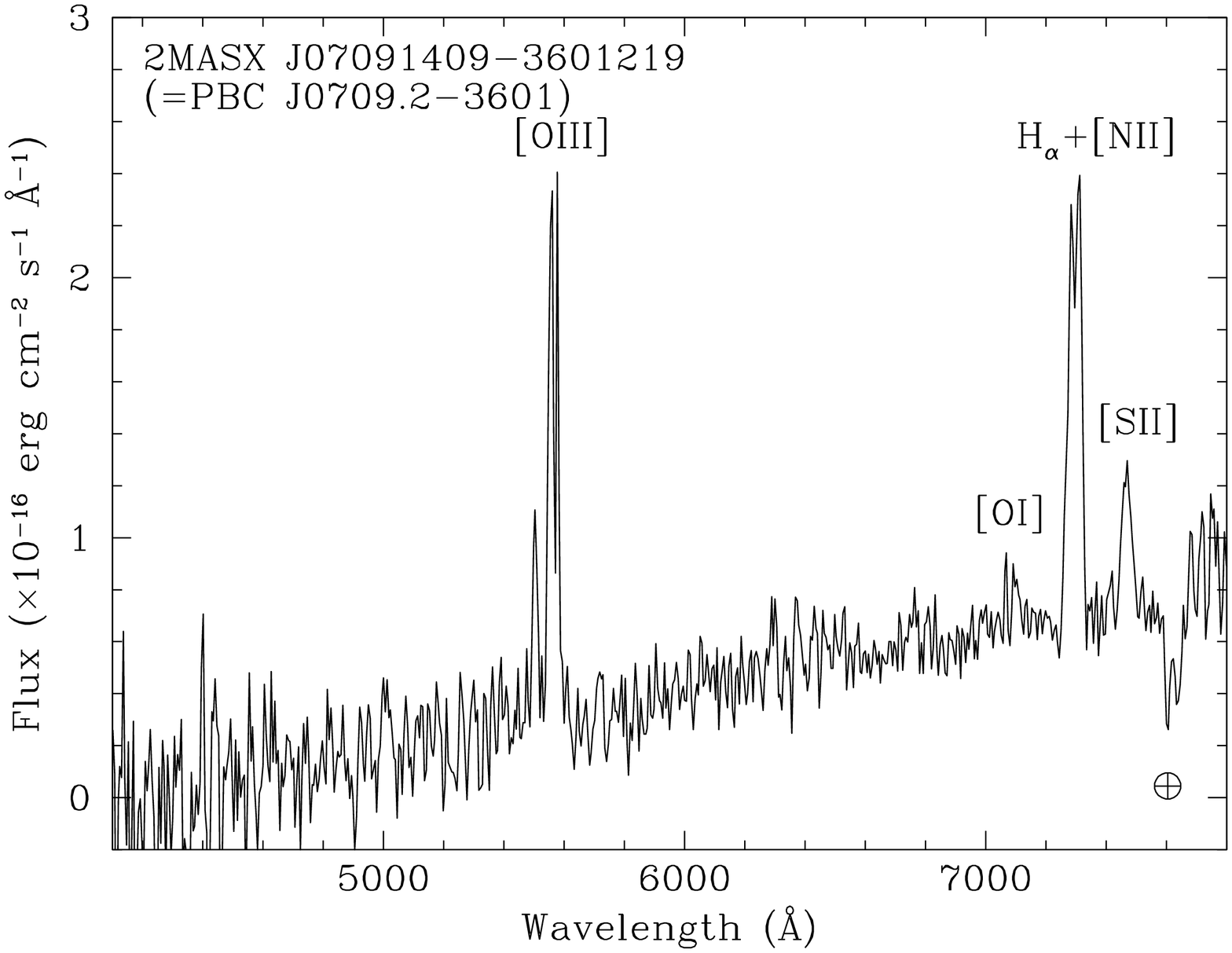,width=6.0cm,angle=270}}}
\centering{\mbox{\psfig{file=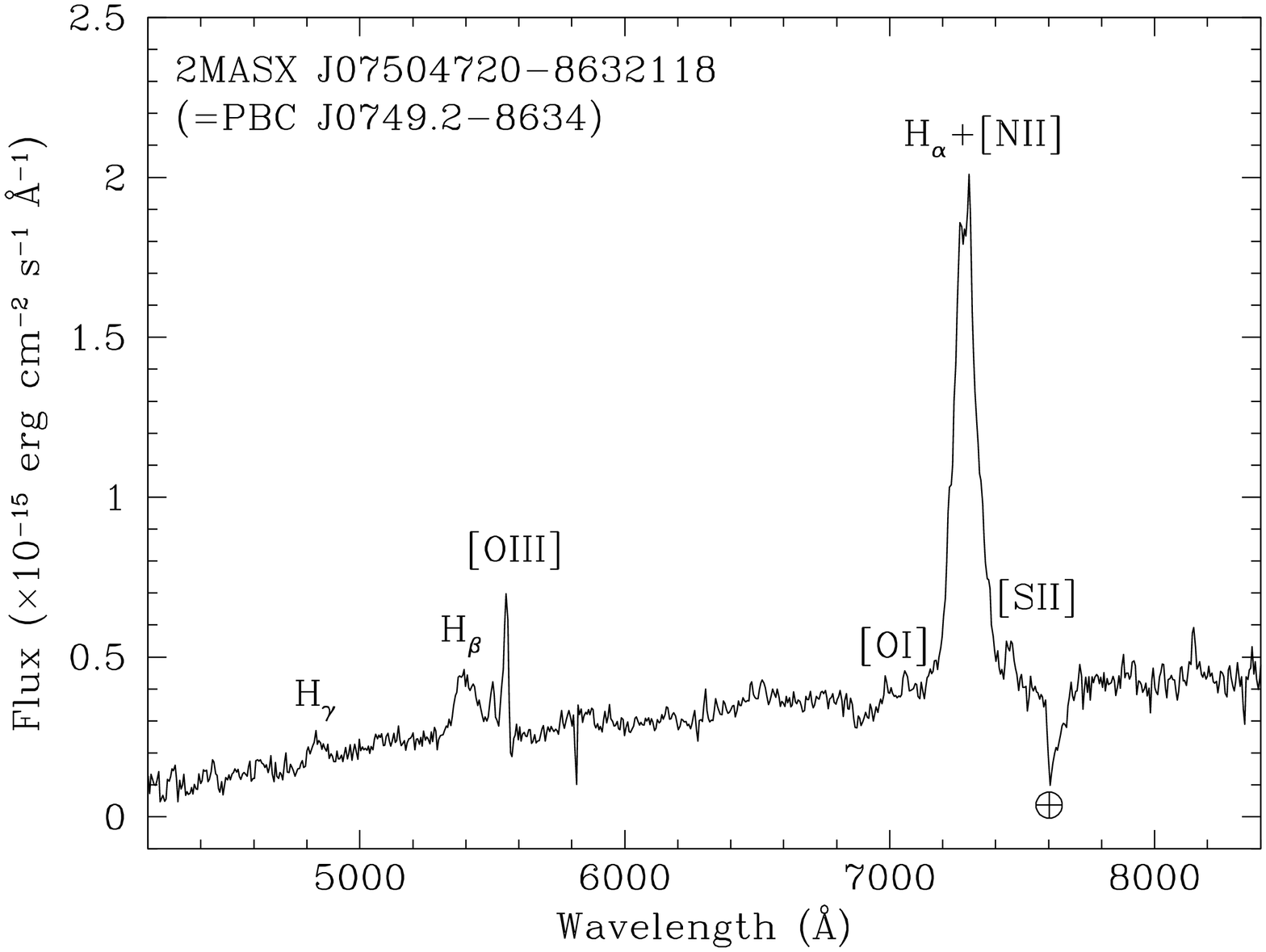,width=6.0cm,angle=270}}}
\centering{\mbox{\psfig{file=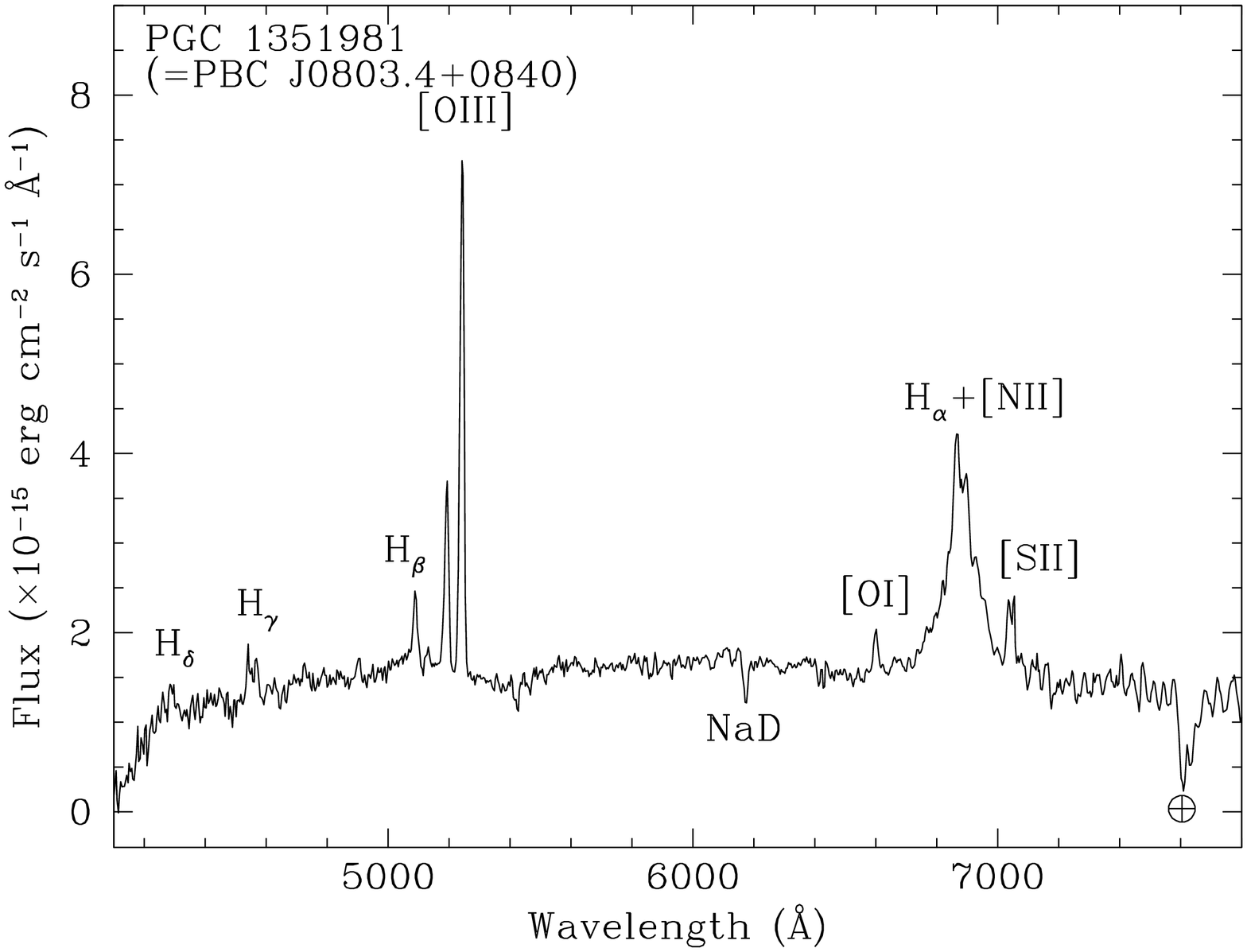,width=6.0cm,angle=270}}}
\centering{\mbox{\psfig{file=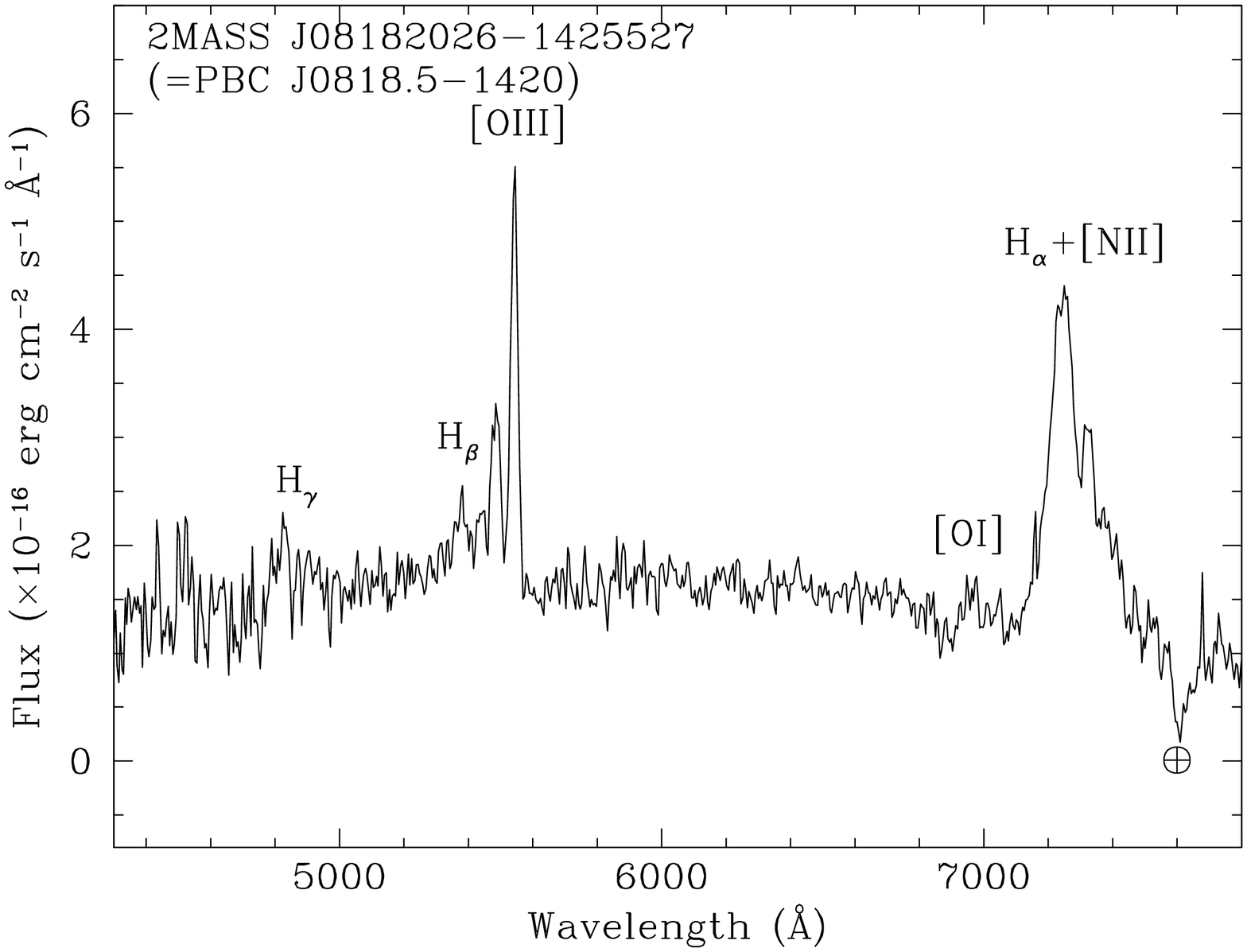,width=6.0cm,angle=270}}}
\centering{\mbox{\psfig{file=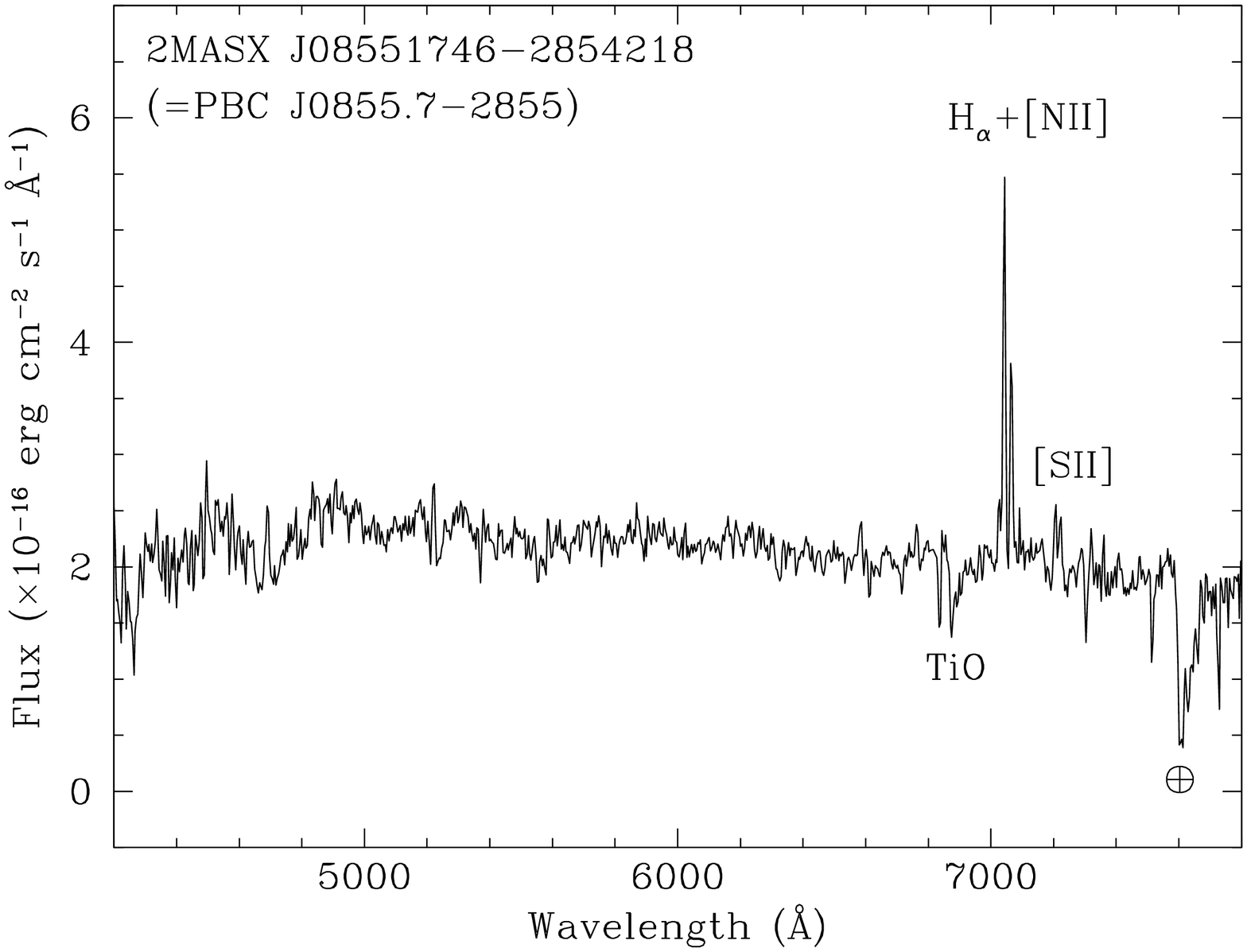,width=6.0cm,angle=270}}}
\caption{-- \emph{continued}}
\label{spectra2}
%\end{center}
\end{figure*}

\begin{figure*}
\setcounter{figure}{1}
%\begin{center}
\hspace{-.1cm}
\centering{\mbox{\psfig{file=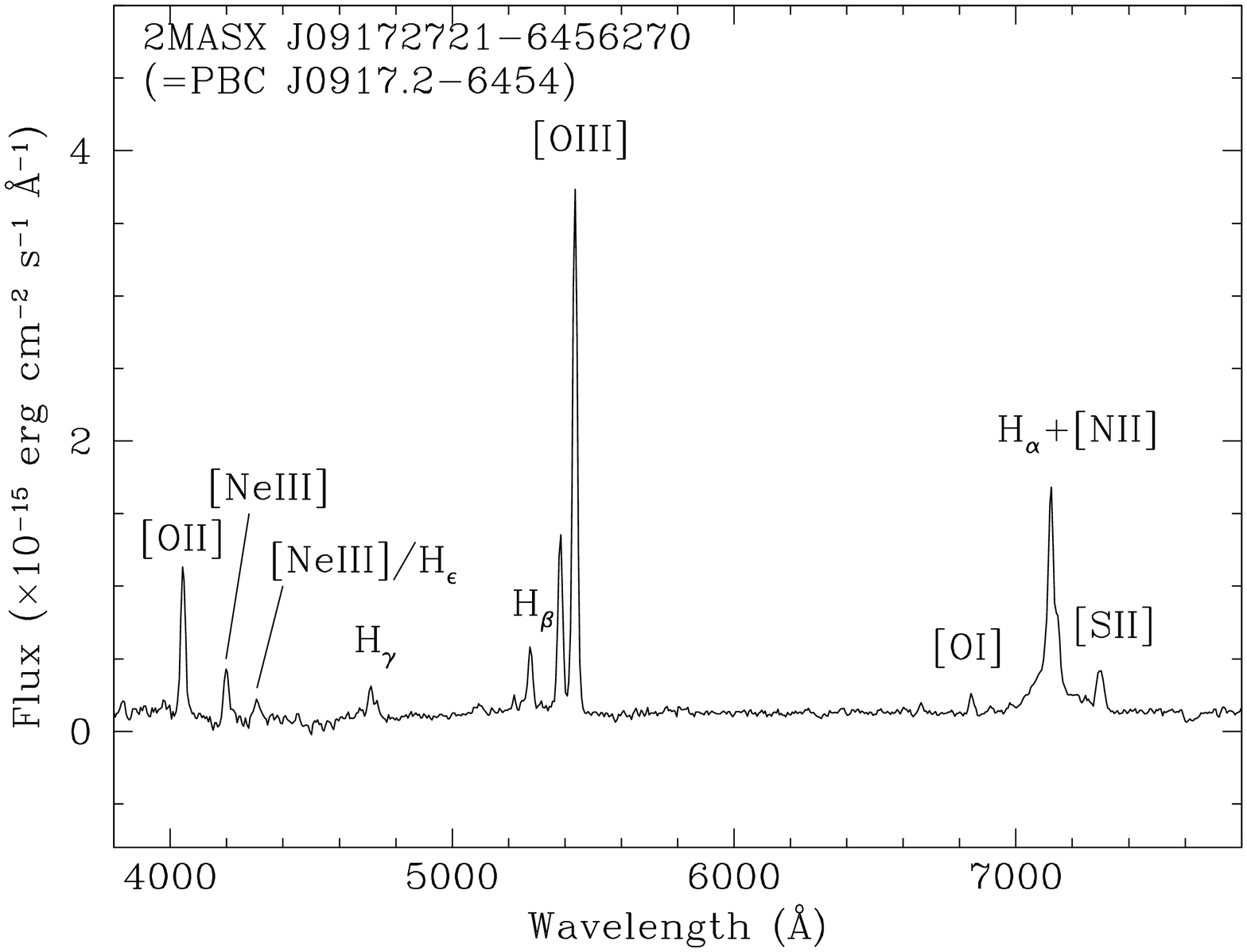,width=6.0cm,angle=270}}}
\centering{\mbox{\psfig{file=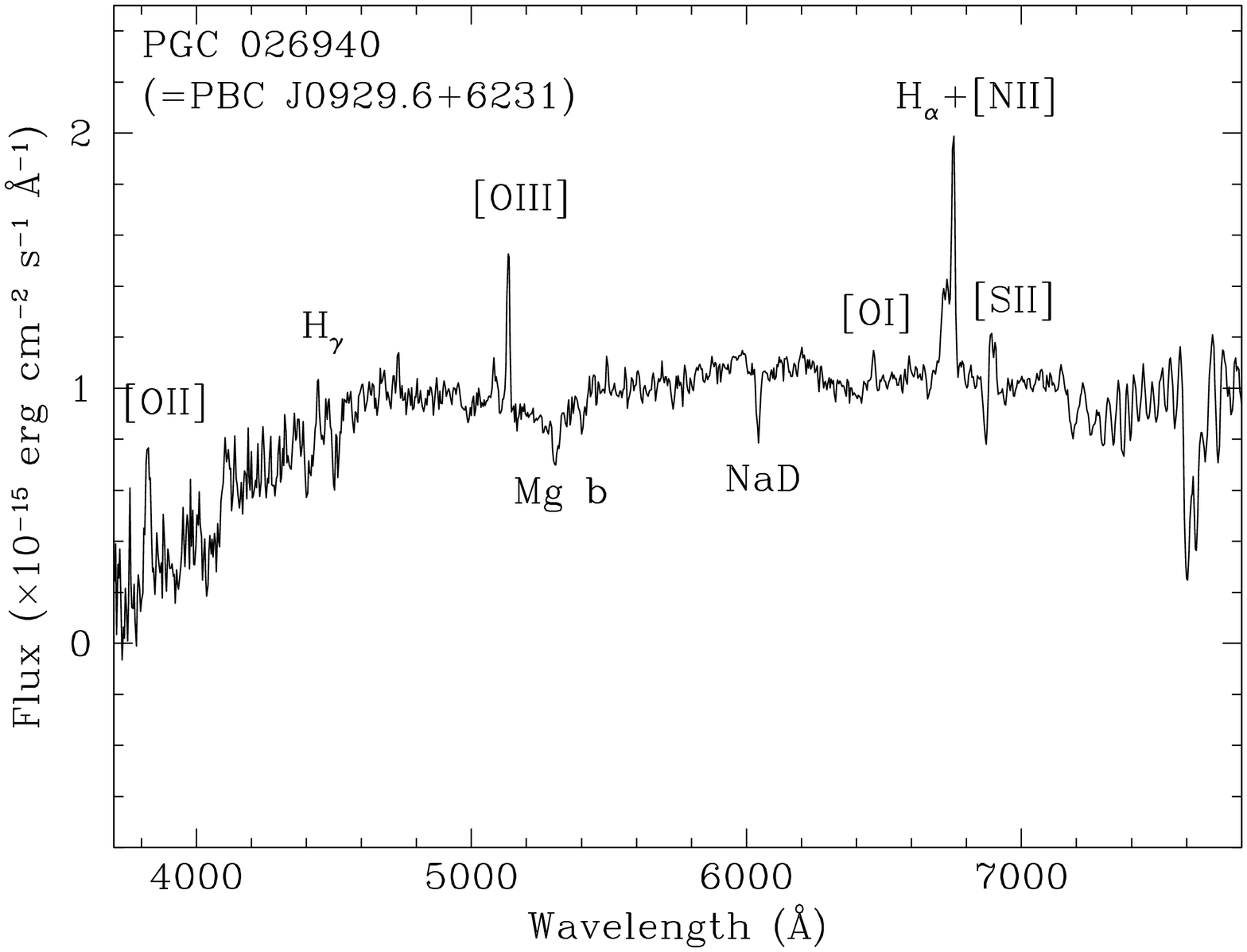,width=6.0cm,angle=270}}}
\centering{\mbox{\psfig{file=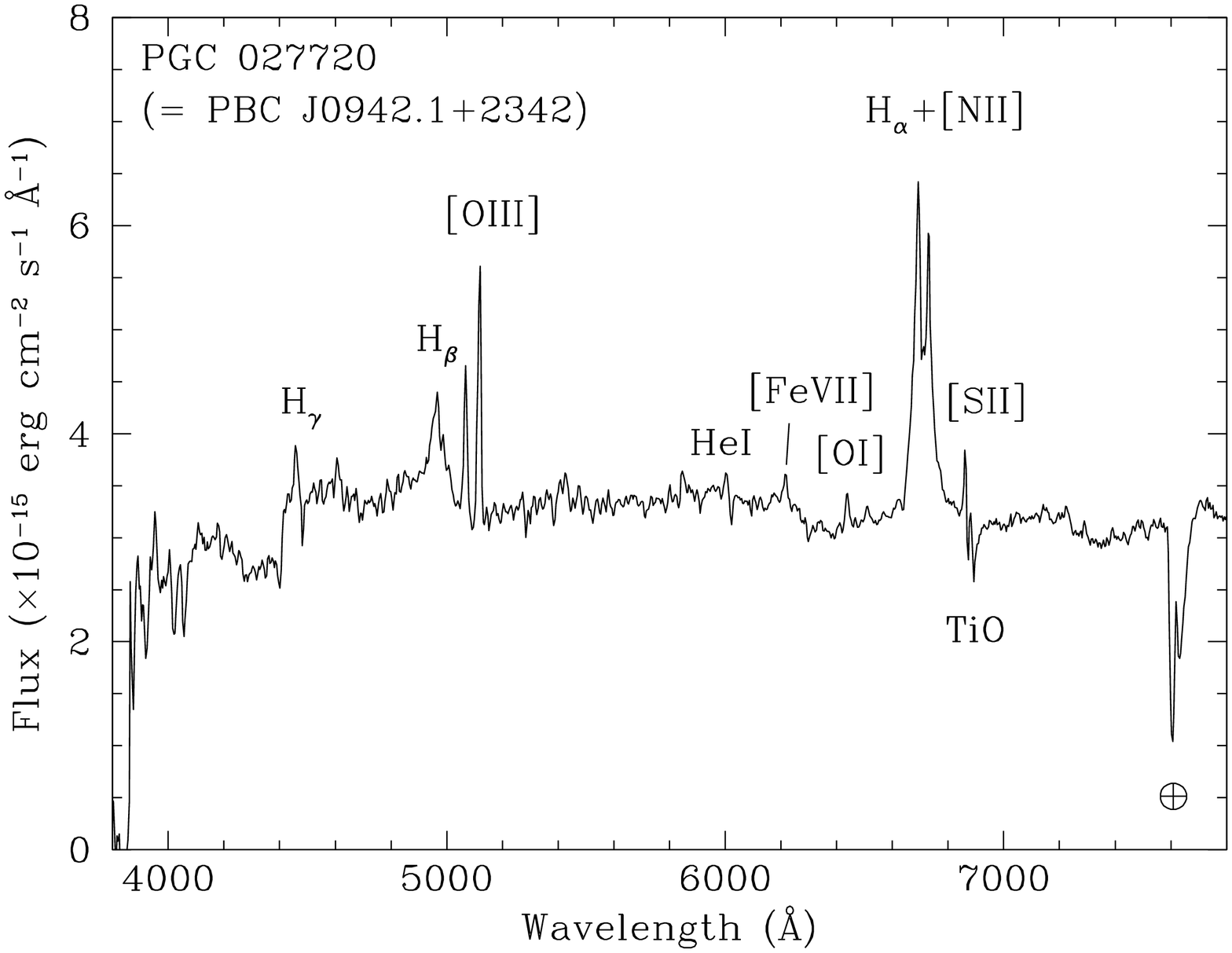,width=6.0cm,angle=270}}}
\centering{\mbox{\psfig{file=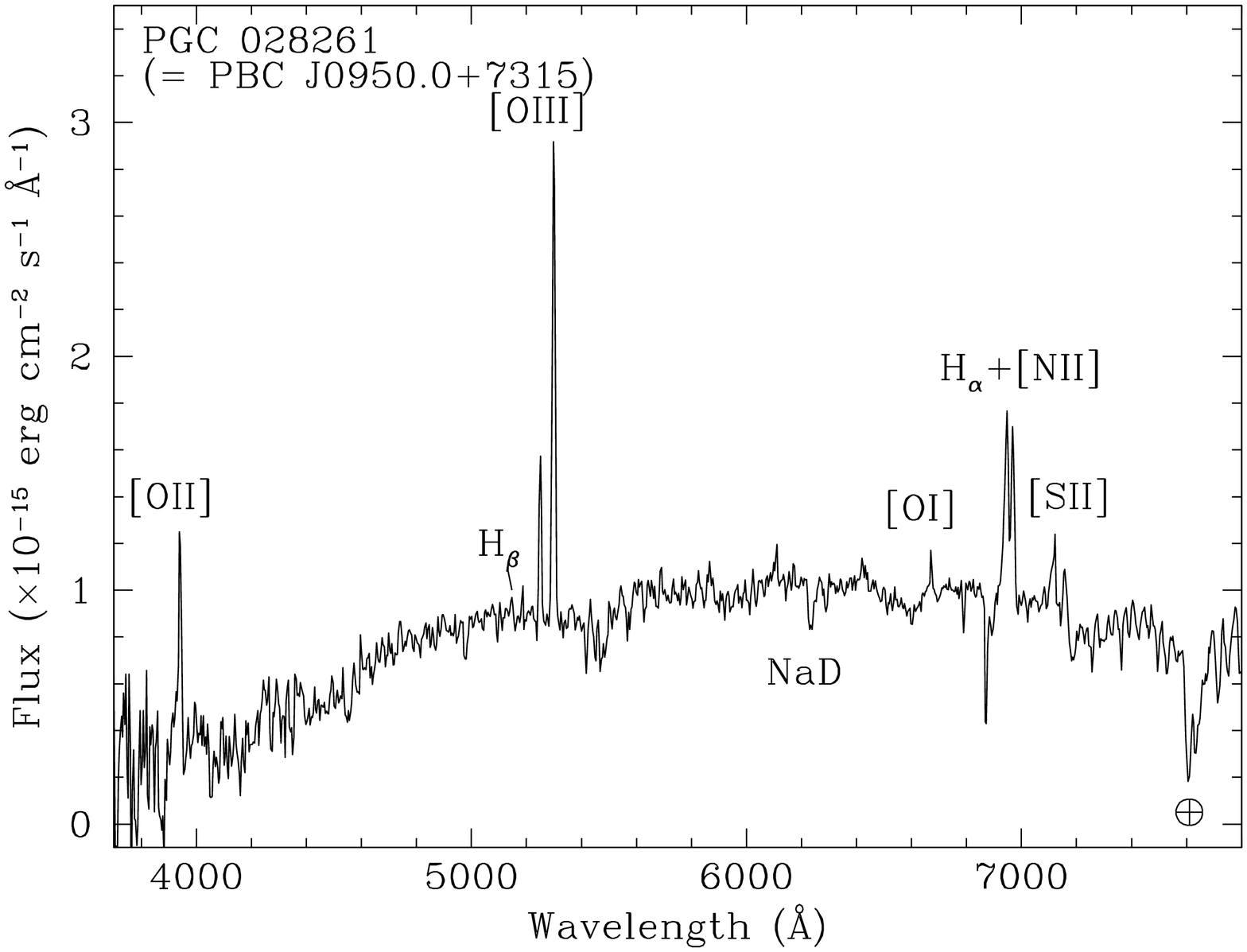,width=6.0cm,angle=270}}}
\centering{\mbox{\psfig{file=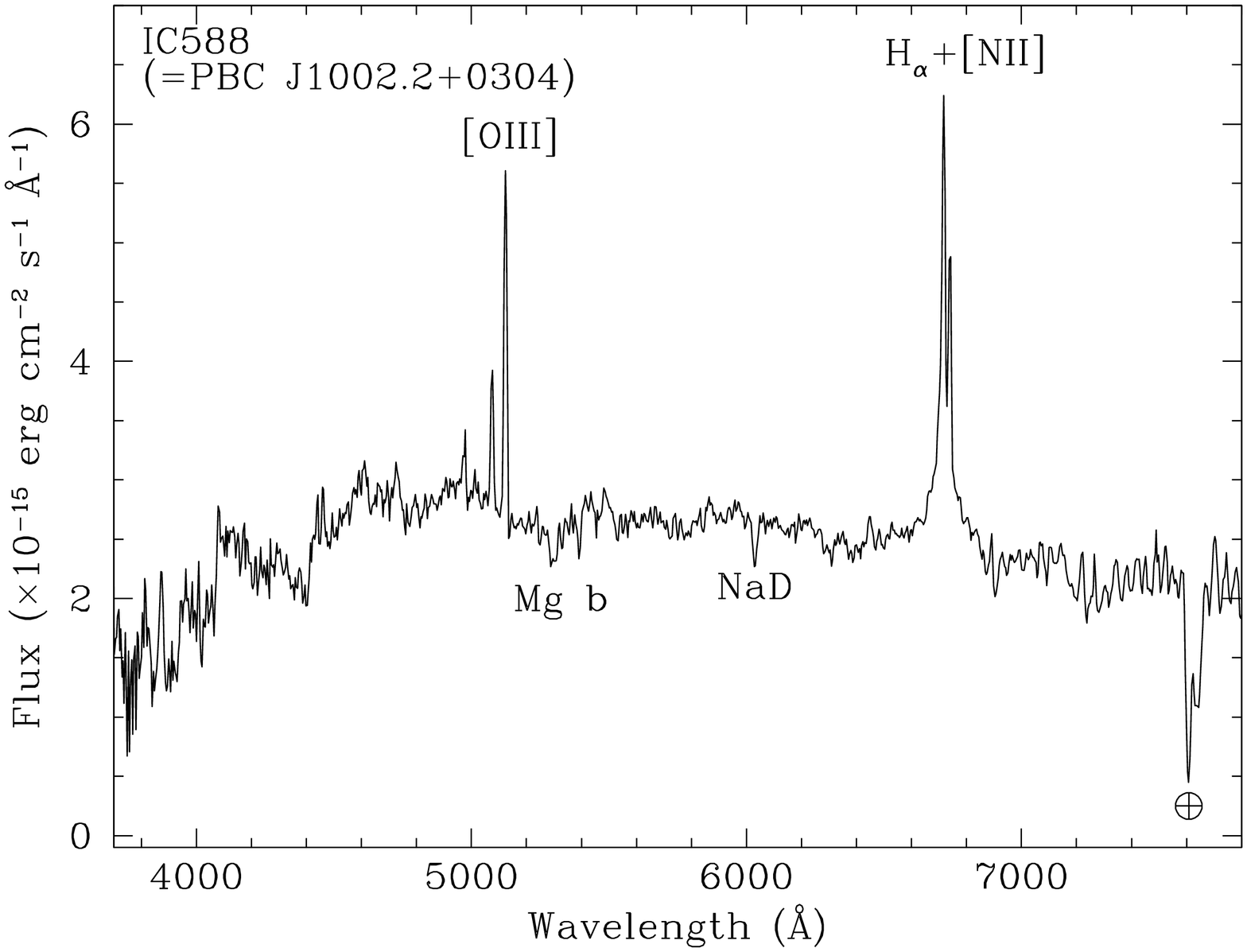,width=6.0cm,angle=270}}}
\centering{\mbox{\psfig{file=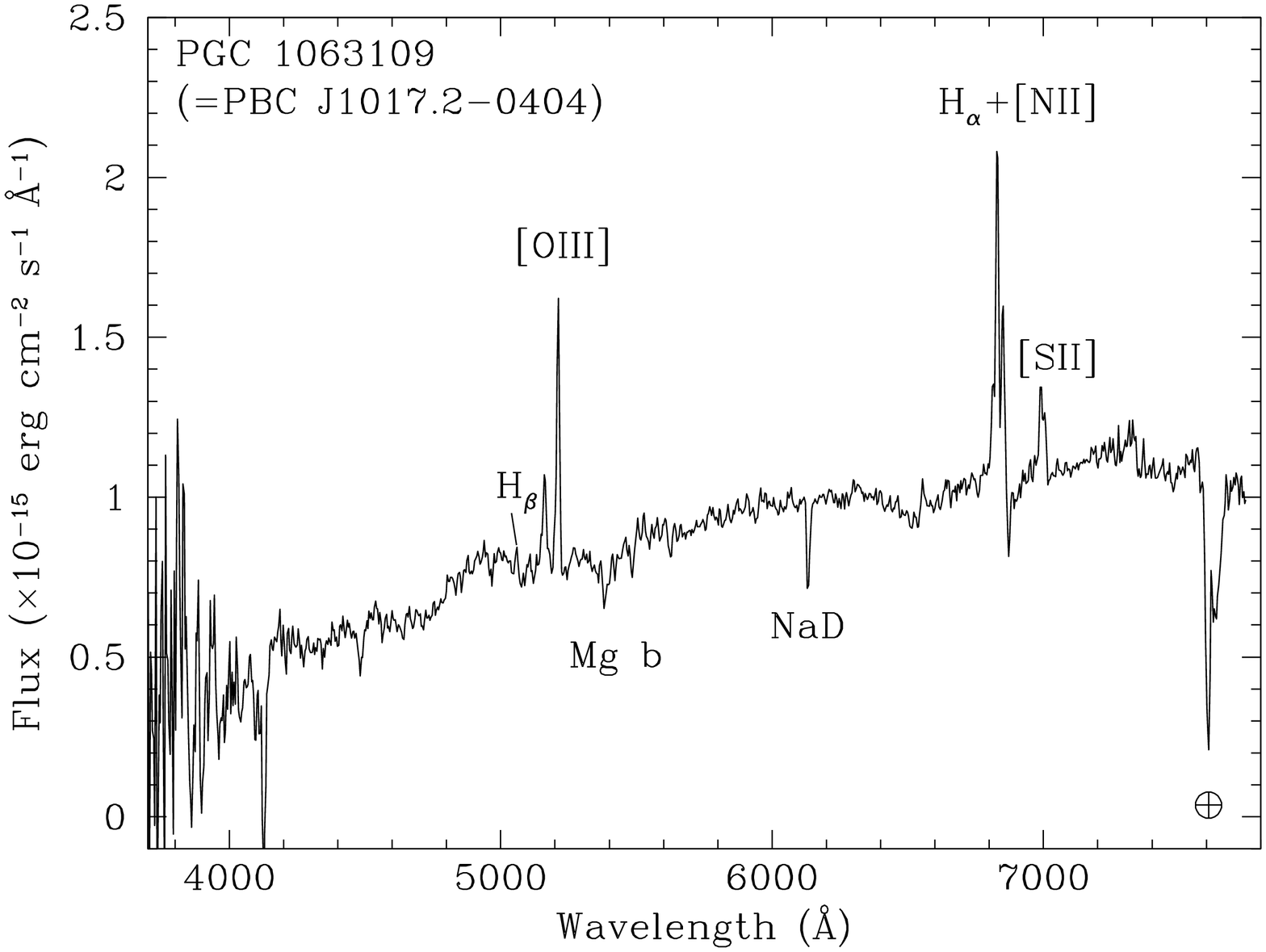,width=6.0cm,angle=270}}}
\centering{\mbox{\psfig{file=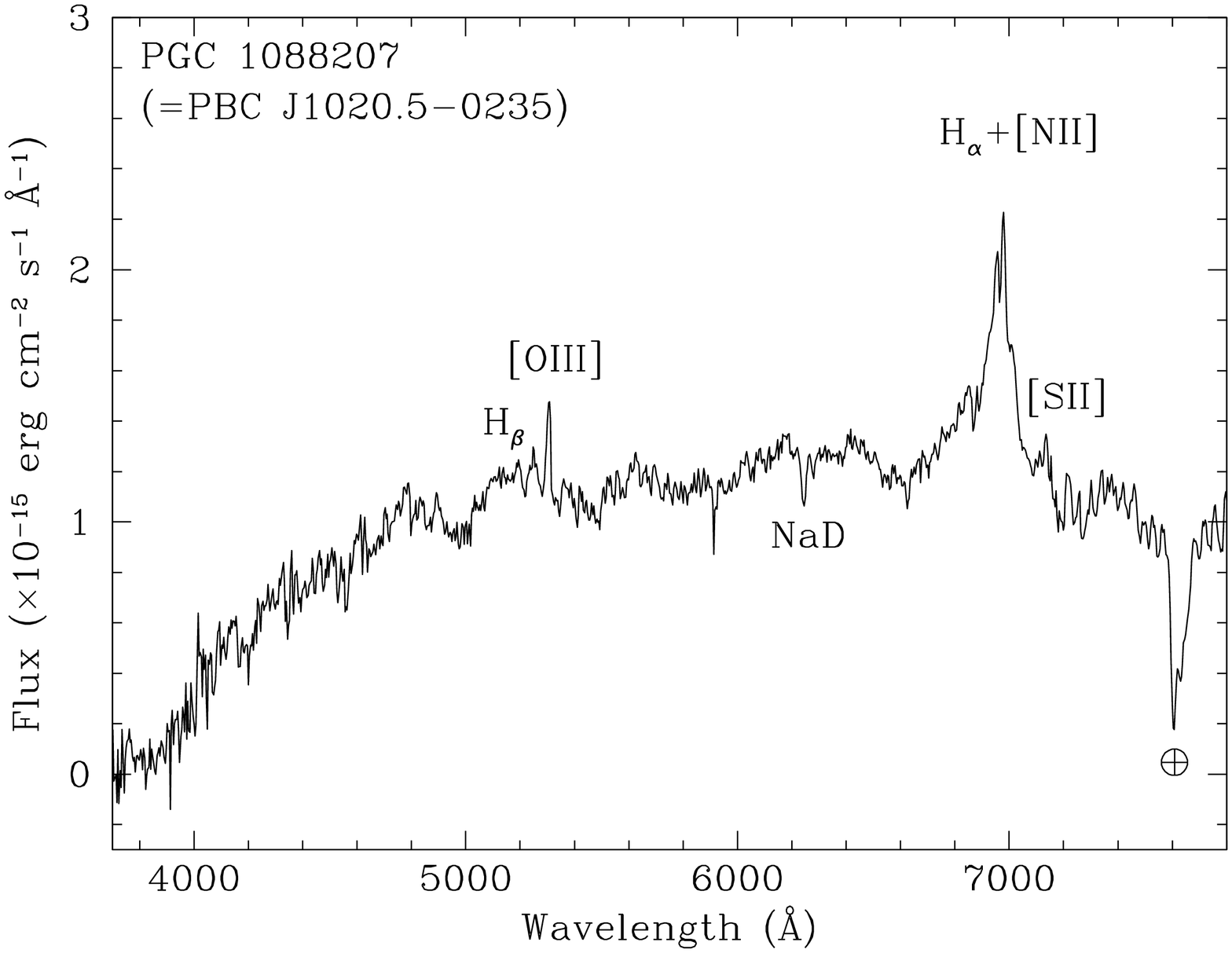,width=6.0cm,angle=270}}}
\centering{\mbox{\psfig{file=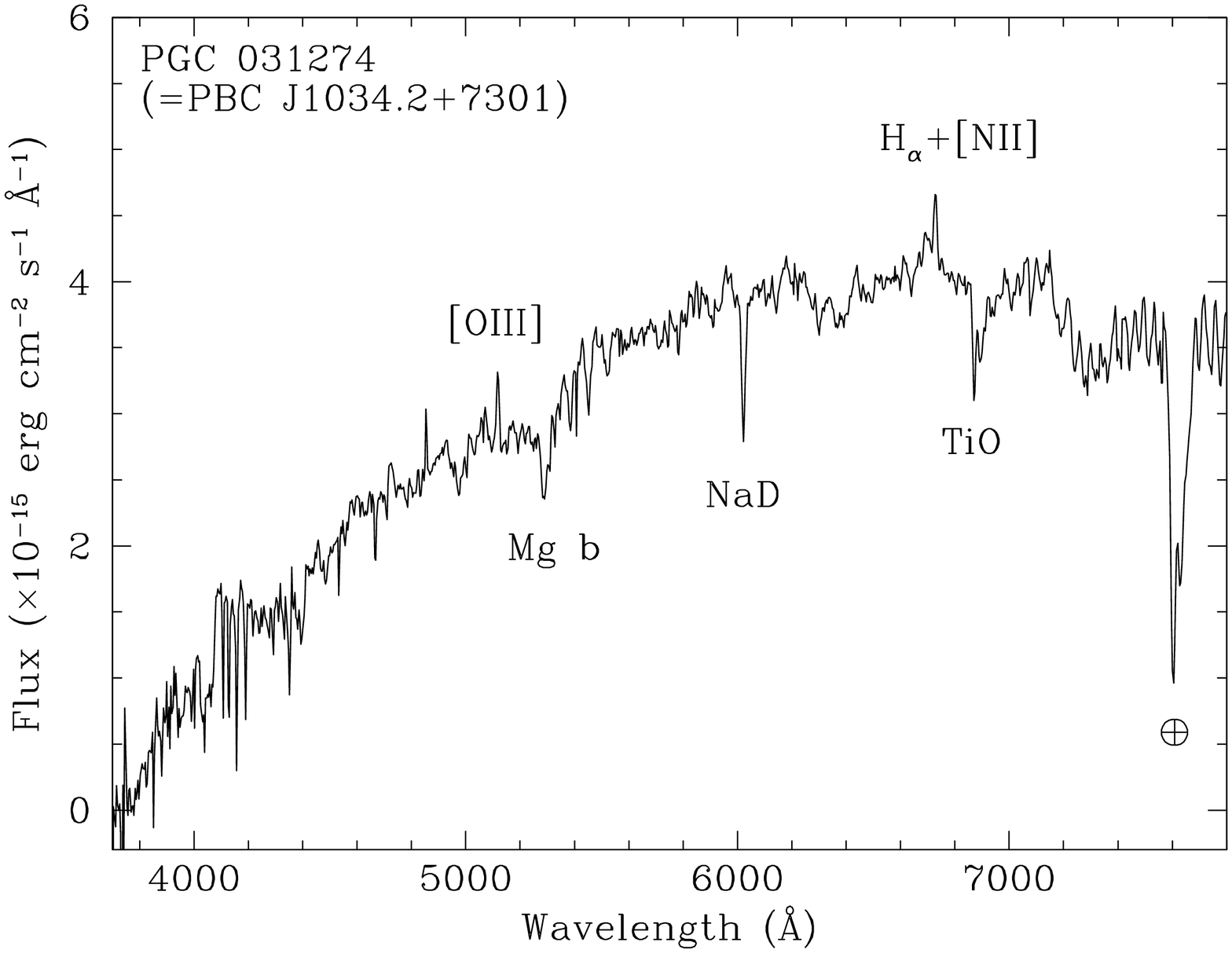,width=6.0cm,angle=270}}}
\centering{\mbox{\psfig{file=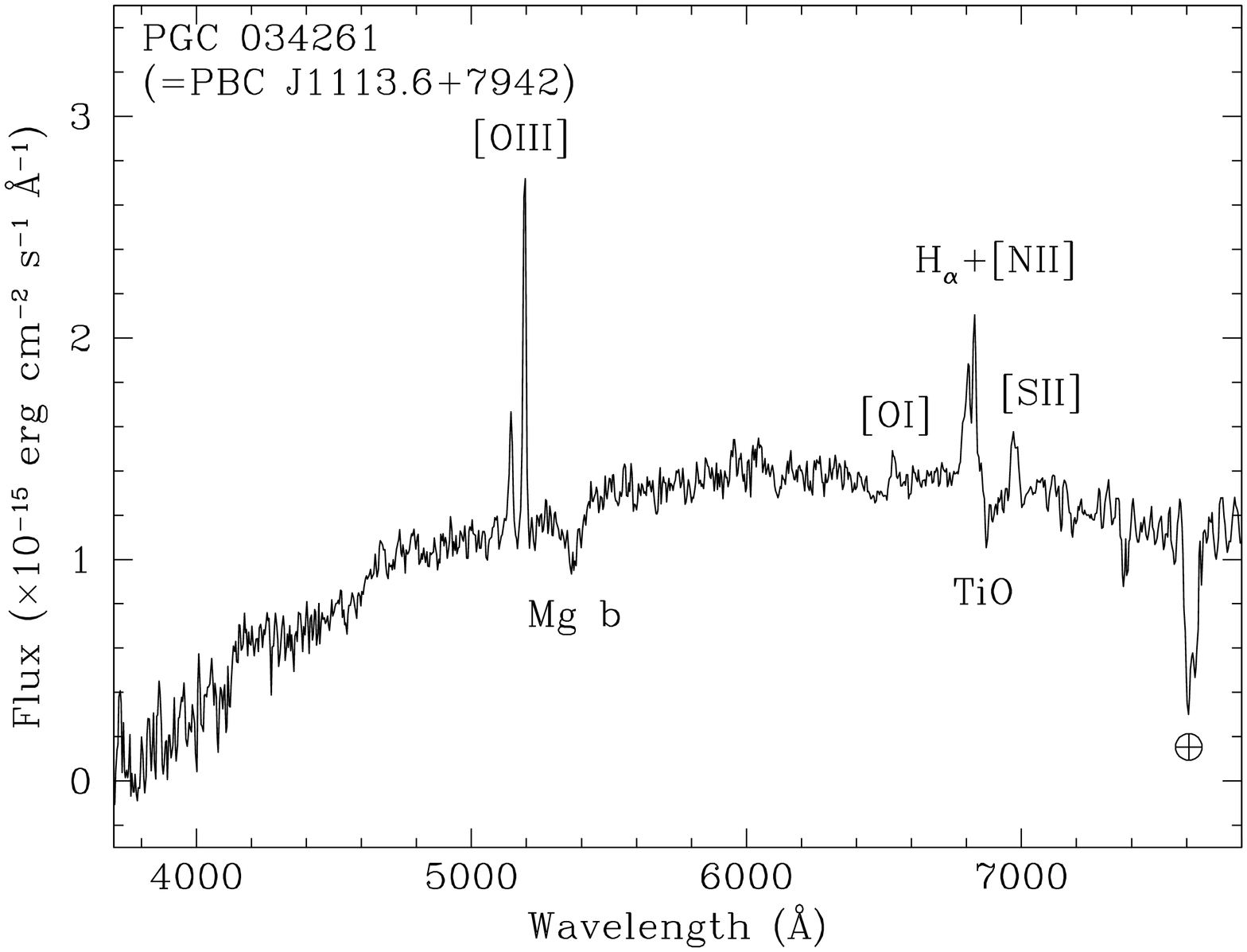,width=6.0cm,angle=270}}}
\centering{\mbox{\psfig{file=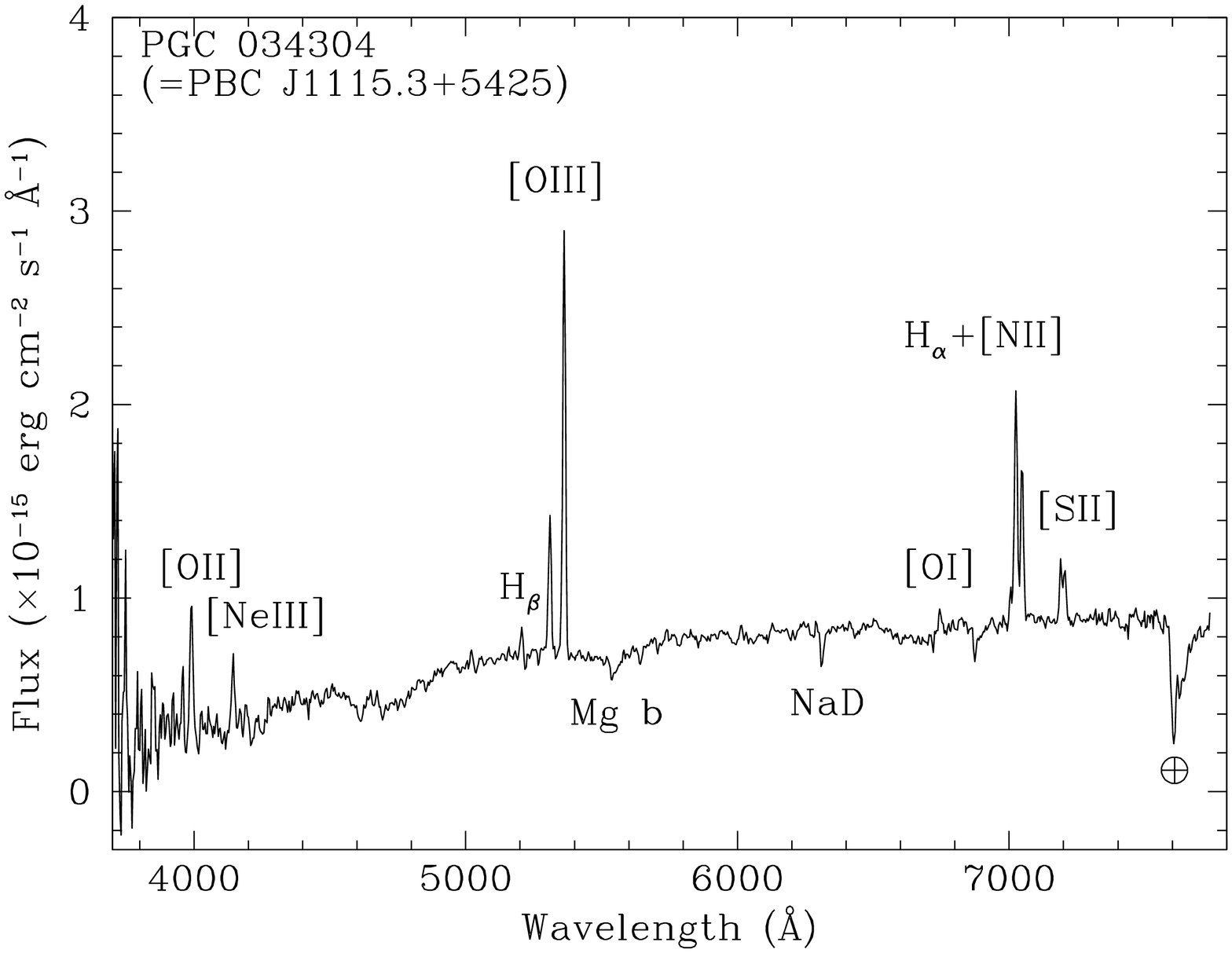,width=6.0cm,angle=270}}}
\centering{\mbox{\psfig{file=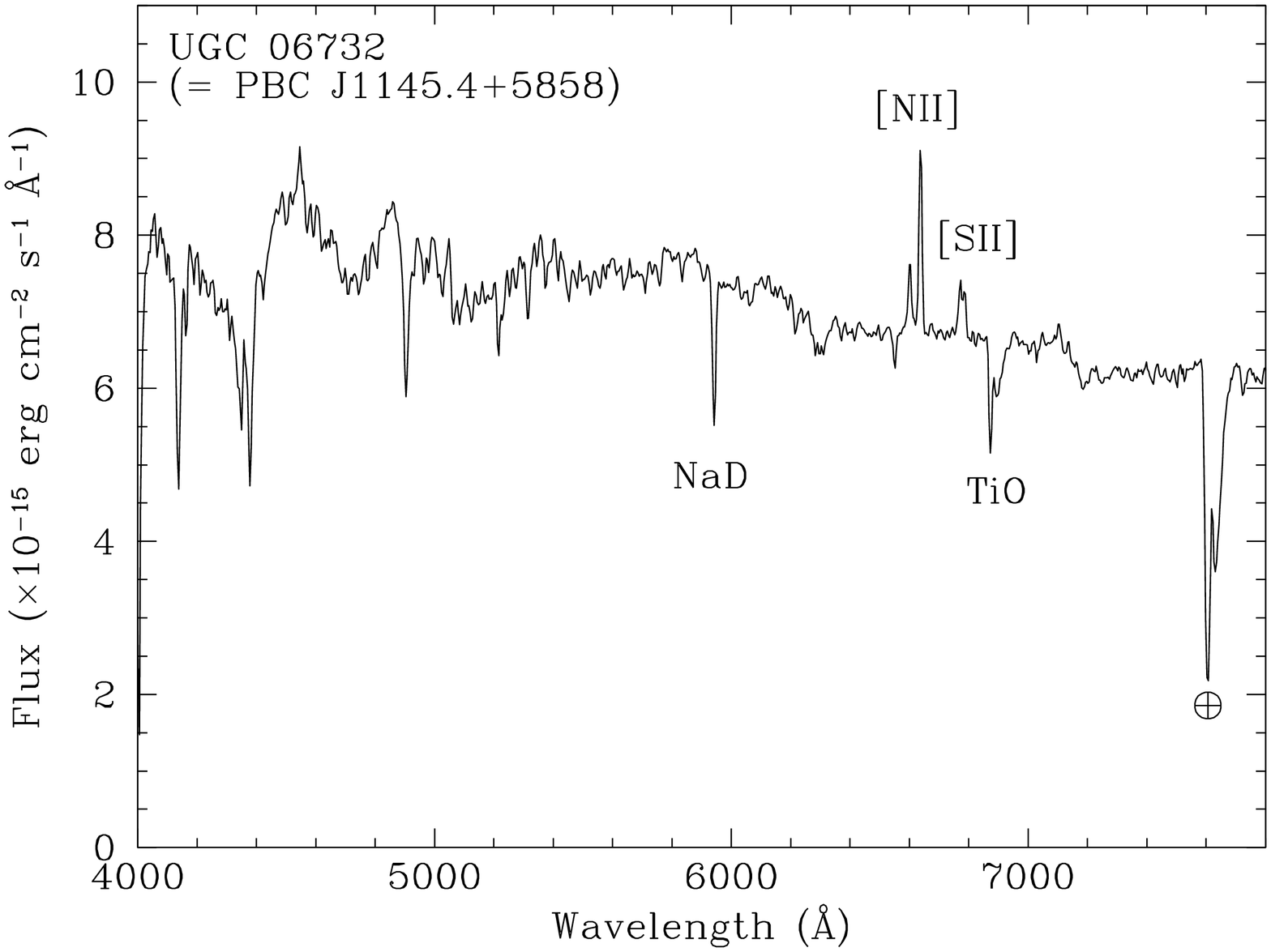,width=6.0cm,angle=270}}}
\centering{\mbox{\psfig{file=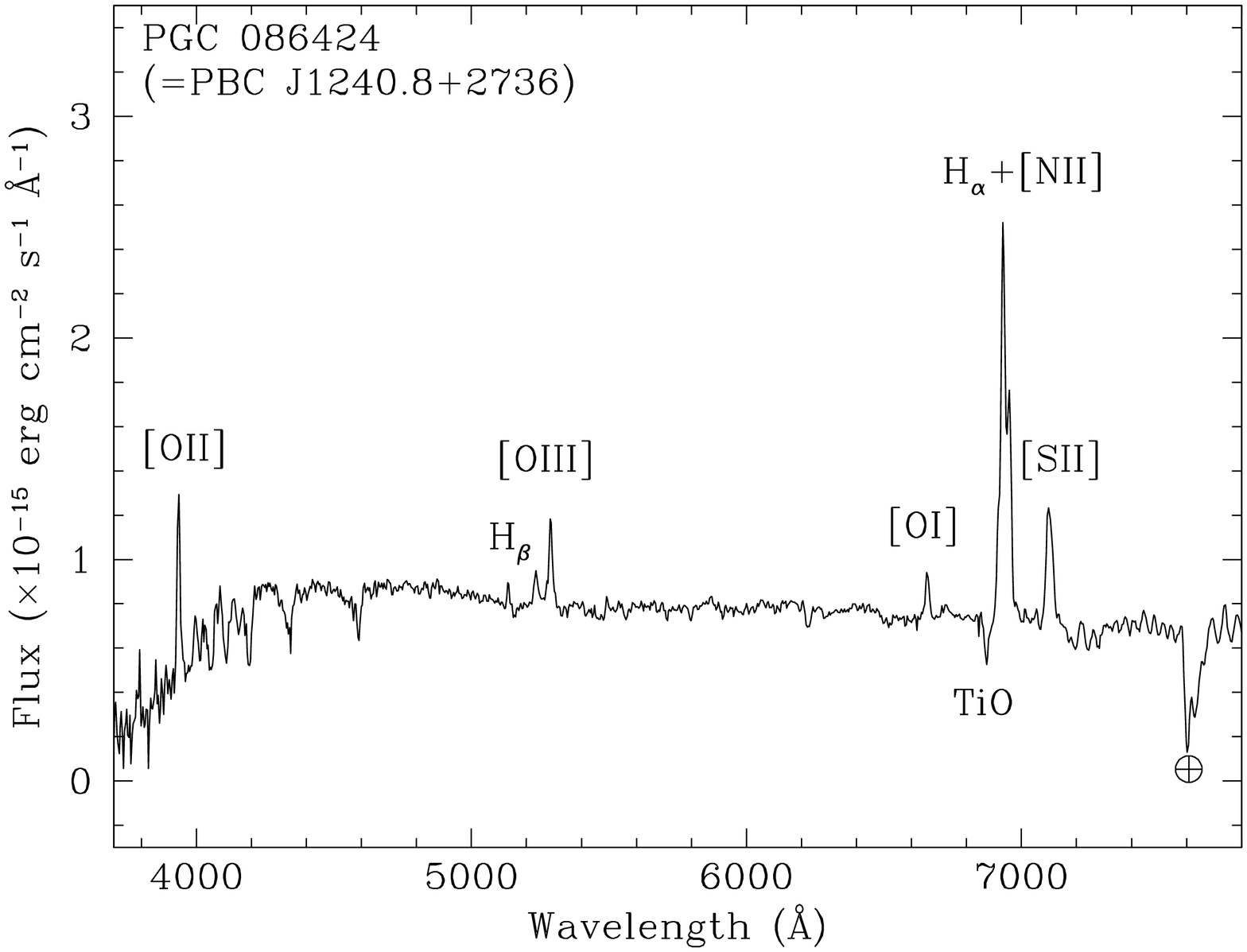,width=6.0cm,angle=270}}}
\centering{\mbox{\psfig{file=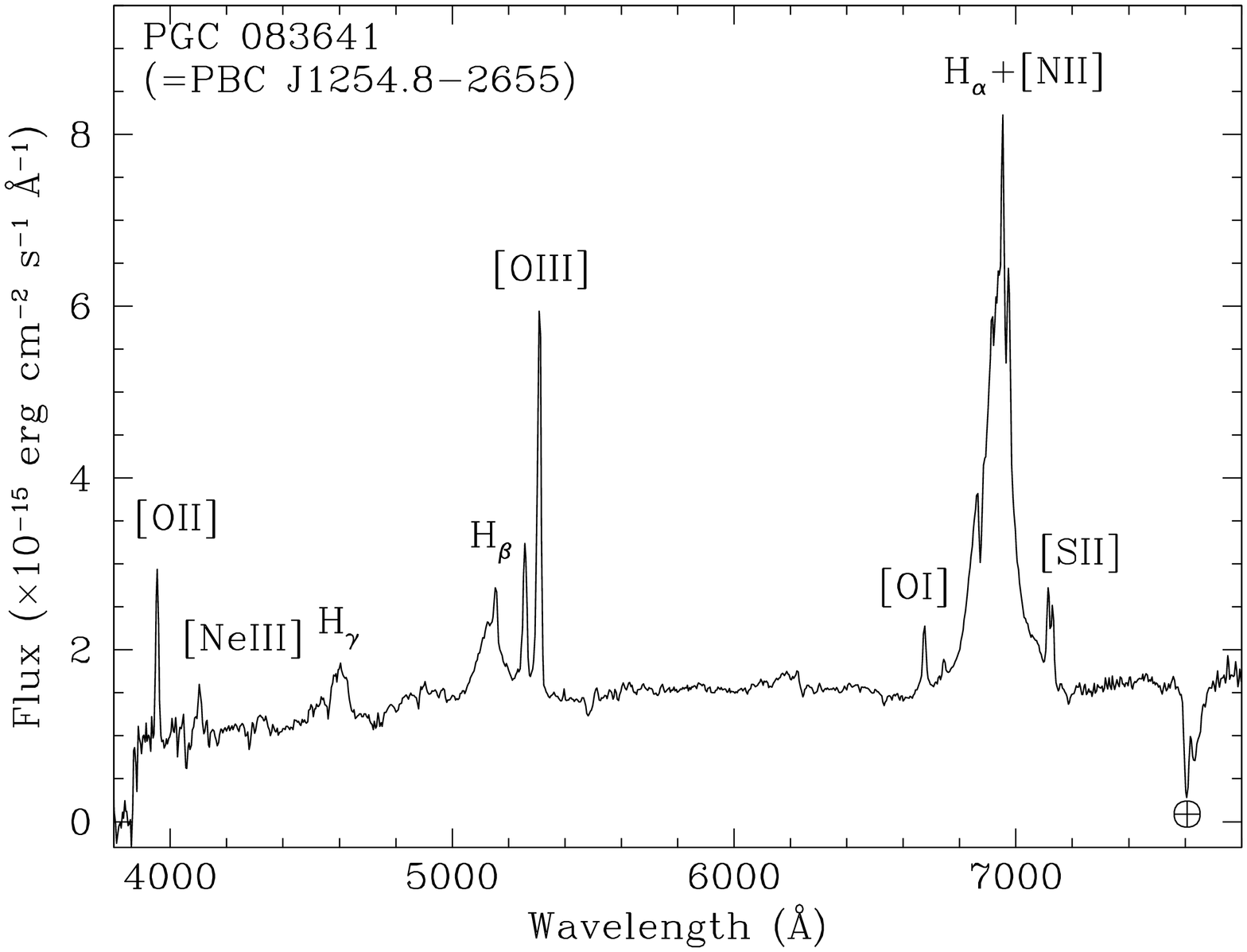,width=6.0cm,angle=270}}}
\centering{\mbox{\psfig{file=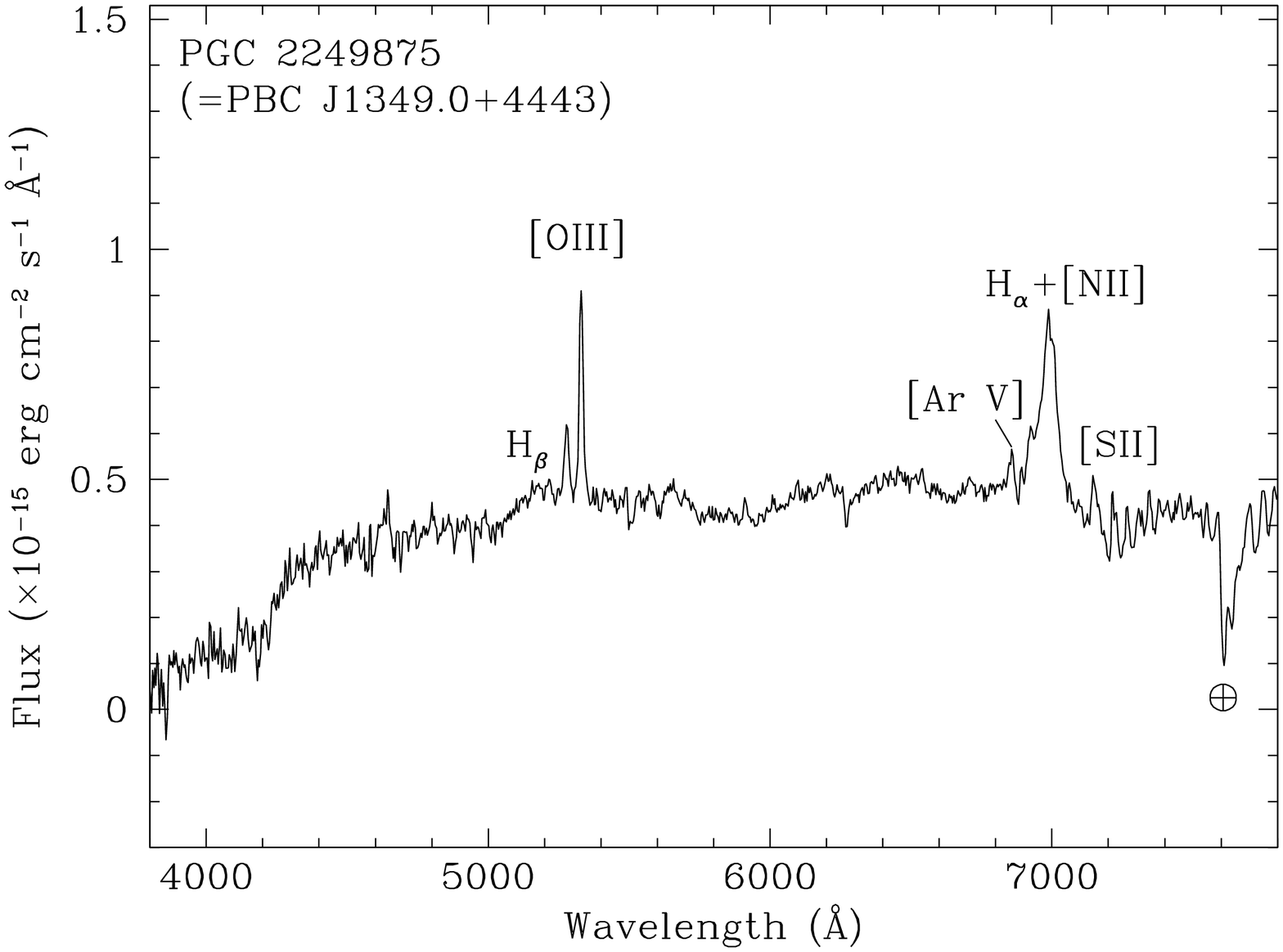,width=6.0cm,angle=270}}}
\centering{\mbox{\psfig{file=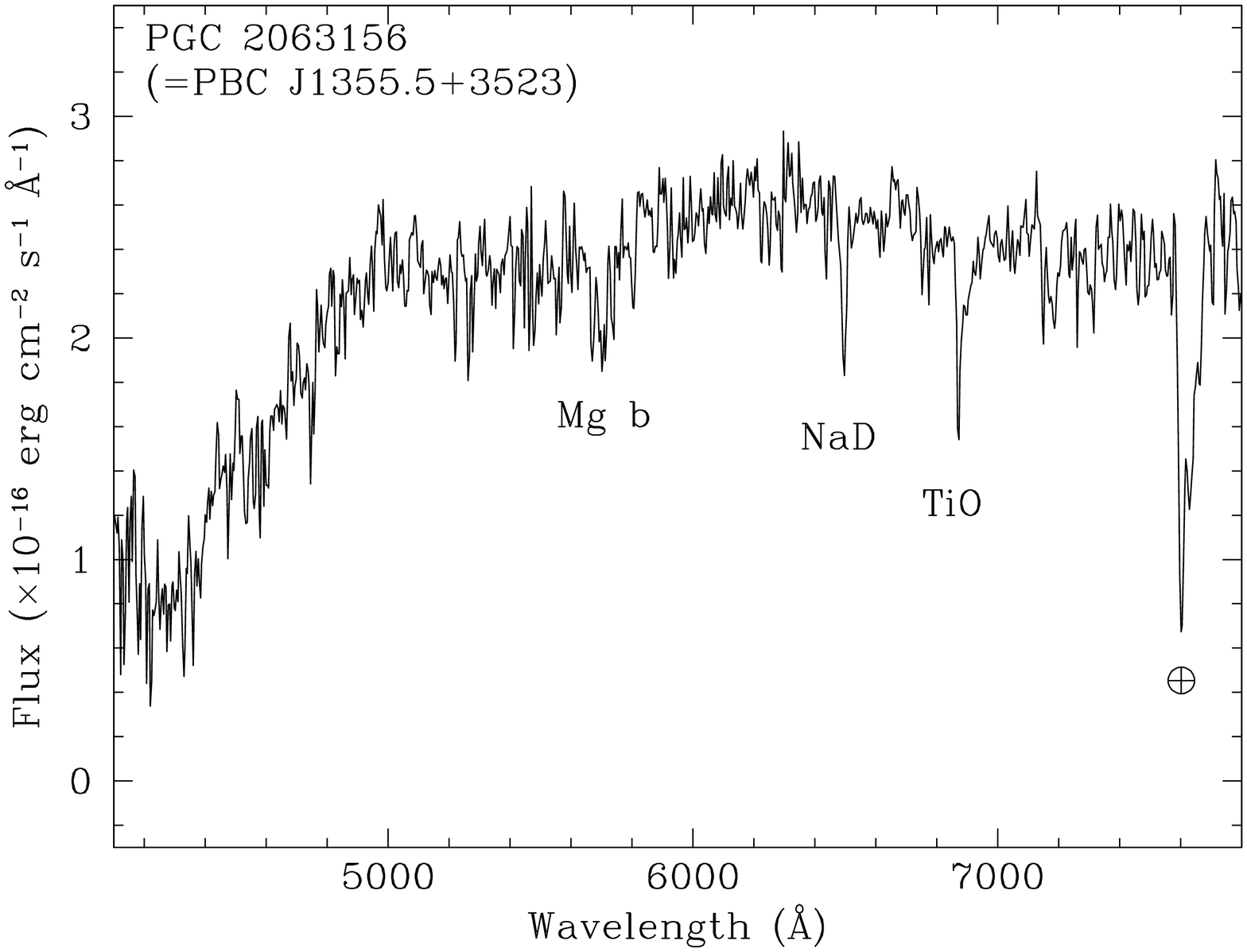,width=6.0cm,angle=270}}}
\caption{-- \emph{continued}}
\label{spectra3}
%\end{center}
\end{figure*}

\begin{figure*}
\setcounter{figure}{1}
%\begin{center}
\hspace{-.1cm}
\centering{\mbox{\psfig{file=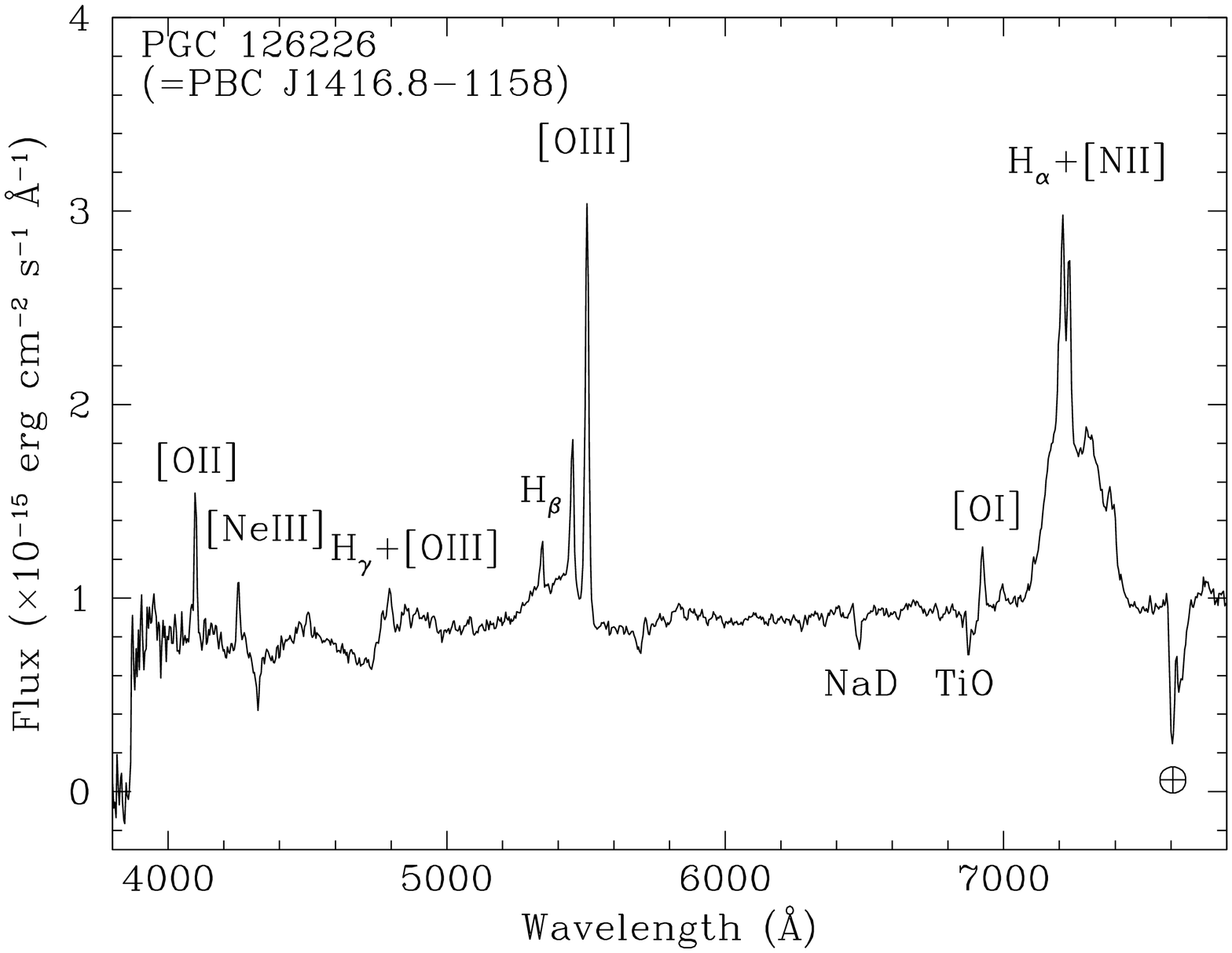,width=6.0cm,angle=270}}}
\centering{\mbox{\psfig{file=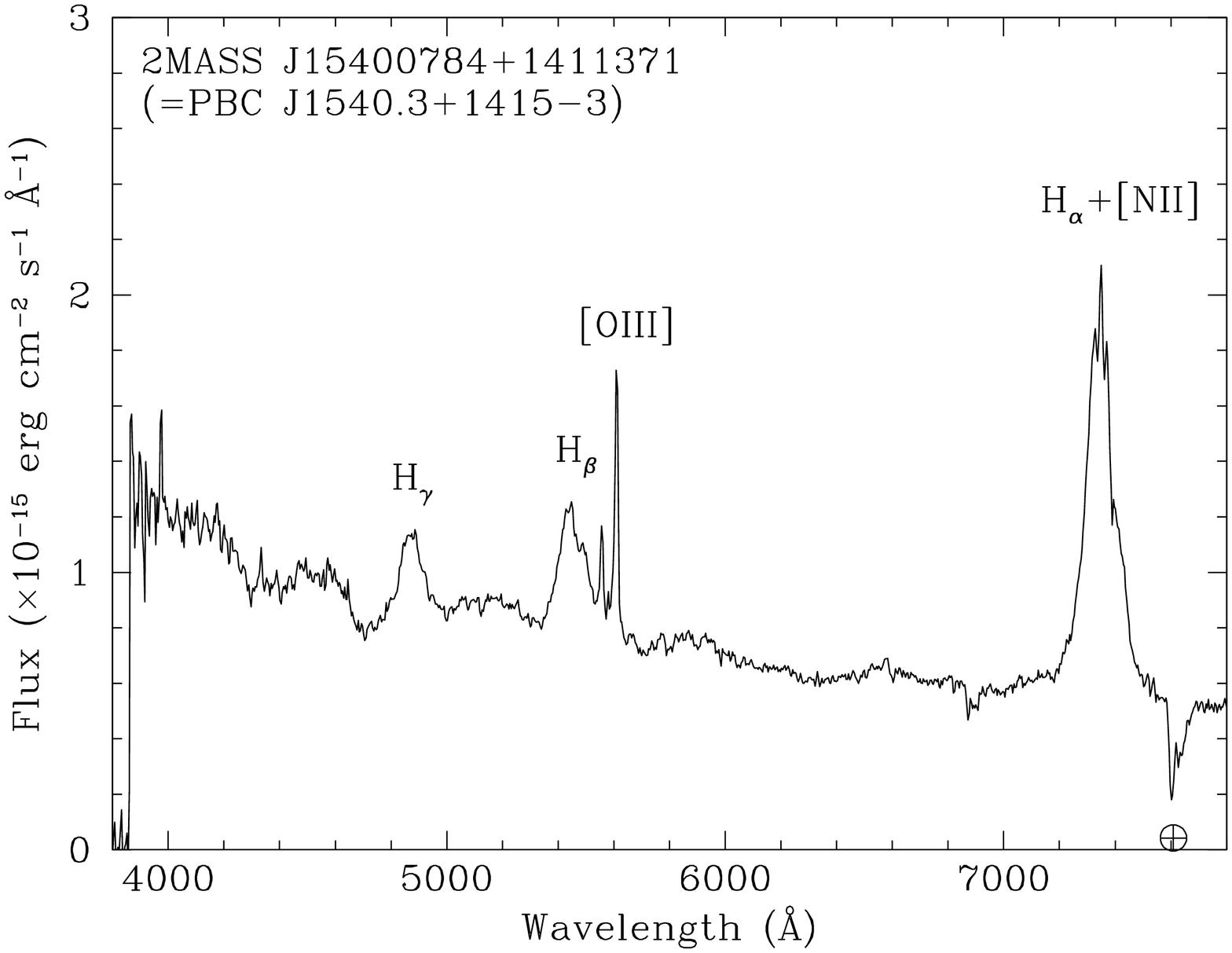,width=6.0cm,angle=270}}}
\centering{\mbox{\psfig{file=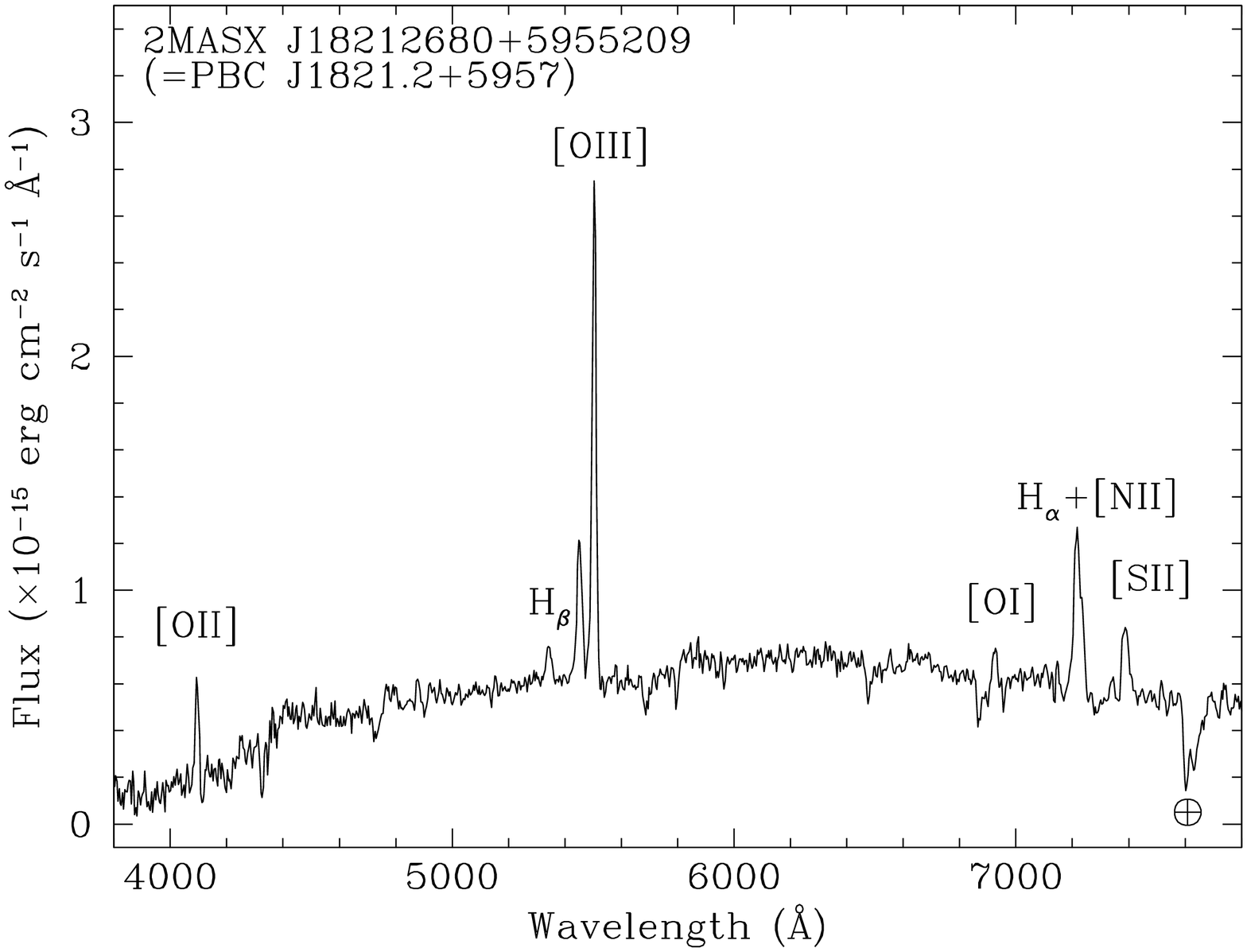,width=6.0cm,angle=270}}}
\centering{\mbox{\psfig{file=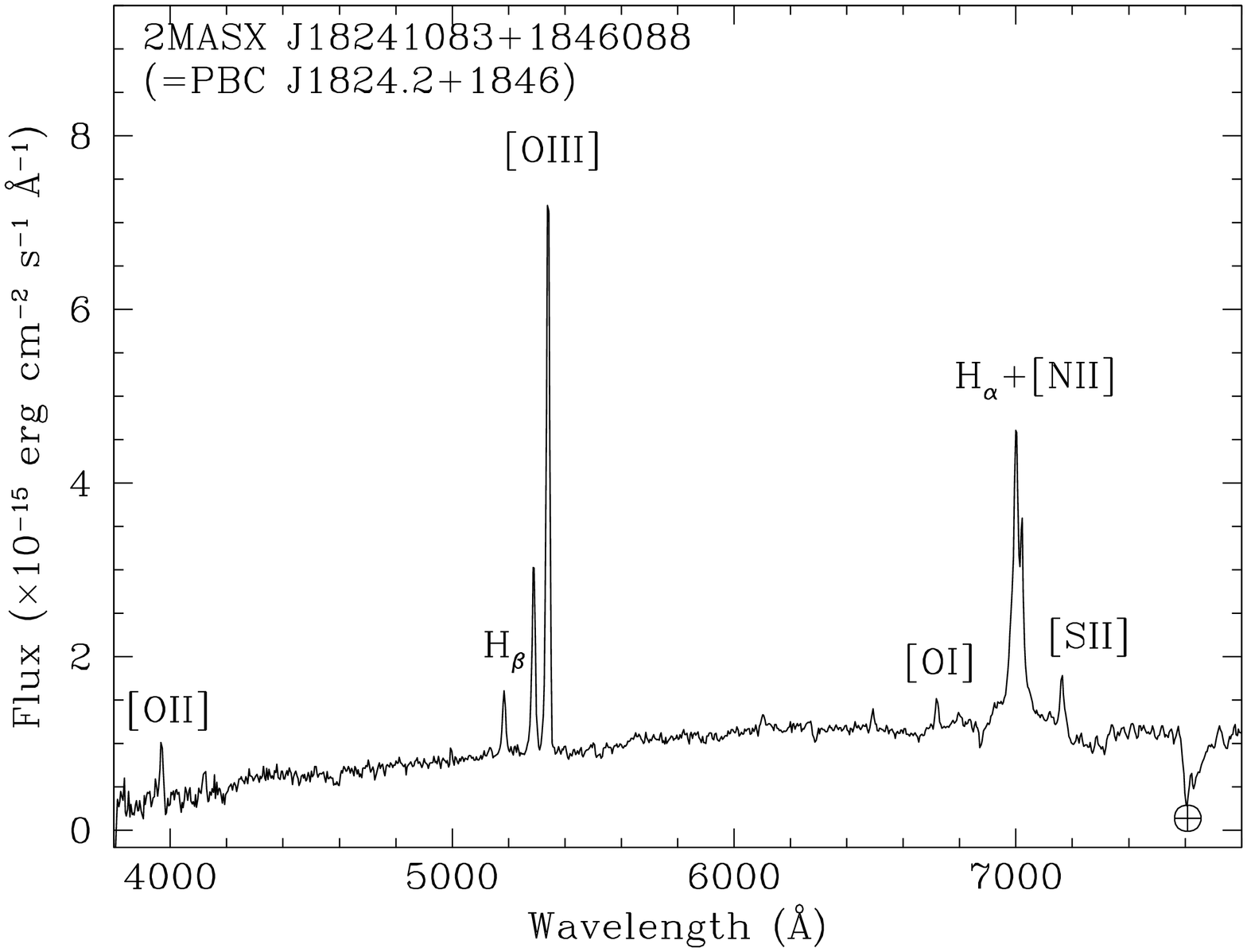,width=6.0cm,angle=270}}}
\centering{\mbox{\psfig{file=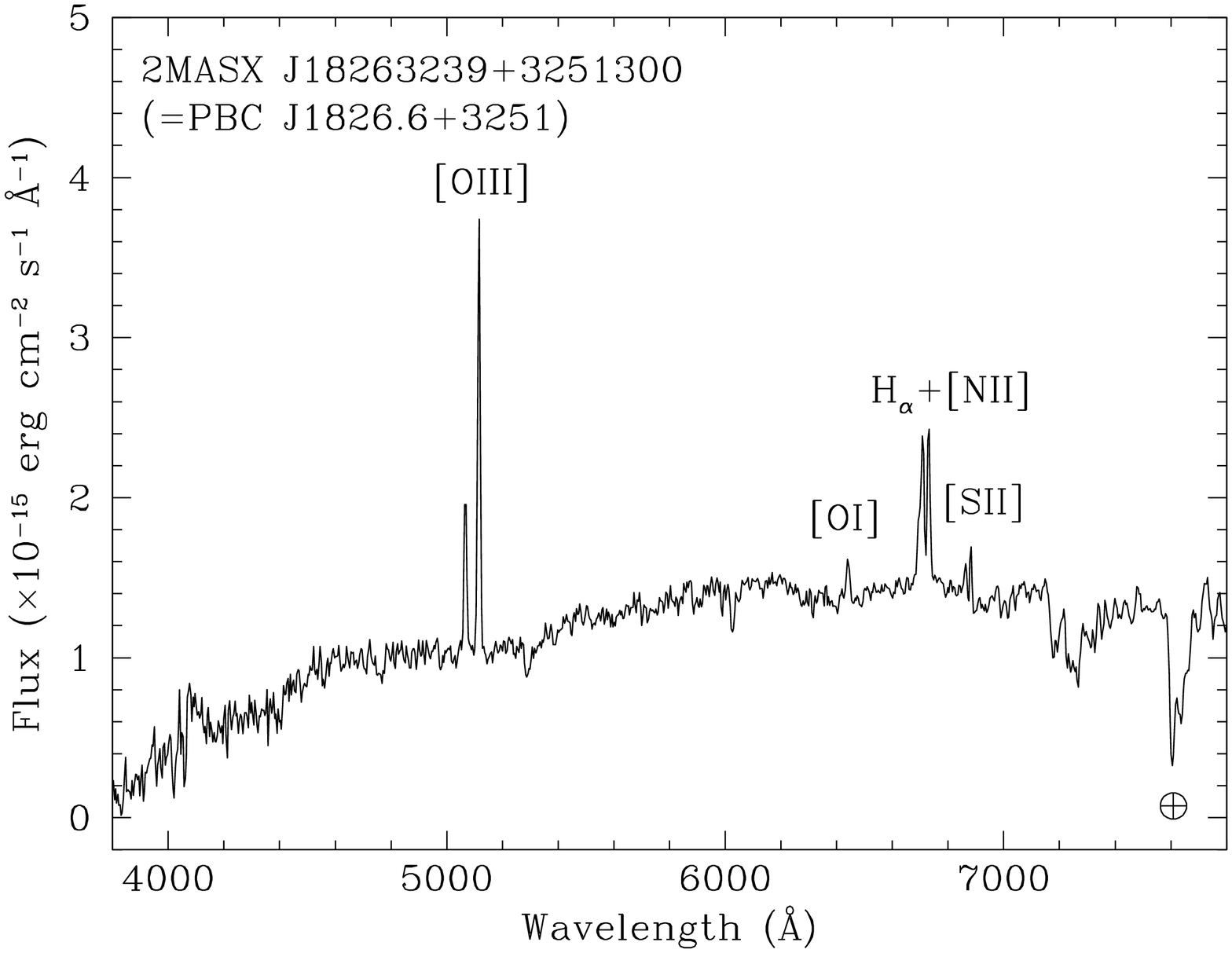,width=6.0cm,angle=270}}}
\centering{\mbox{\psfig{file=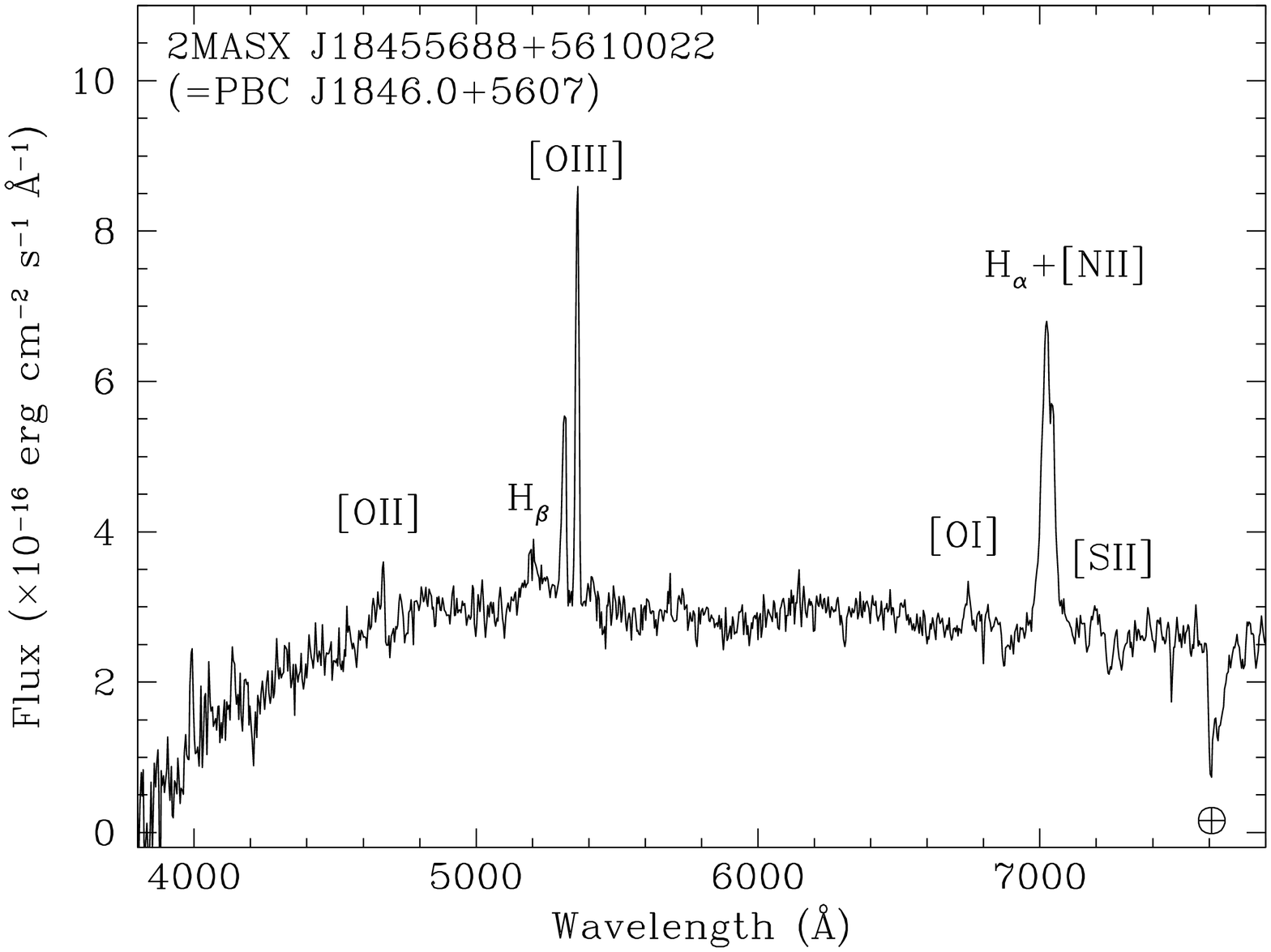,width=6.0cm,angle=270}}}
\centering{\mbox{\psfig{file=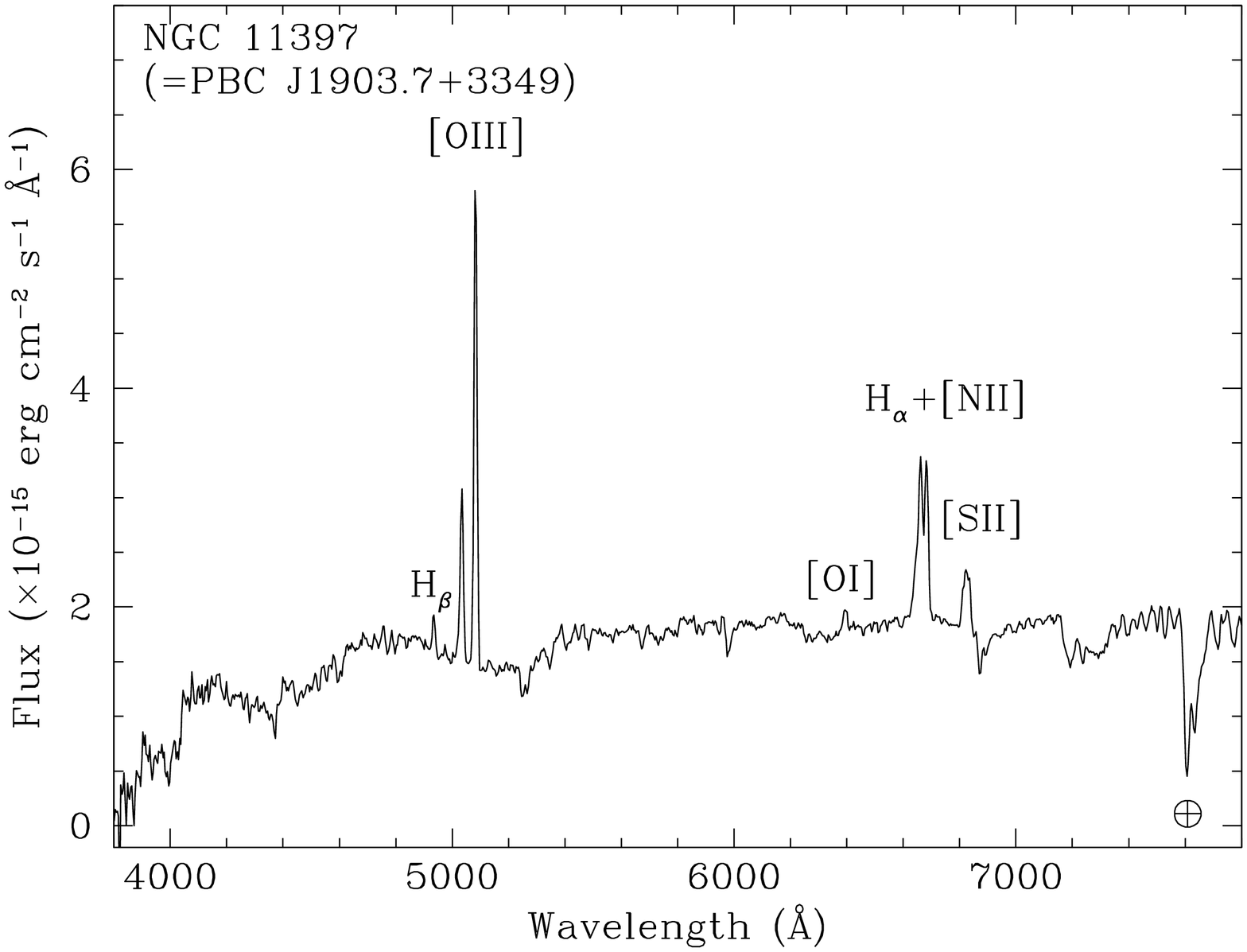,width=6.0cm,angle=270}}}
\centering{\mbox{\psfig{file=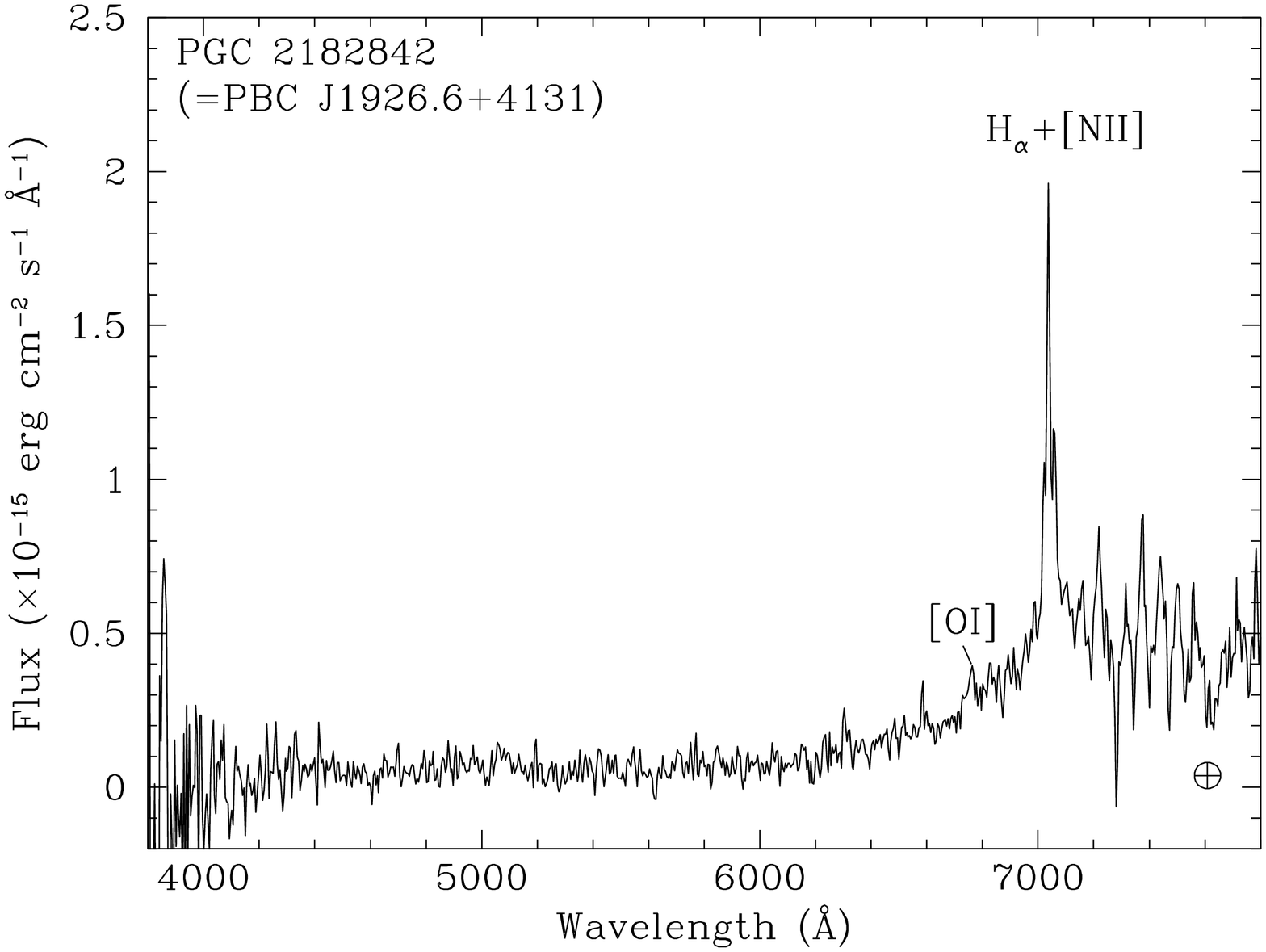,width=6.0cm,angle=270}}}
\centering{\mbox{\psfig{file=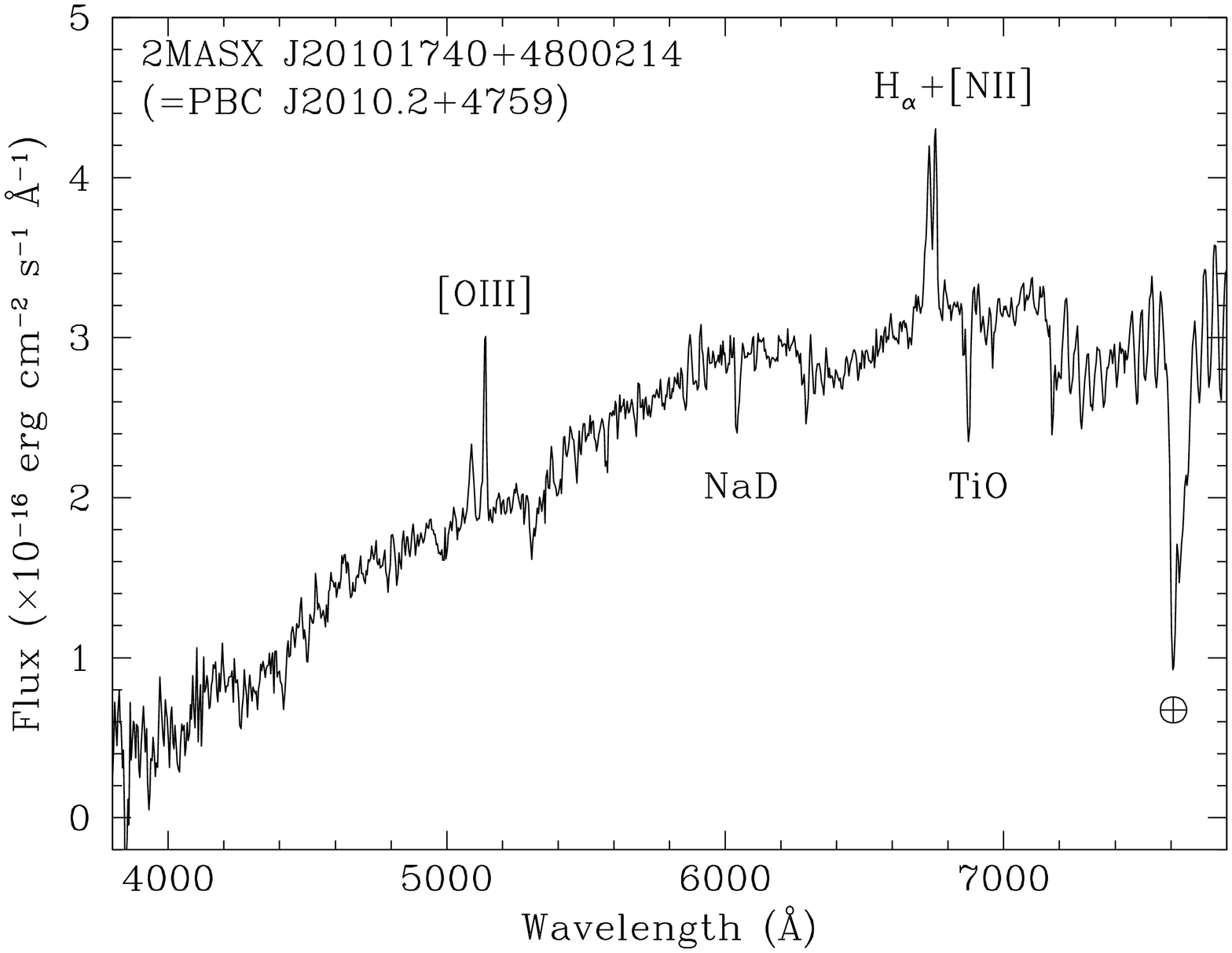,width=6.0cm,angle=270}}}
\centering{\mbox{\psfig{file=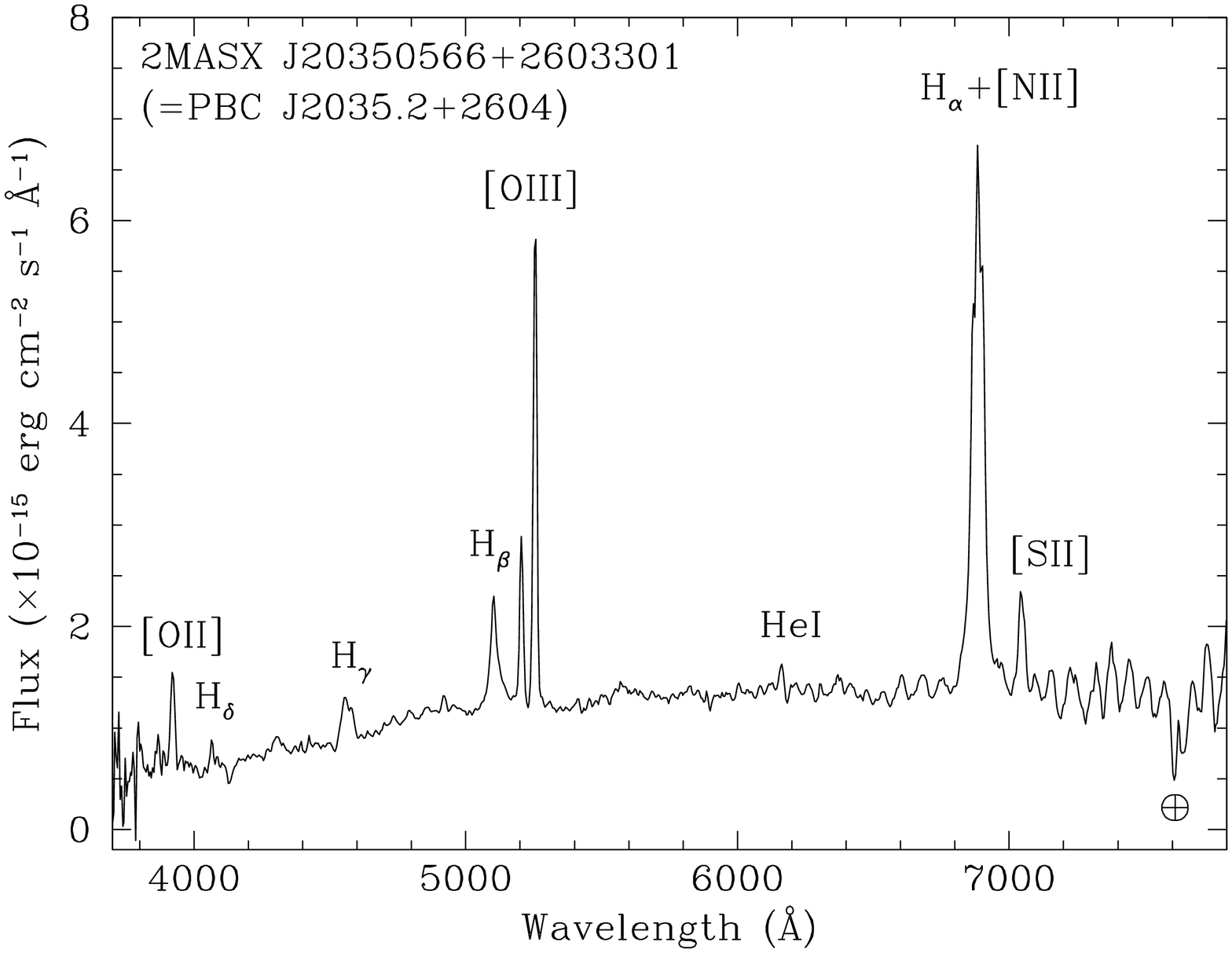,width=6.0cm,angle=270}}}
\centering{\mbox{\psfig{file=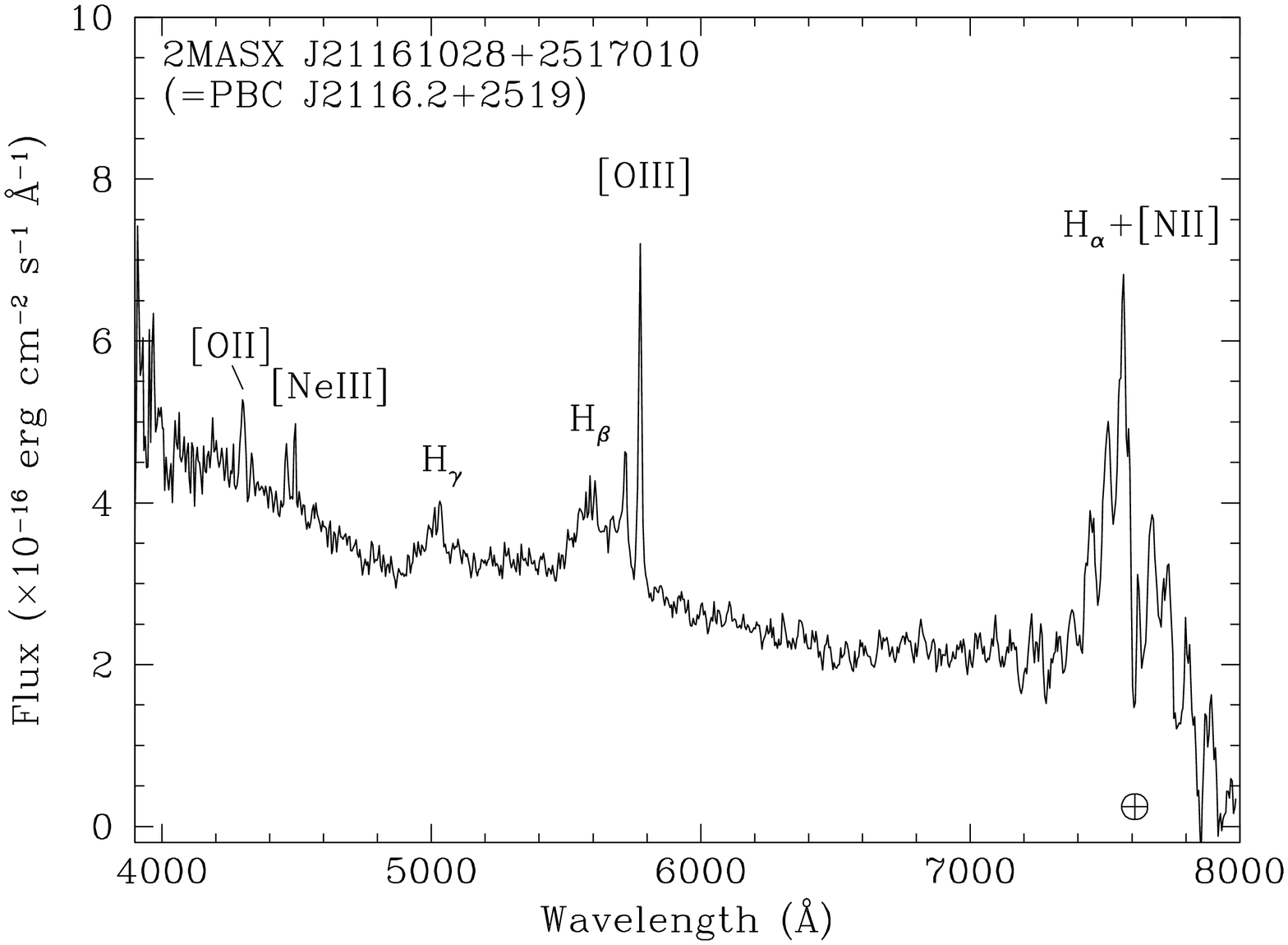,width=6.0cm,angle=270}}}
\centering{\mbox{\psfig{file=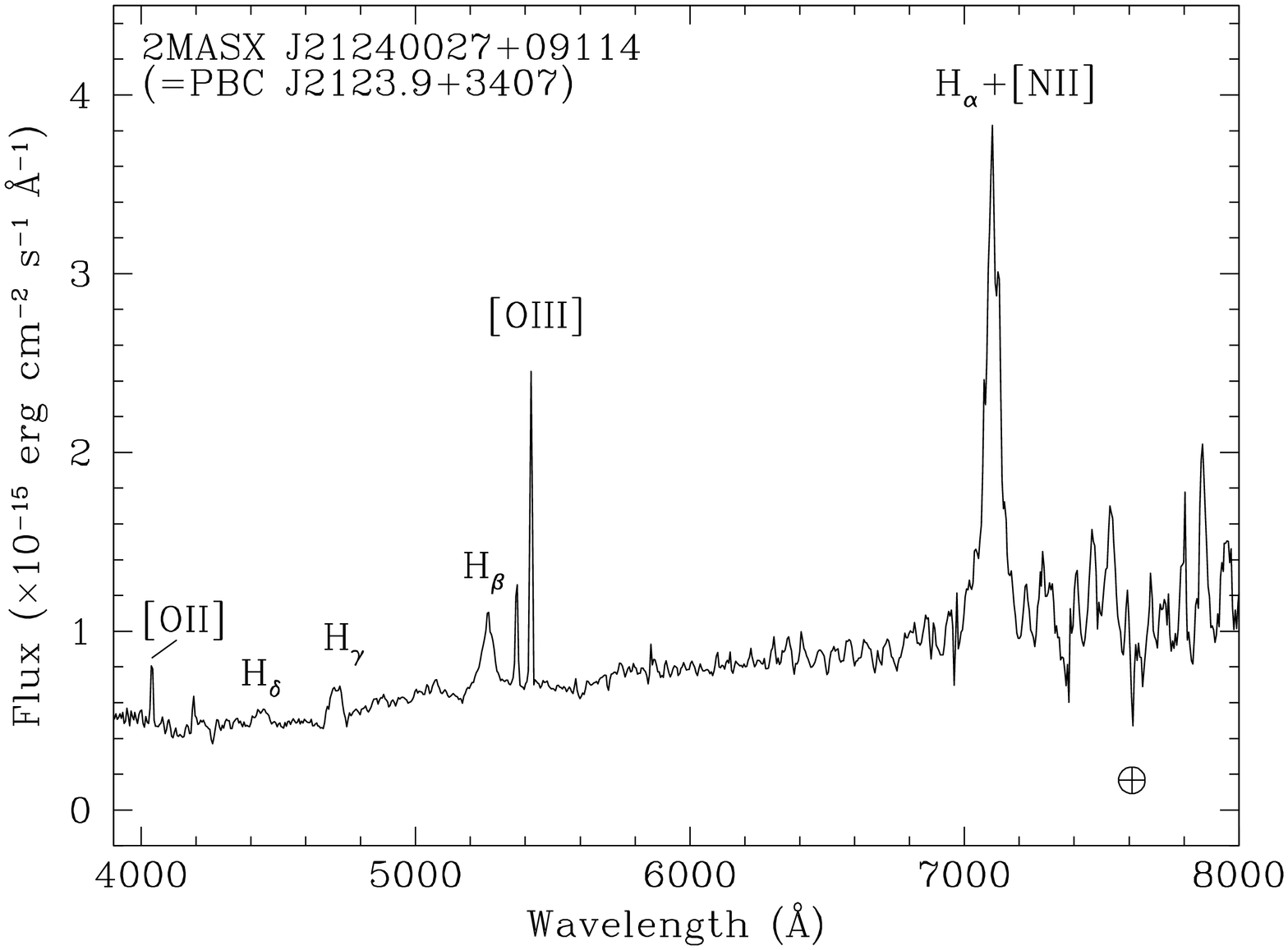,width=6.0cm,angle=270}}}
\centering{\mbox{\psfig{file=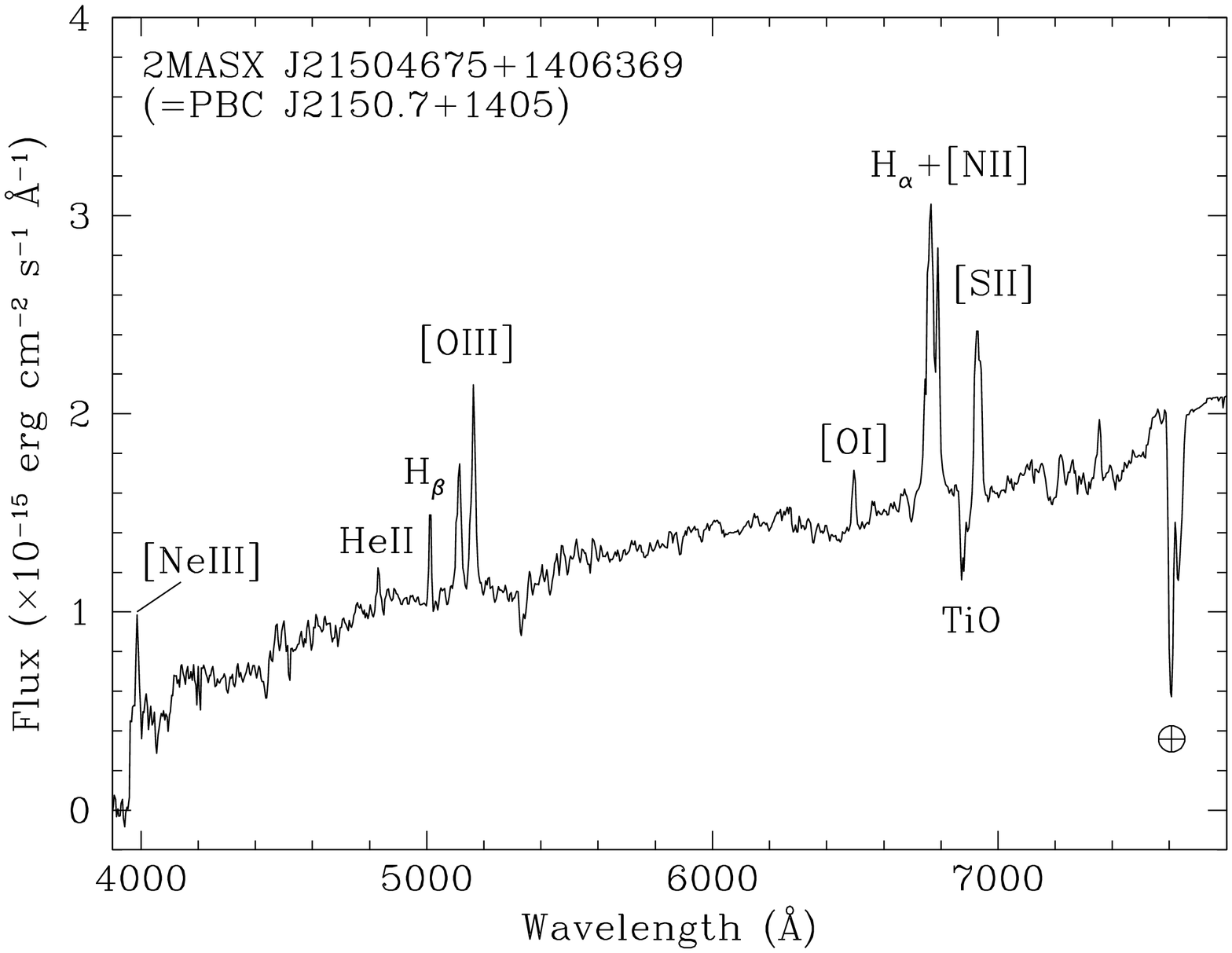,width=6.0cm,angle=270}}}
\centering{\mbox{\psfig{file=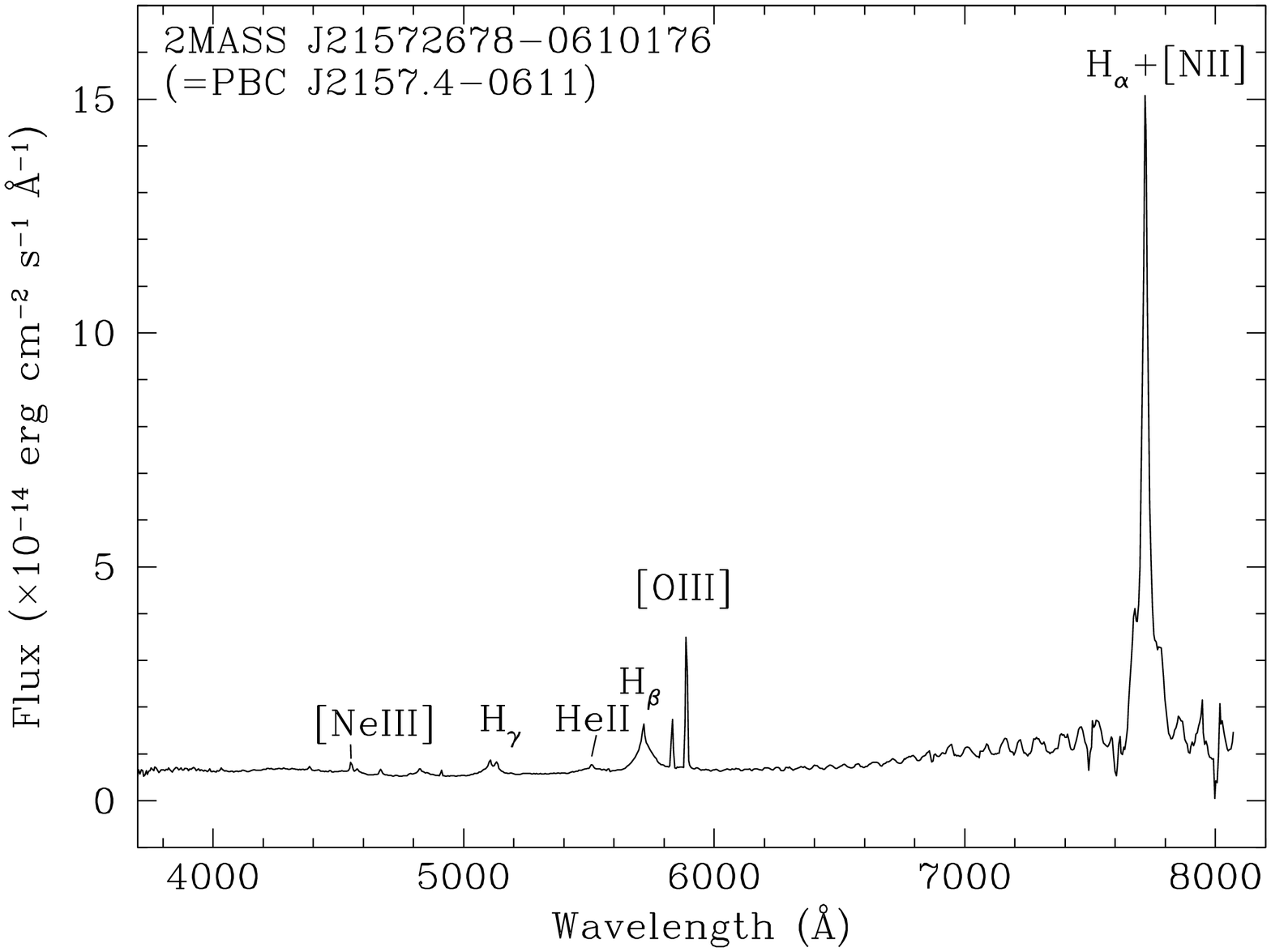,width=6.0cm,angle=270}}}
\centering{\mbox{\psfig{file=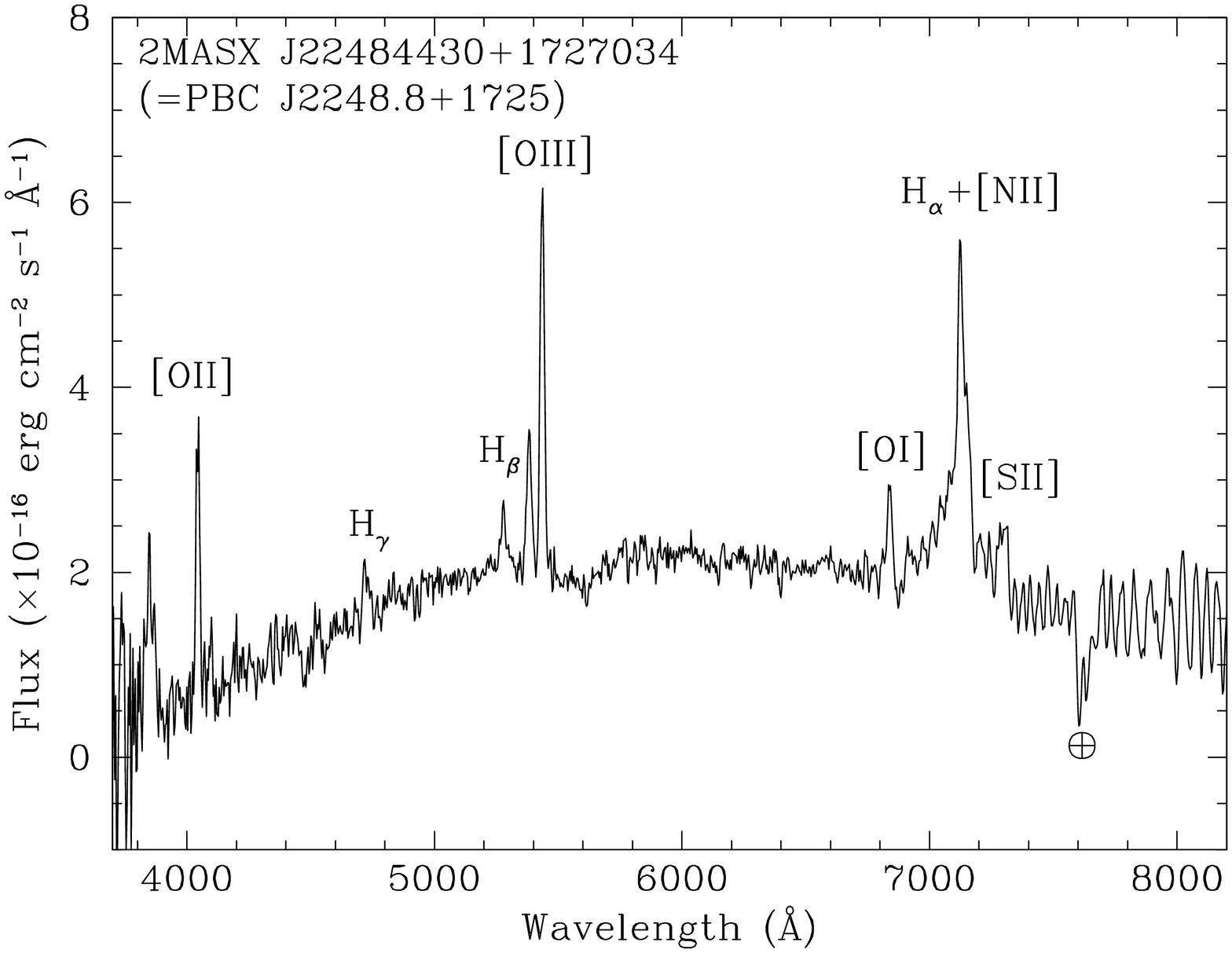,width=6.0cm,angle=270}}}
\caption{-- \emph{continued}}
\label{spectra4}
%\end{center}
\end{figure*}

\begin{figure*}
\setcounter{figure}{1}
%\begin{center}
\hspace{-.1cm}
\centering{\mbox{\psfig{file=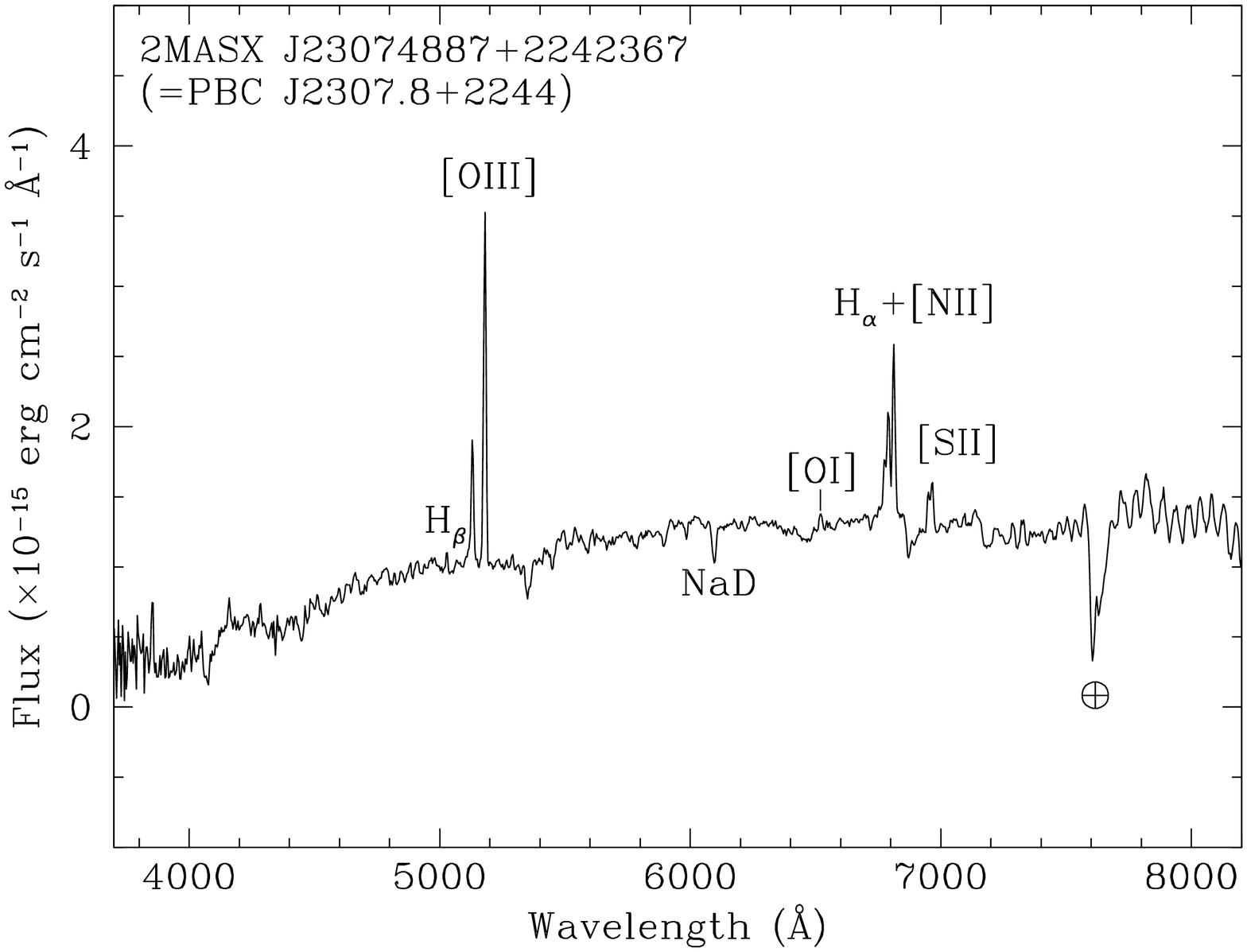,width=6.0cm,angle=270}}}
\centering{\mbox{\psfig{file=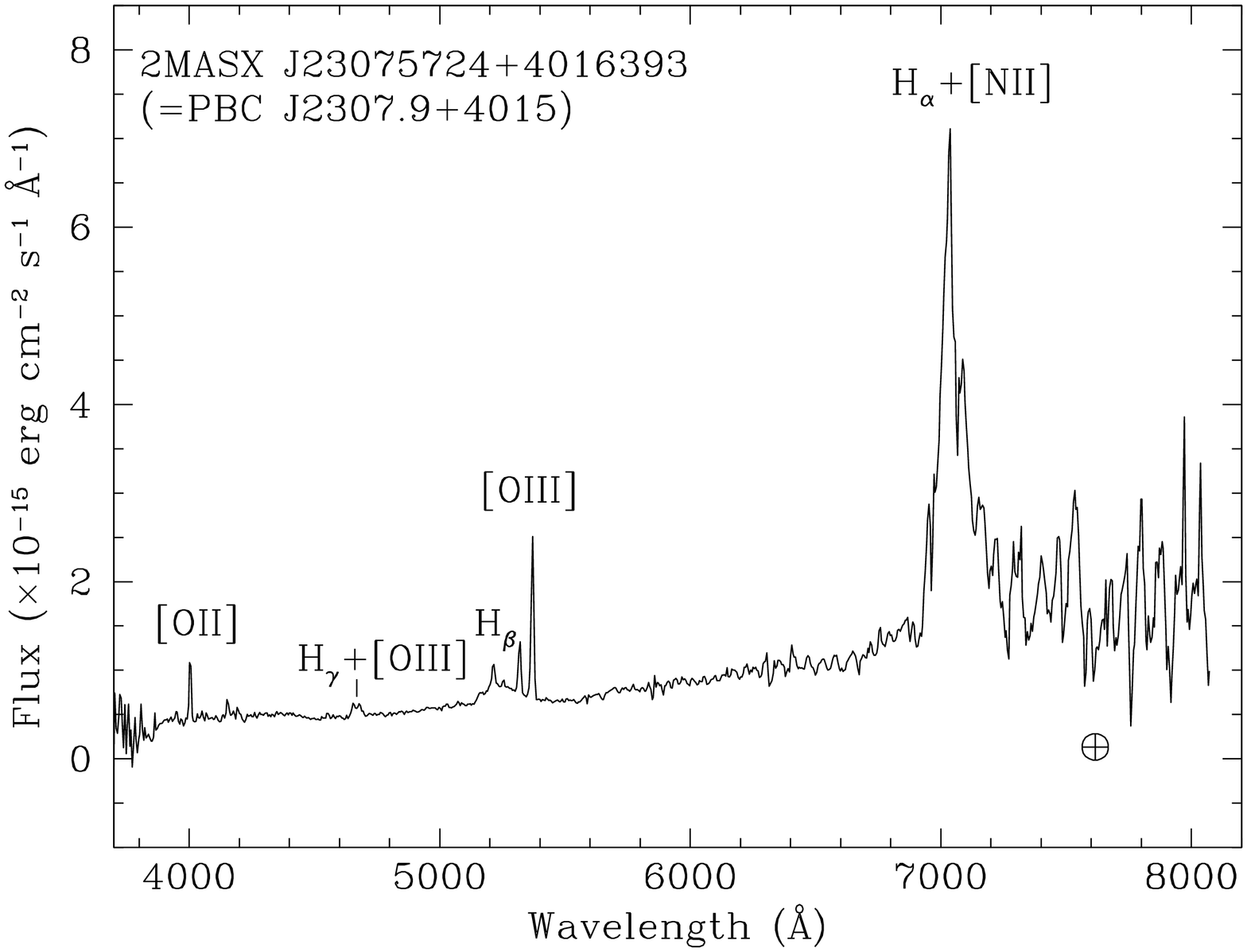,width=6.0cm,angle=270}}}
\centering{\mbox{\psfig{file=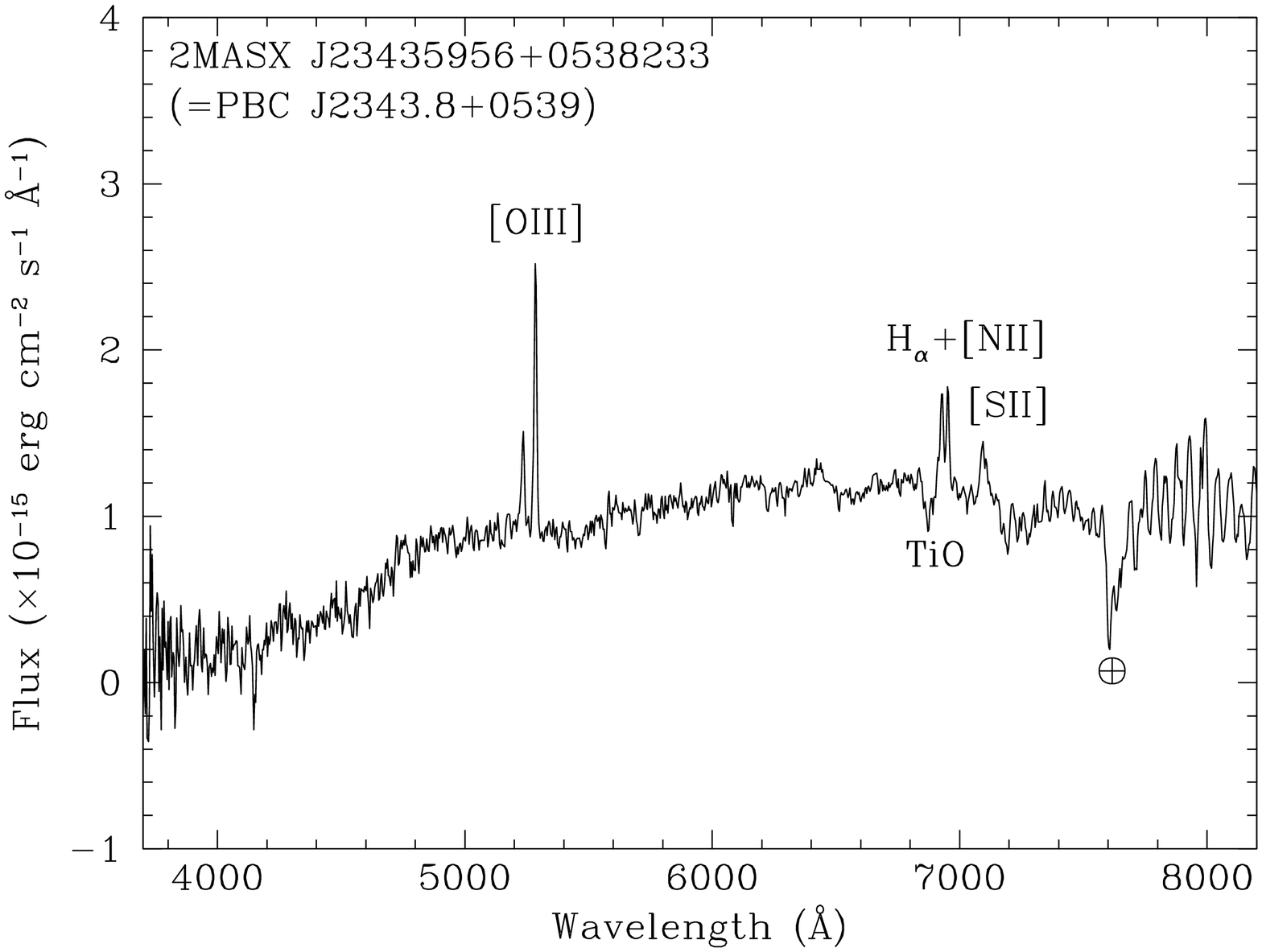,width=6.0cm,angle=270}}}
\caption{-- \emph{continued}}
\label{spectra5}
%\end{center}
\end{figure*}

\begin{table*}[h!]
\caption[]{Broad-line region gas velocities,  
central black hole masses and apparent Eddington ratios for 29 broad line AGNs discussed in this paper.}
\label{blr}
\begin{center}
%\resizebox{9.5cm}{!}{
\begin{tabular}{lcccc}
\noalign{\smallskip}
\hline
\hline
\noalign{\smallskip}
\multicolumn{1}{c}{Object} & $v_{\rm BLR}$ & $M_{\rm BH}$& L$_{15-150\; keV}$ & L$_{15-150}$/L$_{Edd}$ \\
\multicolumn{1}{c}{}& (km s$^{-1}$)&(10$^7$ $M_\odot$) &$\times$ 10$^{43}$ erg s$^{-1}$ & \\
\noalign{\smallskip}
\hline
\noalign{\smallskip}

PBC J0050.8+7648 	  & 3800 & 7.6 & 43 &0.04  \\
PBC J0149.3$-$5017 &4975 & 12 &	  1.7 & 0.001\\
PBC J0157.3+4715	& 2600  & 3.2 &	  4.8 &0.01 \\
PBC J0311.9+5029	& 4370  & 9.4 &  5.9 &0.005 \\
PBC J0429.7$-$6703& 3840  & 7.1 & 4.6 & 0.005\\
PBC J0440.8+2739	& 3300  & 5.3 &  2.8&0.004 \\
PBC J0532.7+1346  & 2000  & 1.9 &  1.1 & $<$0.005 \\
PBC J0535.6+4011	& 1600  & 1.2 &	  0.7 & 0.004\\
PBC J0543.9$-$4325& 1000  & 0.5 & 2.2 &0.4 \\
PBC J0609.4$-$6243& 2900  & 4.1 &  31 & 0.06\\
PBC J0635.0$-$7441& 3800  & 7.6 & 15 &0.02 \\
PBC J0654.5+0703	& 1700 & 1.5 &   0.9 &0.005 \\
PBC J0749.2$-$8634& 5500  & 15 &  14&0.008 \\
PBC J0803.4+0840	& 860  & 0.4 &    4.3 & 0.09\\
PBC J0818.5$-$1420& 9400  & 43 &  21&0.004 \\
PBC J0942.1+2342	& 4000 & 7.7 &   0.8 &$\sim$0.001 \\
PBC J1002.3+0304	& 750 & 0.3 &     0.9&0.02 \\
PBC J1020.5$-$0235& 11900 & 69 &  5.7 &$\sim$0.001 \\
PBC J1254.8$-$2655& 4460 & 9.7 &  10 &0.008 \\
PBC J1349.0+4443	& 14500 & 10 &  9.1 &0.007 \\
PBC J1416.8$-$1158& 12400 & 58 &  32& 0.004\\
PBC J1540.3+1415-3& 4500 & 9.9 &   35 &0.03 \\
PBC J1846.0+5607	& 2800 & 3.6 &   11&0.02\\
PBC J2035.2+2604	& 1600 & 1.2 &   7.4&0.05\\
PBC J2116.2+2519	& 11700 & 69&    28&0.003 \\
PBC J2123.9+3407	& 2500 & 3&       17 &0.05 \\
PBC J2157.4$-$0611& 3000 & 6.2 &   52&0.07\\
PBC J2248.8+1725	& 1000 & 0.5&    15&0.2 \\
PBC J2307.9+4015	& 5300 & 14 &    12&0.007\\
\noalign{\smallskip} 
\hline
\hline
\multicolumn{5}{l}{Note: the final uncertainties on the black hole mass estimates are about 50\% of their}\\
\multicolumn{5}{l}{values. The velocities were determined using H$_{\beta}$ emission line, whereas the}\\
\multicolumn{5}{l}{apparent Eddington ratios were computed using the 15-150 keV luminosities.}\\
\end{tabular}
\end{center}
\end{table*}

\end{document}